\documentclass[aps,prd,groupaddress,floatfix,tighten,nofootinbib,showpacs,amsfonts,superscriptaddress]{revtex4}
\usepackage{url}
\usepackage{xcolor}
\usepackage{hyperref}
\usepackage{amssymb}
\usepackage{amsmath,color}
\usepackage{epsfig}
\usepackage{graphicx,epsf,epsfig}
\usepackage{footnote}
\usepackage{multirow}
\usepackage{graphicx}
\usepackage{subfigure}
\usepackage{lipsum}
\usepackage{multirow} 
\usepackage{tabulary}
\usepackage{rotating}
\usepackage{array}
\usepackage[normalem]{ulem}
\usepackage{color}
\colorlet{shadecolor}{gray!15}
\definecolor{greenLinks}{rgb}{0,0.6,0}
\definecolor{blueLinks}{rgb}{0,0,0.6}
\definecolor{redLinks}{rgb}{0.6,0,0}
\definecolor{tempText}{rgb}{0.55,0.10,0.67}
\definecolor{eprintLinks}{rgb}{0.4,0.4,0.4}
\definecolor{journalLinks}{rgb}{0.6,0,0}
\newcommand{\MYhref}[3][redLinks]{\href{#2}{\color{#1}{#3}}}%

\newcommand{\red}{\color{red}}
\def\slc#1{
 \setbox0=\hbox{$#1$}                      
 \dimen0=\wd0                              
 \setbox1=\hbox{/} \dimen1=\wd1            
 \ifdim\dimen0>\dimen1                     
 \rlap{\hbox to \dimen0{\hfil/\hfil}}      
 #1                                        
 \else                                     
 \rlap{\hbox to \dimen1{\hfil$#1$\hfil}}   
 /                                         
\fi}

\def\be{\begin{equation}}
\def\ee{\end{equation}}
\def\gs{\mathrel{\rlap{\raise 0.511ex \hbox{$>$}}{\lower 0.511ex \hbox{$\sim$}}}}
\def\ls{\mathrel{\rlap{\raise 0.511ex \hbox{$<$}}{\lower 0.511ex \hbox{$\sim$}}}}

\newcommand{\ba}{\begin{array}{c}}
\newcommand{\baz}{\begin{array}{cc}}
\newcommand{\barrr}{\begin{array}{rrr}}
\newcommand{\bad}{\begin{array}{ccc}}
\newcommand{\bav}{\begin{array}{cccc}}
\newcommand{\baf}{\begin{array}{ccccc}}
\newcommand{\bea}{\begin{equation} \begin{array}{c}}
\newcommand{\eea}{ \end{array} \end{equation}}
\newcommand{\ea}{\end{array}}

\usepackage[normalem]{ulem}

\def\21{$\mathrm{SU(2)_L \otimes U(1)_Y}$ }

\newcommand{\ignore}[1]{}

\allowdisplaybreaks \allowdisplaybreaks[2]
 \newcommand{\AddrFCFMBUAP}{Facultad de Ciencias F\'{\i}sico Matem\'aticas, 
  Benem\'erita Universidad Aut\'onoma de Puebla,\\
  Apdo. Postal 1152, Puebla, Pue.  72000, M\'exico.}
 \newcommand{\AddrFCEBUAP}{Facultad de Ciencias de la Electr\'onica, 
  Benem\'erita Universidad Aut\'onoma de Puebla,\\
  Apdo. Postal 542, Puebla, Pue. 72000, M\'exico.}
 \newcommand{\AddrCIFFU}{Centro Internacional de F\'{\i}sica Fundamental (CIFFU), 
  Benem\'erita Universidad Aut\'onoma de Puebla.}
\begin{document}
%
\title{On the lepton CP violation in a $\nu$2HDM with flavor}
%
%
\author{E. Barradas-Guevara}
 \email{barradas@fcfm.buap.mx}
 \affiliation{\AddrFCFMBUAP}
 \affiliation{\AddrCIFFU}
%
%
\author{O. F\'elix-Beltr\'an}
 \email{olga.felix@correo.buap.mx}
 \affiliation{\AddrFCEBUAP}
 \affiliation{\AddrCIFFU}
%
%
\author{F. Gonzalez-Canales}
 \email{felixfcoglz@gmail.com}
 \affiliation{\AddrFCEBUAP}
 \affiliation{\AddrCIFFU}
%
%
\author{M. Zeleny-Mora}
\email{moiseszeleny@gmail.com} 
\affiliation{\AddrFCFMBUAP}
\affiliation{\AddrCIFFU}

%
%
\date{\today}
%
\pacs{14.60.Pq, 12.15.Ff, 12.60.Fr, 12.60.-i, 11.30.Hv}
%
\begin{abstract} 
In this work we propose an extension to the Standard Model in which we consider the 
model 2HDM type-III plus massive neutrinos and the horizontal flavor symmetry 
$S_{3}$ $(\nu$2HDM$\otimes S_3)$.
In the above framework and with the explicit breaking of flavor symmetry $S_{3}$, 
the Yukawa matrices in the flavor adapted basis are represented by means of a 
matrix with two texture zeroes. 
Also, the active neutrinos are considered as 
Majorana particles and their masses are generated through type-I seesaw mechanism. 
The unitary matrices that diagonalize the mass matrices, as well as the flavor 
mixing matrices, are expressed in terms of fermion mass ratios. 
Consequently, in the mass basis the entries of the Yukawa matrices naturally acquire the form 
of the so-called {\it Cheng-Sher ansatz}. 
For the leptonic sector of $\nu$2HDM$\otimes S_3$, we compare, through a $\chi^{2}$ 
likelihood test, the theoretical expressions of the flavor mixing angles with the masses and 
flavor mixing leptons current experimental data. 
The results obtained in this $\chi^{2}$ analysis are in very good agreement with 
the current experimental data. 
We also obtained an allowed value ranges for the ``Dirac-like'' phase factor, as well as for 
the two Majorana phase factors. 
Furthermore, we study the phenomenological implications of these numerical values of the 
CP-violation phases on  the neutrinoless double beta decay, and for Long Base-Line neutrino 
oscillation experiments such as T2K, NO$\nu$A, and DUNE.
\end{abstract}
%
%
%
\begin{flushright}
CIFFU-17-02
\end{flushright}
%
%
\maketitle
%
%
%
\newpage
%
%
\section{Introduction}\label{sec:introduction}
According to the most recent neutrino physics 
literature~\cite{King:2017guk,Capozzi:2016rtj}, there are several issues still unresolved.  
Among others: whether the neutrinos are Dirac or Marjorana fermions; the absolute neutrino 
mass scale; the posible sources of Charge-Parity [CP] violation [CPV] in leptons.  
Nowadays, answers to these questions are being searched by means of the experimental 
results concerning KamLAND reactor 
neutrinos~\cite{Decowski:2016axc,Gando:2010aa,Gando:2013nba}, 
in each one of the current high-statistics Short Base-Line (SBL) reactor neutrino 
experiments RENO~\cite{Seo:2014xei,RENO:2015ksa}, 
Double Chooz~\cite{Abe:2014bwa} and Daya Bay~\cite{An:2015nua}.
Also, one of the most interesting effects related to the neutrino oscillation in matter is, 
that these periodic transformations of neutrinos from one flavor to another, can induce a 
fake CP violating effect. 
Therefore, the Long Base-Line (LBL) neutrino oscillation experiments are good candidates for 
determining the ``Dirac-like'' CP violation phase 
as well as resolving the mass hierarchy problem~\cite{Feldman:2013vca}.
The recent measures reported by T2K~\cite{Abe:2017vif,Batkiewicz:2017xoh,Abe:2017uxa,Abe:2015awa}, 
NO$\nu$A~\cite{Kolupaeva:2016fxw,Bian:2015opa} and Super-Kamiokande experiments~\cite{Cresst} 
suggest a nearly maximal CP violation. 
In these experiments the ``Dirac-like'' CP phase takes the value 
$\delta_{\text{CP}} \simeq 3 \pi /2$, with a statistical significance below 3$\sigma$ level. 
Moreover, the data obtained in the global fits of neutrino oscillations agree with a nonzero 
$\delta_{CP}$ phase, whereby the previous value is confirmed~\cite{Forero:2014bxa,Gonzalez-Garcia:2015qrr,Capozzi:2016rtj,Esteban:2016qun,Ge:2017qqv}.

The flavor and mass generation are two concepts strongly intertwined. 
In order to know the flavor dynamic in models beyond the Standard Model (SM), we 
need to understand the mechanism of flavor and mass generation arising in the 
standard theory. 
In the later, the Yukawa matrices are of great interest because its eigenvalues define the 
fermion masses. 
Moreover, for multi-Higgs models the Flavor Changing Neutral Current (FCNC) arises naturally 
from  impediment to diagonalize simultaneously the mass and Yukawa matrices.
In particular, models like the Two Higgs Doublet Model type-III (2HDM-III), in which the 
two Higgs doublets are coupled to all fermions, allow the presence of FCNC at tree 
level mediated by Higgs~\cite{Atwood:1996vj,DiazCruz:2009ek,DiazCruz:2004pj,Krawczyk:2007ne,BarradasGuevara:2010xs,Crivellin:2013wna}. 
The 2HDM predicts three neutral states $H^{0}_{1,2,3}$ and a pair of charged states 
denoted as $H^{\pm}_{1,2}$~\cite{Krawczyk:2005zy,Branco:2011iw}.
The Higgs-fermions couplings ($H^0 \bar{f}f$) in the 2HDM are given 
as~\cite{Atwood:1996vj,DiazCruz:2009ek,Krawczyk:2007ne,BarradasGuevara:2010xs,Crivellin:2013wna}:
\begin{equation}
 {\cal L} _Y = \bar{f}_i \left( S_{ij} + \gamma^5 P_{ij} \right) f_j H^0_a, 
 \qquad (a=1,2,3).
\end{equation}
In 2HDM-III the FCNC's are kept under control by imposing some texture zeroes in the Yukawa 
matrices. This fact reproduce the observed fermion masses and mixing 
angles~\cite{Deppisch:2012vj}. 
Using texture shapes allow a direct relation between the Yukawa matrix entries and the 
parameters used to compute the decay widths and cross section, without losing the terms 
proportional to the light fermion masses.
Specifically, considering a zero texture Yukawa matrix, one obtains the 
{\it Cheng-Sher ansatz} for flavor mix couplings, widely used in literature, where 
flavored couplings are considered proportional to the involved fermion 
masses~\cite{Dorsner:2002wi}.

The matter content in the 2HDM is divided among the quarks and leptons sectors. 
In turn these sectors are subdivided in two sectors, the up- and down-type for quarks sector, 
while charged leptons and neutrinos for the leptons sector.
The fermions in each one of these subsectors are analogous each other, because they have 
completely identical couplings to all gauge bosons, although their mass values are not 
the same. 
Therefore, before of the spontaneous symmetry breaking (SSB), the Yukawa Lagrangian in 
the above subsectors is invariant under permutations of flavor indices. 
In other words, each one of these subsectors is invariant under the action of a  
$S_{3}$ symmetry group. 
This symmetry group has only three irreducible representations that 
correspond to two singlets and a doublet~\cite{Georgi:1999wka,Ishimori:2012zz}. 

On the other hand, obtained from the experimental data, the mass spectrum for Dirac fermions 
obeys the following strong hierarchy~\cite{Olive:2016xmw}:
\begin{equation}
 \begin{array}{cc}
  \widehat{m}_{e} \sim 10^{-6} , \;\; \widehat{m}_{\mu} \sim 10^{-3}, &
  \quad \widehat{m}_{\tau} \sim 1 , \\
  \widehat{m}_{u} \sim 10^{-5} , \;\; \widehat{m}_{c} \sim 10^{-3}, &
  \quad \widehat{m}_{t} \sim 1 , \\
  \underbrace{ \widehat{m}_{d} \sim 10^{-3} , \;\; \widehat{m}_{s} \sim 10^{-2},   
   }_{\bf 2} &
  \quad \underbrace{ \widehat{m}_{b} \sim 1 . }_{\bf 1}
 \end{array}
\end{equation}
In the above expression $\widehat{m}_{l} = m_{l}/m_{\tau}$ ($l=e,\mu, \tau$) stands for charged leptons, $\widehat{m}_{U}= m_{U}/m_{t}$ ($U = u,c,t$) for up-type quarks, 
while $\widehat{m}_{D} = m_{D}/m_{b}$ ($D = d,s,b$) for down-type quarks. 
The behavior of these mass ratios in terms of irreducible representations of some symmetry 
group, can be interpreted as follows:  
the two lighter particles are associated with a doublet representation ${\bf 2}$, 
whereas for the heaviest particle it is assigned a singlet representation ${\bf 1}$. 
The smallest non-abelian group with irreducible representations of singlet and 
doublet, is the group of permutations of three objects. 
Hence, we expect the hierarchical nature of the Dirac fermion mass matrices to have 
its origin in the representation structure ${\bf 3} = {\bf 2} \oplus {\bf 1}$ of $S_{3}$. 
In the theoretical framework of SM, as well as for 2HDM, the neutrinos are massless 
particles, fact that is in disagree with the results obtained in the neutrino oscillation 
experiments.

Therefore, in this work we will study the flavor dynamics through Yukawa matrices in the 
specific scenario of 2HDM-III plus massive neutrinos and a horizontal flavor symmetry 
$S_3$ $(\nu$2HDM$\otimes S_3)$. 
In this context, under the action of $S_{3}$ flavor symmetry group the right-handed neutrinos 
as well as the two Higgs fields transform as singlets, 
while the active neutrinos are considered as Majorana particles and their masses are generated through type-I seesaw mechanism.
Hence, it is necessary to consider the following hybrid mass 
term, which involves the Dirac and Majorana neutrino mass terms~\cite{Xing:2010zzf}, 
\begin{equation}
 {\cal L}_{\mathrm{M+D}} = 
   - \frac{1}{2} \bar{ \eta }_{L} \, {\bf M}_{\mathrm{M+D}} \, (\eta_{L})^{c} + 
  \textrm{h. c.},
\end{equation}
where $\eta = \left(\, \nu_{L}, \, (N_{R})^{c} \, \right)^{\top} $ and
\begin{equation}
 {\bf M}_{\mathrm{M+D}}  = 
 \left( \begin{array}{cc}
  {\bf 0} & {\bf M}_{\nu_{D}} \\
  {\bf M}_{\nu_{D}}^{\top} & {\bf M}_{R} 
 \end{array}  \right).
\end{equation}
In the above expression ${\bf M}_{\nu_{D}}$ and ${\bf M}_{R}$ are the Dirac and right-handed 
neutrino mass matrix, respectively. 
In the special limit  ${\bf M}_{R} \gg {\bf M}_{\nu_{D}}$, the effective mass 
matrix of left-handed neutrinos is given by the type-I seesaw mechanism whose expression 
is\footnote{Analysis of heavy Majorana neutrinos implications in LHC is dealt in 
Refs.~\cite{Das:2012ze,Das:2017nvm}.}:
\begin{equation}\label{eq:seesaw}
 {\bf M}_{ \nu } = {\bf M}_{\nu_{D}} {\bf M}_{R}^{-1} {\bf M}_{\nu_{D}}^{\top}.
\end{equation}
If the fermion mass matrices do not have any element equal to zero, on one hand, the mass 
matrix of active neutrinos has twelve free parameters, since ${\bf M}_{\nu}$ is a complex 
symmetric matrix because this matrix comes from a Majorana mass term.  
On the other hand, the Dirac fermion mass matrices do not have any special 
feature, {\it i.e.}, these matrices are not Hermitian, nor symmetric. This is mainly due to 
the fact that the Yukawa matrices are represented through a $3 \times 3$ complex 
matrix. Hence, for Dirac mass matrices we have eighteen free parameters.

After the explicit sequential breaking of flavor symmetry according to the chain 
$S_{ 3L }^{ \texttt{j} } \otimes S_{ 3R }^{ \texttt{j} } \supset 
S_{3}^{\mathrm{diag}} \supset S_{2}^{\mathrm{diag}}$, all Yukawa matrices in
the flavor adapted basis are represented by means of a matrix with two texture zeroes.
Therefore, all fermion mass matrices in the model have the same generic form with two 
texture zeroes. 

The difference between 2HDM and $\nu$2HDM$\otimes S_{3}$ lays on the Yukawa structure, on the 
symmetries of the Higgs sector and on the possible appearance of new CPV sources.
This CPV can arise from the same phase appearing in the 
Cabibbo-Kobayashi matrix, as in the SM, or some extra phase which arises from the Yukawa 
field or from the Higgs potential, either explicitly or spontaneously. The Higgs potential 
preserves CP symmetry, whereby CPV comes from the Yukawa matrices. 

In order to validate our hypothesis where the $S_{3}$ horizontal flavor symmetry is 
explicitly breaking, hence all  fermion mass matrices are represented through a matrix 
with two texture zeroes, we make a likelihood test where the $\chi^{2}$ function is 
defined in terms of leptonic flavor mixing angles. Afterwards, we shall investigate the 
phenomenological implications of these results on the neutrinoless double beta decay 
and the CPV in neutrino oscillations in matter.

The organization of this work is as follows. In section~\ref{sec:Yukawa_Lag} we present the 
Yukawa Lagrangian in the $\nu$2HDM$\otimes S_3$, the form of the Dirac and Majorana fermion 
mass matrices in terms of its eigenvalues. In this way, we derive explicit and 
analytical expressions for the leptonic flavor mixing angles and Higgs-fermions 
couplings. In section~\ref{sec:numericalanal} we present a detailed likelihood test 
where the $\chi^{2}$ function is defined in terms of leptonic mixing angles. 
Also, in section~\ref{sec:PenImp} we explore the phenomenological implications of the 
numerical values obtained for the CP violating phase factors, on the neutrinoless double beta 
decay and the neutrino oscillations in matter. Finally, in the section~\ref{sec:conclusions} 
we present the conclusions and remarks of the present work.
%
%

\section{The Yukawa Lagrangian in the $\nu$2HDM$\otimes S_{3}$}\label{sec:Yukawa_Lag}
%
%

In the fermion matter content of the SM, which is the same for the 2HDM, there are no 
right-handed neutrinos, consequently in both models a neutrino mass term is not allowed. 
This latter fact gainsay the results obtained in the neutrino oscillation experiments which 
requires neutrinos to have nonzero masses~\cite{Olive:2016xmw}.  
In order to include a Majorana neutrino mass term to the 2HDM, we need to increase its  
matter content. 
For this reason, we consider six neutrino fields: three left-handed 
$\nu_{L} = \left( \nu_{e L}, \nu_{\mu L}, \nu_{\tau L} \right)^{\top}$ and  
three right-handed 
$N_{R} = \left( N_{1 R}, N_{2 R}, N_{3 R} \right)^{\top}$. 
The right-handed neutrinos must be uncharged under the weak and electromagnetic 
interactions, which means that this kind of neutrinos are singlets under 
$G_{\mathrm{EW}} \equiv SU(2)_L \otimes U(1)_Y$. 
In other words, only the left-handed neutrinos take part in the electroweak interaction. 
In this theoretical framework, we have all SM matter content plus massive neutrinos and an 
extra Higgs boson, whereby it is called as $\nu$2HDM.
In the weak basis, the Yukawa interaction Lagrangian for Dirac fermions in the $\nu$2HDM is 
given by~\cite{Branco:2011iw,Barradas-Guevara:2016gda,Wang:2016vfj}:
\begin{equation}\label{2HDMlagrangian}
 {\cal L}_{\mathrm{Y}}^{\mathrm{w}} = 
  \sum_{k = 1}^{2} 
   \left( 
    {\bf Y}_{k}^{\mathrm{w}, u} \, \bar{Q} \tilde{\Phi}_{k} u_{R} 
  + {\bf Y}_{k}^{\mathrm{w}, d} \, \bar{Q} \Phi_{k} d_{R} 
  + {\bf Y}_{k}^{\mathrm{w}, \nu_{D}} \, \bar{L} \tilde{\Phi}_{k} N_{R} 
  + {\bf Y}_{k}^{\mathrm{w}, l} \,  \bar{L} \Phi_{k} l_{R} \right) 
  + \textrm{h. c.,}
\end{equation} 
where 
$Q = \left( \, u , \, d \, \right)^{\top}_{L}$ and  
$L = \left( \, \nu_{l} , \, l \, \right)^{\top}_{L}$ are the left-handed doublets of 
$SU(2)_{L}$; $u_{R}$, $d_{R}$ and $l_{R}$ are the right-handed singlets of the electroweak 
gauge group.
In this expression, the $\mathrm{w}$ superscript indicates that we are working in the weak 
basis, while the indices $l$, $u$ and $d$ represent the charged leptons, 
$u$- and $d$-type quarks, respectively. 
Also, $\Phi_{k} = \left( \phi^{+}_{k}, \, \phi^{0}_{k} \right)^{\top}$ denotes the
two Higgs fields which are doublets of $SU(2)_{L}$, with
$\tilde{\Phi}_{k} = i \sigma_2 \Phi_{k}^{*}$. 
Finally, the ${\bf Y}_{k}^{\mathrm{w},\texttt{j}}$ are the Yukawa matrices in 
the weak basis, where the $\texttt{j}$ superscript denotes the Dirac fermions, 
($\texttt{j} = u,d,l,\nu_{D}$)~\cite{Branco:2011iw}. 
In general, after the SSB and in the context of $\nu$2HDM, the Dirac fermion mass matrix in 
the weak basis can be written 
as~\cite{Branco:2011iw,Felix-Beltran:2013tra,Barradas-Guevara:2016gda}: 
\begin{equation}\label{masa-fermiones}
 {\bf M}_{ \texttt{j} }^{\mathrm{w}} =  
 \frac{1}{ \sqrt{2} } 
 \sum_{k=1}^{2} v_{k} \, {\bf Y}_{k}^{\mathrm{w},\texttt{j}} ,
\end{equation} 
where $v_{k}$ are the vacuum expectation values (vev's) of the two Higgs bosons $\Phi_{k}$, 
with $k=1,2$.  
%
\subsection{Mass matrices from the $S_{3}$ flavor symmetry}
%
To reduce the free parameters in the fermion mass matrices, we will consider 
a horizontal symmetry which correlates the particle flavor indices each one with the other, 
thus the Yukawa matrices could be represented by means of a $3\times3$ Hermitian matrix.
Consequently, for three families or generations of quarks and leptons, 
we propose\footnote{As other authors have done in the 
SM~\cite{Canales:2012dr,Canales:2013cga,Barranco:2010we} (and references in there).} 
that after the SSB, the Yukawa Lagrangian in the $\nu$2HDM presents the permutations group 
$S_{3}$ as horizontal flavor symmetry. 
In this context, the right-handed neutrinos, as well as the two Higgs bosons, transform as 
singlets under the action of $S_{3}$ flavor symmetry.
In other words, the right-handed neutrinos and the two scalar fields $\Phi_{k}$ are 
flavorless particles, whereby these fields are treated as scalars with respect to the 
$S_{3}$~symmetry transformations.      
The general way to implement the flavor symmetry is considering that under the action of 
$S_{3}$ symmetry,  the left- and right-handed spinors transform as~\cite{Canales:2013cga}:
\begin{equation}
 \begin{array}{l}
  \psi^{ \mathrm{\,s} }_{ \texttt{j} L } = {\bf g}_{\mathrm{a}}^{ \texttt{j} } \, 
   \psi_{\texttt{j} L} 
  \quad \textrm{and} \quad
  \psi^{ \mathrm{\,s} }_{ \texttt{j} R } = \widetilde{\bf g}_{\mathrm{b}}^{ \texttt{j} } \, 
   \psi_{ \texttt{j} R }, 
  \qquad \mathrm{a,b} = 1, \ldots, 6. 
 \end{array}
\end{equation}
At this point, the proposed flavor symmetry for the Yukawa Lagrangian is the
$S_{ 3L }^{ \texttt{j} } \otimes S_{ 3R }^{ \texttt{j} }$ group, whose 
elements are the pairs 
$\left( {\bf g}_{\mathrm{a}}^{ \texttt{j} }, \widetilde{\bf g}_{\mathrm{b}}^{ \texttt{j} }
\right)$, 
where 
${\bf g}_{\mathrm{a}}^{ \texttt{j} } \in S_{3L}^{ \texttt{j} }$ and 
$\widetilde{\bf g}_{\mathrm{b}}^{ \texttt{j} } \in S_{ 3R }^{ \texttt{j} }$,
while the superscript ``$\mathrm{s}$'' means that fields are in the flavor symmetry 
adapted basis.
However, the mass terms 
${\cal L}_{m} \sim \bar{ \psi }_{\texttt{j} L} \, {\bf M}_{\texttt{j}} \,
\psi_{\texttt{j} R} $ and charged currents
${\cal J}_{\mu} \sim \bar{ \psi }_{\texttt{j} L} \ \gamma_{\mu} \psi_{\texttt{j} L}  W^{\mu}$
in the Lagrangian are not invariant under 
$S_{ 3L }^{ \texttt{j} } \otimes S_{ 3R }^{ \texttt{j} }$ group.
In order to make ${\cal L}_{m}$ and ${\cal J}_{\mu}$ invariant under the flavor group 
transformations, the elements of the 
$S_{ 3L }^{ \texttt{j} } \otimes S_{ 3R }^{ \texttt{j} }$ group must satisfy the relation
${\bf g}_{\mathrm{a}} \equiv {\bf g}_{\mathrm{a}}^{\texttt{j}}  
= \widetilde{\bf g}_{\mathrm{b}}^{\texttt{j}}$. 
The latter condition implies that the flavor group is reduced according to the chain:
$S_{ 3L }^{ \mathrm{j} } \otimes S_{ 3R }^{ \mathrm{j} } \supset 
S_{3}^{\mathrm{diag}}$~\cite{Canales:2013cga}. 
This flavor symmetry breaking chain should be interpreted as: all fermions in the model must 
be transformed with the same flavor group and the same element thereof. 
In the above, the flavor group is called $S_{3}^{\mathrm{diag}}$
because its elements are the pairs 
$\left({\bf g}_{\mathrm{a}} , {\bf g}_{\mathrm{a}} \right)$, where 
${\bf g}_{\mathrm{a}} \in S_{3L}$~\cite{Canales:2013cga}.

Finally, it is easy conclude that $S_{3}^{\mathrm{diag}}$ is the horizontal 
flavor symmetry which conserves the invariance of ${\cal L}_{\mathrm{Y}}^{\mathrm{w}}$ 
under the action of electroweak gauge group.
Also, from the invariance of ${\cal L}_{m}$ Lagrangian under the action of 
$S_{3}^{\mathrm{diag}}$ group, we obtain that fermion mass matrices commute with all 
elements of the flavor group.

In the weak basis, the two Yukawa matrices in Eq.~(\ref{masa-fermiones}) 
are represented by means of a matrix with the exact $S_{3}^{\mathrm{diag}}$ symmetry. 
Therefore, these Yukawa matrices are expressed as
\begin{equation}\label{eq:Ykj0}
 {\bf Y}_{k}^{\mathrm{w},\texttt{j}} \equiv {\bf Y}_{k}^{\texttt{j} 3}
 = \alpha^{\texttt{j}}_{k} \,  {\bf P}_{\bf 1} ,
\end{equation}
where $\alpha^{\texttt{j}}_{k}$ are real constants associated with the flavor 
symmetry.
The explicit form of ${\bf P}_{\bf 1}$ matrix is given by Eq.~(\ref{eq:A-P1}), and 
corresponds to the projector associated with the symmetric singlet representation of $S_{3}$. 
Hence, the Dirac fermion mass matrix is
\begin{equation}
 {\bf M}_{ \texttt{j} }^{\mathrm{w}} = 
 \frac{1}{\sqrt{2}} \sum_{k=1}^{2} v_{k} {\bf Y}_{k}^{\texttt{j} 3} = 
 m_{\texttt{j}3} {\bf P}_{\bf 1} ,
\end{equation} 
where 
$m_{\texttt{j}3} = \frac{1}{\sqrt{2}}\left( v_{1} \, \alpha^{\texttt{j}}_{1} 
+ v_{2} \, \alpha^{\texttt{j}}_{2} \right)$. 
In the flavor adapted basis, the Dirac fermion mass matrices 
are~\cite{Felix-Beltran:2013tra}
\begin{equation}\label{eq:MS3_exact}
 {\bf M}_{ \texttt{j} }^{\mathrm{s}} = 
 {\bf U}_{\mathrm{s}}^{\dagger} {\bf M}_{ \texttt{j} }^{\mathrm{w}} {\bf U}_{\mathrm{s}} =
 m_{\texttt{j}3} {\bf U}_{\mathrm{s}}^{\dagger} {\bf P}_{\bf 1} {\bf U}_{\mathrm{s}} =
 \textrm{diag} \left( 0, 0, m_{\texttt{j}3} \right) ,
\end{equation}
where 
\begin{equation}\label{eq:Us}
 {\bf U}_{\mathrm{s}} = 
 \frac{1}{\sqrt{6}}
 \left( \begin{array}{ccc}
  \sqrt{3} & 1 & \sqrt{2} \\
 - \sqrt{3} & 1 & \sqrt{2} \\
  0  & -2  & \sqrt{2} 
 \end{array}  \right) .
\end{equation}
The interpretation of Eq.~(\ref{eq:MS3_exact}) is that under an exact 
$S_{3}^{\mathrm{diag}}$ symmetry, the mass spectrum for Dirac fermions consists of a 
massive particle and two massless particles~\cite{PhysRevD.59.093009}. 
The only massive particle in each of fermion mass spectrum corresponds to the heaviest 
fermion. 
However, this result disagrees with the experimental data on quarks and leptons 
masses~\cite{Olive:2016xmw}. 

Since the two Higgs fields are invariant under flavor symmetry transformations, these 
are naturally assigned to $S_{3}$ flavor singlets. If the Yukawa Lagrangian is exactly 
invariant under $S_{3}$ flavor transformations, the two scalar fields can only couple with the 
$S_{3}$-singlet component of fermion fields. Consequently, only the $S_{3}$-singlet component 
of fermion fields acquires a mass non zero. As the third family is the heaviest, here we assign 
the fermion fields in the third family to singlet irreducible representation of $S_{3}$.

So, with the aim to generate a non zero mass for all fermions in the model, here we will break the flavor symmetry in an explicit sequential way,  according to the chain  $S_{ 3L }^{ \texttt{j} } \otimes S_{ 3R }^{ \texttt{j} }\supset S_{3}^{\mathrm{diag}} \supset S_{2}^{\mathrm{diag}}$. 
Respectively, the first two fermion families and third one are assigned to the doublet and  
singlet irreducible representations of $S_{3}^{\mathrm{diag}}$.  
The mass of the second fermion family is generated when the $S_{3}^{\mathrm{diag}}$ flavor 
symmetry is explicitly breaks into the $S_{2}^{\mathrm{diag}}$ group.
This symmetry breaking is carried out when we add the following term to the 
${\bf Y}_{k}^{\texttt{j3}}$ matrix in Eq.~(\ref{eq:Ykj0}) :
\begin{equation}\label{eq:Ykj2}
 {\bf Y}_{k}^{\texttt{j2} } = \beta_{k}^{\texttt{j}} \, {\bf T}_{z1}^{+} + 
 \gamma_{k}^{\texttt{j}} \, {\bf T}_{z2}^{+},
\end{equation}
where $\beta_{k}^{\texttt{j}}$ and $\gamma_{k}^{\texttt{j}}$ are real constant 
parameters. 
The explicit form of the tensors ${\bf T}_{z1}^{+}$ and ${\bf T}_{z2}^{+}$ is given by 
Eq.~(\ref{eq:Tz12+}). The ${\bf Y}_{k}^{\texttt{j2} }$ matrix mixes the 
symmetric component of the doublet with the singlet. 
Finally, the first fermion family's mass is generated by adding the term
\begin{equation}\label{eq:Ykj1}
 {\bf Y}_{k}^{\texttt{j1} } = 
  \epsilon_{k}^{\texttt{j}} \, {\bf T}_{x}^{+} + 
  \rho_{k}^{\texttt{j}} \, {\bf T}_{x}^{-} 
\end{equation}
to matrices in Eqs.~(\ref{eq:Ykj0}) and~(\ref{eq:Ykj1}),
where $\epsilon_{k}^{\texttt{j}}$ and $\rho_{k}^{\texttt{j}}$ are real constant parameters. Thus,
the explicit form of the tensors ${\bf T}_{x}^{+}$ and ${\bf T}_{x}^{-}$ is given by
Eq.~(\ref{eq:Ty+-}). The ${\bf Y}_{k}^{\texttt{j1} }$ matrix mixes the components of the 
doublet representation between each other in the weak basis. 
So, in the weak basis and under the explicit sequential breaking of flavor symmetry according 
to the chain; 
$S_{ 3L }^{ \texttt{j} } \otimes S_{ 3R }^{ \texttt{j} } \supset 
S_{3}^{\mathrm{diag}} \supset S_{2}^{\mathrm{diag}}$, we obtain the Yukawa matrices which 
produce three massive fermions.
These Yukawa matrices are the sum of the three expressions given by
Eqs.~(\ref{eq:Ykj0}), ~(\ref{eq:Ykj2}), and~(\ref{eq:Ykj1}). 
Then,
\begin{equation}\label{eq:Yuka-w}
 \begin{array}{l}\vspace{2mm}
  {\bf Y}_{k}^{\mathrm{w}, \texttt{j}} = 
   {\bf Y}_{k}^{\texttt{j3} } + 
   {\bf Y}_{k}^{\texttt{j2} } + 
   {\bf Y}_{k}^{\texttt{j1} } = 
   \alpha^{\texttt{j}}_{k} \, {\bf P}_{\bf 1}   + 
   \beta_{k}^{\texttt{j}}  \, {\bf T}_{z1}^{+}  + 
   \gamma_{k}^{\texttt{j}} \, {\bf T}_{z2}^{+}  + 
   \epsilon_{k}^{\texttt{j}} \, {\bf T}_{x}^{+} + 
   \rho_{k}^{\texttt{j}} \, {\bf T}_{x}^{-} ,  \\
  {\bf Y}_{k}^{\mathrm{w}, \texttt{j} } = 
  \left( \begin{array}{ccc}
   e^{\mathrm{w} , \texttt{j} }_{k}     & a^{\mathrm{w} , \texttt{j} }_{k} & 
    f^{\mathrm{w} , \texttt{j} }_{k} \\
   a^{\mathrm{w} , \texttt{j} \,* }_{k} & b^{\mathrm{w} , \texttt{j} }_{k} & 
    c^{\mathrm{w} , \texttt{j} }_{k} \\
   f^{\mathrm{w} , \texttt{j} \,* }_{k} & c^{\mathrm{w} , \texttt{j} \,* }_{k} & 
    d^{\mathrm{w} , \texttt{j} }_{k}
  \end{array}  \right) , 
 \end{array}
\end{equation}
where 
\begin{equation}
 \begin{array}{lll} \vspace{2mm}
  a^{\mathrm{w} , \texttt{j} }_{k} = 
   \frac{ 
    \alpha^{\texttt{j}}_{k} + 3 \left( \beta_{k}^{\texttt{j}} + i \rho_{k}^{\texttt{j}} 
    \right) }{ 3 }, &
  b^{\mathrm{w} , \texttt{j} }_{k} = 
   \frac{ 
    \alpha^{\texttt{j}}_{k} + 3 \left( \beta_{k}^{\texttt{j}} + \epsilon_{k}^{\texttt{j}} 
    \right) }{ 3 } , & 
  c^{\mathrm{w} , \texttt{j} }_{k} = 
   \frac{ 
    \alpha^{\texttt{j}}_{k} + 3 \left( \gamma_{k}^{\texttt{j}} - \epsilon_{k}^{\texttt{j}} 
    + i \rho_{k}^{\texttt{j}} 
    \right) }{ 3 } ,       \\ \vspace{2mm}
  d^{\mathrm{w} , \texttt{j} }_{k} = 
   \frac{
    \alpha^{\texttt{j}}_{k} - 6 \beta_{k}^{\texttt{j}}
    }{ 
     3
    }, &
  e^{\mathrm{w} , \texttt{j} }_{k} = 
   \frac{ 
    \alpha^{\texttt{j}}_{k} + 3 \left( \beta_{k}^{\texttt{j}} - \epsilon_{k}^{\texttt{j}} 
    \right) }{ 3 } , &
  f^{\mathrm{w} , \texttt{j} }_{k} = 
   \frac{ 
    \alpha^{\texttt{j}}_{k} + 3 \left( \gamma_{k}^{\texttt{j}} + \epsilon_{k}^{\texttt{j}} 
    - i \rho_{k}^{\texttt{j}} 
    \right) }{ 3 } .       
 \end{array}
\end{equation} 
In this same basis, now the fermion mass matrices ${\bf M}_{\texttt{j}}^{\mathrm{w}}$ take the 
form:
\begin{equation}\label{eq:Mwj}
 \begin{array}{l}
  {\bf M}_{ \texttt{j} }^{\mathrm{w}} = 
    {\displaystyle \frac{ 1 }{ \sqrt{2} } \sum_{k=1}^{2} } \, v_{k}
   {\bf Y}_{k}^{\mathrm{w} , \texttt{j}} 
  = {\displaystyle \frac{ 1 }{ \sqrt{2} } \sum_{k=1}^{2} } \, v_{k} 
   \left( 
    \alpha^{\texttt{j}}_{k} \, {\bf P}_{\bf 1}   + 
    \beta_{k}^{\texttt{j}}  \, {\bf T}_{z1}^{+}  + 
    \gamma_{k}^{\texttt{j}} \, {\bf T}_{z2}^{+}  + 
    \epsilon_{k}^{\texttt{j}} \, {\bf T}_{x}^{+} + 
    \rho_{k}^{\texttt{j}} \, {\bf T}_{x}^{-}  
   \right).  
 \end{array}
\end{equation}
Thus, with help of previous expression and Eq.~(\ref{eq:Yuka-w}), it is easy to conclude that 
the Dirac fermion mass matrices are Hermitian matrices without any of their elements equal to 
zero. 
However, in the flavor adapted basis the mass matrices in Eq.~(\ref{eq:Mwj}) acquire the 
following form:
\begin{equation}\label{MMM}
 \begin{array}{l}\vspace{2mm}
  {\bf M}_{ \texttt{j} }^{\mathrm{s}} = 
   {\bf U}_{\mathrm{s}}^{\dagger} {\bf M}_{ \texttt{j} }^{\mathrm{w}} {\bf U}_{\mathrm{s}} =
   {\displaystyle \frac{1}{ \sqrt{2} } \sum_{k=1}^{2} } \,
    v_{k} {\bf U}_{\mathrm{s}}^{\dagger} {\bf Y}_{k}^{\mathrm{w} , \texttt{j}} 
   {\bf U}_{\mathrm{s}} , \\
  {\bf M}_{\texttt{j}}^{\mathrm{s}} = 
  {\bf P}_{ \texttt{j} } 
  \left( \begin{array}{ccc}
   0 & \left| A_{ \texttt{j} } \right| & 0 \\
   \left| A_{ \texttt{j} } \right| & B_{ \texttt{j} }  & C_{ \texttt{j} } \\
   0 & C_{ \texttt{j} } & D_{ \texttt{j} }
  \end{array}\right) 
  {\bf P}_{ \texttt{j} }^{\dagger}= 
  \frac{ v \cos \beta }{\sqrt{2}}  \times 
  \left[ \left( \begin{array}{ccc}
   0 & A^{\texttt{j} }_{1} & 0 \\
   A^{\texttt{j} \,* }_{1} & B^{\texttt{j} }_{1} & C^{\texttt{j} }_{1} \\
  0 & C^{\texttt{j}  }_{1} & D^{\texttt{j} }_{1}
 \end{array} \right) 
  + \tan \beta \left( \begin{array}{ccc}
   0 & A^{\texttt{j} }_{2} & 0 \\
   A^{\texttt{j} \,* }_{2} & B^{\texttt{j} }_{2} & C^{\texttt{j} }_{2} \\
  0 & C^{\texttt{j} }_{2} & D^{\texttt{j} }_{2}
 \end{array} \right)  \right],
 \end{array} 
\end{equation}
where 
${{\bf P}_{ \texttt{j} } } = \textrm{diag} 
\left( 1 , e^{ - i \phi_{ \texttt{j} } }, e^{ - i \phi_{ \texttt{j} } } \right)$ with 
$\phi_{ \texttt{j} } = \arg \left\{ A_{ \texttt{j} } \right\}$, and
\begin{equation}\label{Eq:Yuk:FB}
 \begin{array}{llll}
  A^{\texttt{j} }_{k} = 
    - \sqrt{3} \left( \epsilon_{k}^{\texttt{j}} - i \rho_{k}^{\texttt{j}}  \right) , &
  B^{\texttt{j} }_{k} = 
   - \frac{ 2 }{ 3 } \left( \beta_{k}^{\texttt{j}} + 2 \gamma_{k}^{\texttt{j}} \right) , & 
  C^{\texttt{j} }_{k} = 
    \frac{ \sqrt{2} }{ 3 } 
    \left( 4 \beta_{k}^{\texttt{j}} -  \gamma_{k}^{\texttt{j}} \right) , & 
  D^{\texttt{j} }_{k} = 
   \alpha^{\texttt{j}}_{k} 
    + \frac{ 2 }{ 3 } \left( \beta_{k}^{\texttt{j}} + 2 \gamma_{k}^{\texttt{j}} \right), 
 \end{array}
\end{equation} 
with $k=1,2$. The parameters in Eq.~(\ref{Eq:Yuk:FB}) are the Yukawa matrices elements 
expressed in the flavor adapted basis. 
Finally, in the Higgs sector $\tan \beta = \dfrac{ v_{2} }{ v_{1} }$ with 
$v^{2} = v_{1}^{2} + v_{2}^{2} = \left( 246.22~\textrm{GeV}\right)^{2}$.

Here, we consider that the active neutrinos acquire their mass through the type-I seesaw 
mechanism, Eq.~(\ref{eq:seesaw}), where the Dirac fermion mass matrix is given by
Eq.~(\ref{MMM}), while we suppose that in the flavor adapted basis the right-handed 
neutrino mass matrix has the form: 
\begin{equation}
 {\bf M}_{R}^{\mathrm{s}} = \textrm{diag} \left( A_{R}, A_{R}, D_{R} \right) {\bf D}^{(3)} 
 \left( A_{1} \right).
\end{equation} 
In the latter expression $A_{R}$ and $D_{R}$ are real parameters, and  the form of 
${\bf D}^{(3)}\left( A_{1} \right)$ matrix is given by Eq.~(\ref{eq:A-1}). 
So, in the flavor adapted basis the active neutrinos mass matrix is
\begin{equation}\label{eq:Mnu}
 \begin{array}{ccc}
  {\bf M}_{\nu}^{\mathrm{s}} =
  {\bf P}_{\nu}^{\dagger} 
  \left( \begin{array}{ccc}
   0 & a_{\nu} & 0 \\
   a_{\nu}  & \left| b_{\nu} \right| & \left| c_{\nu} \right|  \\
   0 & \left| c_{\nu} \right| &  d_{\nu}
  \end{array}  \right)
  {\bf P}_{\nu}^{\dagger} ,
 \end{array}
\end{equation} 
where ${\bf P}_{\nu} = e^{ i \phi_{ \nu } } \textrm{diag} 
\left( 1, e^{ - i 2 \phi_{ \nu } }, e^{ -i \phi_{ \nu } } \right)$ with 
$\phi_{ \nu } = \arg \left\{ C_{ \nu } \right\}$, and 
$\arg \left\{ C_{ \nu } \right\} = 2 \arg \left\{ B_{ \nu } \right\}$,
\begin{equation}
 \begin{array}{l}
  a_{\nu} = \frac{ \left| A_{ \nu_{D} } \right|^{2} }{ A_{R} }, \quad
  b_{\nu} = \frac{ C_{ \nu_{D} }^{2} }{ D_{R} } + 
  \frac{ 2 B_{ \nu_{D} } A_{ \nu_{D} }^{*} }{ A_{R} } , \quad
  c_{\nu} = \frac{ C_{ \nu_{D} } D_{ \nu_{D} } }{ D_{R} } + 
  \frac{ C_{ \nu_{D} } A_{ \nu_{D} }^{*} }{ A_{R} } ,  
  \quad \textrm{and} \quad 
  d_{\nu} = \frac{ D_{ \nu_{D} }^{2} }{ D_{R} }.  
 \end{array}
\end{equation}
In this work, we study the flavor dynamics through the Yukawa matrices in the 2HDM-III plus 
massive neutrinos and a horizontal flavor symmetry $S_3$. This theoretical framework is 
called $\nu$2HDM$\otimes S_3$.
%
\subsection{The mass and mixing matrices as function of fermion masses}
\label{sec:PMNS_Masses}
%
For a normal [inverted] hierarchy\footnote{The inverted hierarchy is only valid for 
neutrinos.} in the mass spectrum, the real symmetric matrices in Eqs.~(\ref{MMM}) 
and~(\ref{eq:Mnu}), which are associated with the fermion mass matrices, can be 
reconstructed through the orthogonal transformation\footnote{The superscript 
$\mathrm{n[i]}$ denote the normal [inverted] hierarchy in the neutrino mass spectrum.} 
\begin{equation}\label{eq:DMSym}
 \bar{\bf M}_{f}^{ \mathrm{n[i]} } = m_{f 3[2] } \,
  {\bf O}_{f}^{ \mathrm{n[i]} } {\bf \Delta}_{f}^{ \mathrm{n[i]} } 
  \left( { \bf O}_{f}^{ \mathrm{n[i]} } \right)^{\top}, \qquad f = u, d, l, \nu,
\end{equation}
where ${\bf \Delta}_{f}^{ \mathrm{n[i]} } = \textrm{diag} 
\left( \widehat{m}_{f 1[3]}, -\widehat{m}_{f 2[1] }, 1 \right) $,
in this expression 
$\widehat{m}_{ f 1[3] } = m_{ f 1[3] } /  m_{ f 3[2] }$ and
$\widehat{m}_{ f 2[1] } = | m_{ f 2[1] } | /  m_{ f 3[2] }$. 
Here, $m_{f 2[1] } = -| m_{f 2[1] } |$ and the $m_{f}$'s are the eigenvalues of fermion mass matrices, {\it i.e.}, the particle masses. 
From algebraic invariants of the expression in Eq.~(\ref{eq:DMSym}) we have
\begin{equation}
 \begin{array}{ll}\vspace{2mm}
  a_{f}^{ \mathrm{n[i]} } \equiv 
   \dfrac{ \left( {\bf M}_{f} \right)_{12} }{ m_{f 3[2] } } 
   = \sqrt{ \frac{ \widehat{m}_{f 1[3]} \widehat{m}_{f 2[1]} }{ 1 - \delta_{f} } } , &
  b_{f}^{ \mathrm{n[i]} } \equiv 
   \dfrac{ \left( {\bf M}_{f} \right)_{22} }{ m_{f 3[2] } } = 
   \widehat{m}_{f 1[3]} - \widehat{m}_{f 2[1]} + \delta_{f}, \\ 
  c_{f}^{ \mathrm{n[i]} } \equiv 
   \dfrac{ \left( {\bf M}_{f} \right)_{23} }{ m_{f 3[2] } } = 
   \sqrt{ \frac{ \delta_{f} }{ 1 - \delta_{f} }  \xi_{ f 1[3] } \xi_{ f 2[1] } } ,&    
  d_{f}^{ \mathrm{n[i]} } \equiv 
   \dfrac{ \left( {\bf M}_{f} \right)_{33} }{ m_{f 3[2] } } = 
   1 - \delta_{f} ,   
 \end{array}
\end{equation}
where $ \xi_{ f 1[3] } = 1 - \widehat{m}_{ f 1[3] } - \delta_{f}$ and
$\xi_{ f 2[1] } = 1 + \widehat{m}_{ f 2[1] } - \delta_{f}$. 
Also, the free parameter $\delta_{f}$ must 
satisfy the relation $1 - \widehat{m}_{ f 1[3] } > \delta_{f} > 0$~\cite{Canales:2012dr}. 
The real orthogonal matrix given in Eq.~(\ref{eq:DMSym}) written in terms of fermion masses 
has the shape~\cite{Barradas-Guevara:2016gda,Barranco:2010we}
\begin{equation}\label{eq:Real-O}
 {\bf O}_{f}^{ \mathrm{n[i]} } = 
  \left( \begin{array}{ccc} \vspace{2mm}
   \sqrt{ \frac{ \widehat{m}_{ f 2 [1] } 
    \xi_{ f 1[3] } }{ {\cal D}_{ f 1[3] } } } & 
  -\sqrt{ \frac{ \widehat{m}_{ f 1 [3] } 
    \xi_{ f 2[1] } }{ {\cal D}_{ f 2[1] } } } &
   \sqrt{ \frac{ \widehat{m}_{ f 1 [3] } \widehat{m}_{ f 2[1] } 
    \delta_{f} }{ {\cal D}_{ f 3[2] } } } \\ \vspace{2mm}
   \sqrt{ \frac{ \widehat{m}_{ f 1 [3] } \left( 1 - \delta_{f} \right) 
    \xi_{ f 1[3] } }{ {\cal D}_{ f 1[3] } } } &
   \sqrt{ \frac{ \widehat{m}_{ f 2 [1] } 
    \left( 1 - \delta_{f} \right) \xi_{ f 2[1] } }{ {\cal D}_{ f 2[1] } } } &
   \sqrt{ \frac{ \delta_{f} \left( 1 - \delta_{f} \right) 
    }{ {\cal D}_{ f 3[2] } } } \\ \vspace{2mm}
  -\sqrt{ \frac{ \widehat{m}_{ f 1[3] } \delta_{f} \xi_{ f 2[1] } }{ 
    {\cal D}_{ f 1[3] } } }  &
  -\sqrt{ \frac{ \widehat{m}_{ f 2[1] } \delta_{f} \xi_{ f 1[3] } }{ 
    {\cal D}_{ f 2[1] } } }  &
   \sqrt{ \frac{ \xi_{ f 1[3] } \xi_{ f 2[1] } }{ {\cal D}_{ f 3[2] } } } 
 \end{array}  \right) .
\end{equation}
In this orthogonal matrix we have
\begin{equation}
 \begin{array}{l}
  {\cal D}_{f 1[3]} = 
   \left( 1 - \delta_{f} \right) 
   \left( \widehat{m}_{ f 1[3] } + \widehat{m}_{ f 2[1] } \right) 
   \left( 1 - \widehat{m}_{ f 1[3] } \right), \\
  {\cal D}_{f 2[1]} = 
   \left( 1 - \delta_{f} \right) 
   \left( \widehat{m}_{ f 1[3] } + \widehat{m}_{ f 2[1] } \right) 
   \left( 1 + \widehat{m}_{ f 2[1] } \right), \\
  {\cal D}_{f 3[2]} = 
   \left( 1 - \delta_{f} \right) \left( 1 - \widehat{m}_{f 1 [3]} \right) 
   \left( 1 + \widehat{m}_{f 2 [1]} \right). 
 \end{array}
\end{equation} 

In the theoretical framework of $\nu$2HDM$\otimes S_{3}$, the unitary matrices that 
diagonalize the mass matrices of charged leptons and active neutrinos  are defined as
\begin{equation}
 {\bf U}_{\ell}^{\mathrm{n[i]}} = {\bf U}_{\mathrm{s}} {\bf P}_{\ell} 
  {\bf O}_{\ell}^{\mathrm{n[i]}},   \qquad \ell = l, \nu.
\end{equation} 
The matrix ${\bf U}_{\mathrm{s}}$ is given by Eq.~(\ref{eq:Us}),
and the diagonal phase matrices can be founded in Eqs.~(\ref{MMM}) and~(\ref{eq:Mnu}). 
Finally, the real orthogonal matrix ${\bf O}_{\ell}^{ \mathrm{n[i]} }$ is given in 
Eq.~(\ref{eq:Real-O}). From the previous unitary matrix, in the mass states basis, 
the Dirac fermion mass matrices take the shape 
\begin{equation}
 \begin{array}{l}\vspace{2mm}
  {\bf \Delta}_{ \texttt{j} } = 
   {\bf U}_{\texttt{j} }^{\dagger} 
   {\bf M}_{ \texttt{j} }^{\mathrm{w}} 
   {\bf U}_{\texttt{j} } =
  {\displaystyle  \frac{ 1 }{ \sqrt{2} }\sum_{k=1}^{2} } \, v_{k} \,
   {\bf U}_{\texttt{j} }^{\dagger}
    {\bf Y}_{k}^{\mathrm{w} , \texttt{j} } 
   {\bf U}_{\texttt{j} } = 
  {\displaystyle \frac{ 1 }{ \sqrt{2} } \sum_{k=1}^{2} } \, v_{k} \, 
    \widetilde{\bf Y}_{k}^{ \texttt{j} }  , \qquad \texttt{j} = u,d,l,\nu_{D},
 \end{array} 
\end{equation}
where $\widetilde{\bf Y}_{k}^{ \texttt{j} } = 
{\bf U}_{\texttt{j} }^{\dagger} {\bf Y}_{k}^{\mathrm{w}, \texttt{j} } {\bf U}_{\texttt{j} }$ 
are the Yukawa matrices in the mass states basis.
Now, with the help of the orthogonal matrix given in Eq.~(\ref{eq:Real-O}) and, as the 
mass spectrum of Dirac particles only has the normal hierarchy, it is easy to obtain that 
the elements of Yukawa matrices  $\widetilde{\bf Y}_{k}^{\texttt{j}}$ can be expressed in 
terms of geometric mean of Dirac fermion masses normalized with respect of electroweak 
scale, {\it i.e.},
\begin{equation}\label{eq:Cheng-Sher}
 \begin{array}{l}
  \left( \widetilde{\bf Y}_{k}^{ \texttt{j} } \right)_{ \texttt{rt} } =  
   \dfrac{ \sqrt{ m_{ \texttt{jr}  } m_{ \texttt{jt}  } } }{ v }
   \left( \widetilde{\bf \chi}_{k}^{ \texttt{j} } \right)_{ \texttt{rt} } \qquad 
   (\texttt{r,t} = 1, 2, 3).
 \end{array}
\end{equation}
Here, $\left( \widetilde{\bf \chi}_{k}^{ \texttt{j} } \right)_{ \texttt{rt} }$ are complex 
parameters, whose the explicit form is given in Appendix~\ref{App:B}. 
In the literature, this last relation is called 
{\it Cheng-Sher ansatz}~\cite{Cheng:1987rs}, which is associated with Higgs-fermions 
couplings. 
Also, the result obtained in Eq.~(\ref{eq:Cheng-Sher}) can be extended to any Hermitian mass 
matrix that may be brought to a two zero texture matrix by means of an unitary 
transformation.   

As one knowns, the 2HDM-III is a generic description of particle physics at a higher energy 
scale ($\gtrsim \textrm{TeV}$), and its imprint at low energies is reflected in Yukawa 
couplings structure.  
A detailed study of Yukawa Lagrangian within the 2HDM-III  is given 
in~\cite{DiazCruz:2009ek,DiazCruz:2004pj,DiazCruz:2004tr,HernandezSanchez:2013xj,Felix-Beltran:2013tra}, 
while the phenomenological implications of this model in scalar sector including Lepton  
Violation (LV) and/or FCNC's are presented 
in~\cite{GomezBock:2009xz,Martinez:2008hu,Martinez:2005sya}.  
In these works, the FCNC's are under control because the authors have been assuming that 
Yukawa matrices in the weak and mass basis, are represented by means of an Hermitian matrix with two 
texture zeroes and the {\it Cheng-Sher ansatz}, respectively.
In our model  $\nu\textrm{2HDM}\otimes S_3$, the two texture zeroes shape 
for the Yukawa matrices is obtained by imposing a flavor symmetry $S_{3}$ and from its 
explicit sequential breaking according to the chain 
$S_{3L}^{ \texttt{j} } \otimes S_{3R }^{ \texttt{j} } \supset S_{3}^{\mathrm{diag}} \supset 
S_{2}^{\mathrm{diag}}$. An immediate consequence from the above is that Yukawa matrices in 
the mass basis naturally take the form of so-called {\it Cheng-Sher ansatz}. 
Therefore, we can say that the FCNC's are under control in the $\nu$2HDM$\otimes S_{3}$.

The lepton flavor mixing matrix is defined as
${\bf U}_{\mathrm{PMNS}} = {\bf U}_{l}^{\dagger} {\bf U}_{\nu}$~\cite{Hochmuth:2007wq}. 
In this context the $\textrm{PMNS}$ matrix takes the expression
\begin{equation}\label{Eq:PMNS_th}
 {\bf U}_{\mathrm{PMNS}} = 
  {\bf O}_{l}^{ \top } {\bf P}^{(\nu - l)} {\bf O}_{\nu}^{ \mathrm{n[i]} } ,
\end{equation}
where 
${\bf P}^{(\nu - l)} = 
 \textrm{diag} \left( 1, e^{i \phi_{\ell 1} }, e^{i \phi_{\ell2} } \right)$ 
with $\phi_{\ell 1} = \phi_{l} - 2 \phi_{\nu}$ and 
$\phi_{\ell 2} = \phi_{l} - \phi_{\nu}$. 
The explicit form of the entries of the PMNS matrix are given in 
Appendix~\ref{App:C}.
The first conclusion of this $\textrm{PMNS}$ matrix is that
${\bf U}_{\mathrm{s}}$ unitary matrix, the one that allows us to pass from the weak basis to 
the flavor symmetry adapted basis, is unobservable in the lepton flavor mixing matrix. 
%
%
\subsection{The mixing angle and CP violation phases}\label{Sec:Angle-Phase}
%
%
In the symmetric parametrization  of lepton flavor mixing matrix, the relation between 
mixing angles and the entries of $\textrm{PMNS}$ matrix is~\cite{Hochmuth:2007wq,Bilenky:2010zza}
\begin{equation}\label{Eq:senoscuadrados}
 \sin^{2} \theta_{13} \equiv 
  \left| \left( {\bf U}_{\mathrm{PMNS}} \right)_{13} \right|^{2}, \quad 
 \sin^{2} \theta_{12} \equiv 
  \frac{
   \left| \left( {\bf U}_{\mathrm{PMNS}} \right)_{12} \right|^{2} 
  }{ 
   1 - \left| \left( {\bf U}_{\mathrm{PMNS}} \right)_{13}  \right|^{2} }, \quad 
 \sin^{2} \theta_{23} \equiv 
  \frac{ 
   \left| \left( {\bf U}_{\mathrm{PMNS}} \right)_{23}  \right|^{2} 
  }{ 
   1 - \left| \left( {\bf U}_{\mathrm{PMNS}} \right)_{13}  \right|^{2} 
  }.
\end{equation}

On the one hand, the lepton Jarlskog invariant which appears in conventional neutrino 
oscillations is defined as: 
\begin{equation}
{\cal J}_{\mathrm CP} = 
 {\cal I}m \left \{ 
 \left( {\bf U}_{\mathrm{PMNS}} \right)_{11}^{*}  
 \left( {\bf U}_{\mathrm{PMNS}} \right)_{23}^{*} 
 \left( {\bf U}_{\mathrm{PMNS}} \right)_{13} 
 \left( {\bf U}_{\mathrm{PMNS}} \right)_{21} 
 \right \}, 
 \end{equation}
and in the symmetric parametrization it has the form
\begin{equation} \label{INV:JCP}
 \begin{array}{l}
  {\cal J}_{\mathrm CP} = 
   \frac{1}{8} 
   \sin 2 \theta_{12} \, \sin 2 \theta_{23} \, \sin 2 \theta_{13} \, \cos \theta_{13} \,
   \sin \delta_{\mathrm CP} ,
 \end{array}
\end{equation}
where $\delta_{\mathrm CP } = \phi_{13} - \phi_{23} - \phi_{12}$. 
Moreover, the invariants 
\begin{equation}
 \begin{array}{l}
  {\cal I}_{1} = {\cal I}m \left \{ 
   \left( {\bf U}_{\mathrm{PMNS}} \right)_{12}^{2} 
   \left( {\bf U}_{\mathrm{PMNS}} \right)_{11}^{*2} 
   \right \} 
  \quad \textrm{and} \quad
  {\cal I}_{2} = {\cal I}m \left \{
   \left( {\bf U}_{\mathrm{PMNS}} \right)_{13}^{2} 
   \left( {\bf U}_{\mathrm{PMNS}} \right)_{11}^{*2}  
   \right \}, 
 \end{array}
\end{equation}
associated with the Majorana phases~\cite{Branco:1986gr,Branco:2011zb,Jenkins:2007ip} 
take the expressions
\begin{equation}\label{INV:I1-I2}
 \begin{array}{l}
  {\cal I}_{1} = \frac{1}{4} \sin^{2} 2 \theta_{12} \cos^{4} \theta_{13} \sin ( - 2 \phi_{12} ) 
   \quad \textrm{and} \quad 
  {\cal I}_{2} = \frac{1}{4} \sin^{2} 2 \theta_{13} \cos^{2} \theta_{12} \sin ( - 2 \phi_{13} ).
 \end{array}
\end{equation}
Then, the phase factors associated with the CP violation can be written as:
\begin{equation}\label{Eq:CP_Phases}
 \begin{array}{l}\vspace{2mm}
  \sin ( \delta_{\mathrm CP } ) = 
   \frac{ {\cal J}_{\rm CP} 
    \left( 1 - \left| \left( {\bf U}_{\mathrm{PMNS}} \right)_{13} \right|^{2} \right) 
   }{ 
    \left| \left( {\bf U}_{\mathrm{PMNS}} \right)_{11} \right| 
    \left| \left( {\bf U}_{\mathrm{PMNS}} \right)_{12} \right| 
    \left| \left( {\bf U}_{\mathrm{PMNS}} \right)_{13} \right| 
    \left| \left( {\bf U}_{\mathrm{PMNS}} \right)_{23} \right| 
    \left| \left( {\bf U}_{\mathrm{PMNS}} \right)_{33} \right| } , \\
  \sin \left( - 2 \phi_{12} \right) = 
   \frac{ 
    {\cal I}_{1} 
   }{ 
    \left| \left( {\bf U}_{\mathrm{PMNS}} \right)_{11} \right|^{2}  
    \left| \left( {\bf U}_{\mathrm{PMNS}} \right)_{12} \right|^{2} 
   } , \quad
  \sin \left( - 2 \phi_{13} \right) = 
   \frac{ 
    {\cal I}_{2} 
   }{ 
    \left| \left( {\bf U}_{\mathrm{PMNS}} \right)_{11} \right|^{2}  
    \left| \left( {\bf U}_{\mathrm{PMNS}} \right)_{13} \right|^{2} 
   } .
 \end{array} 
\end{equation} 
\begin{figure}[!htbp]
 \begin{center}
  \begin{tabular}{cc}
   \includegraphics[width=8.7cm, height=6.3cm]{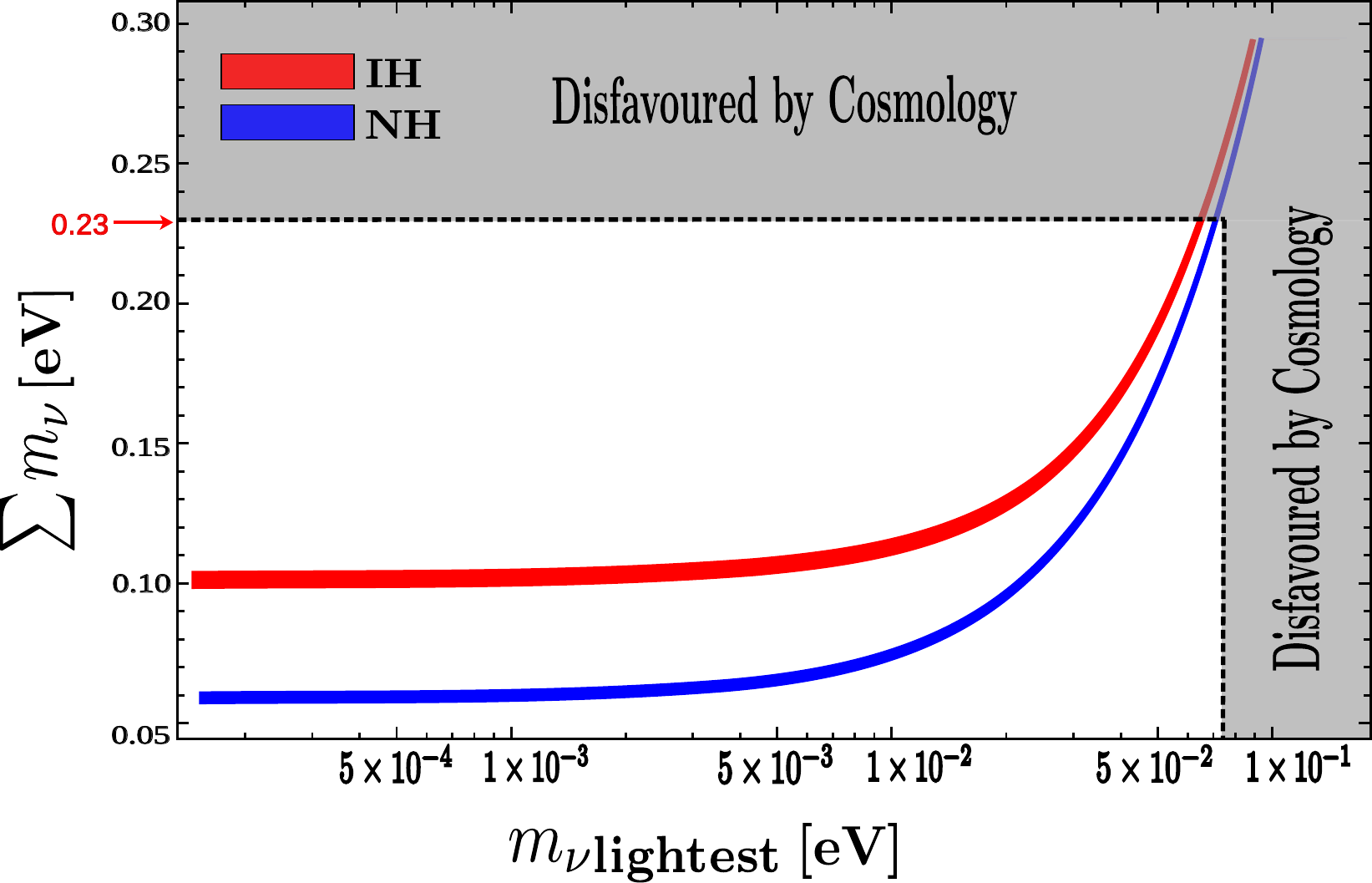} &
   \includegraphics[width=8.7cm, height=6.3cm]{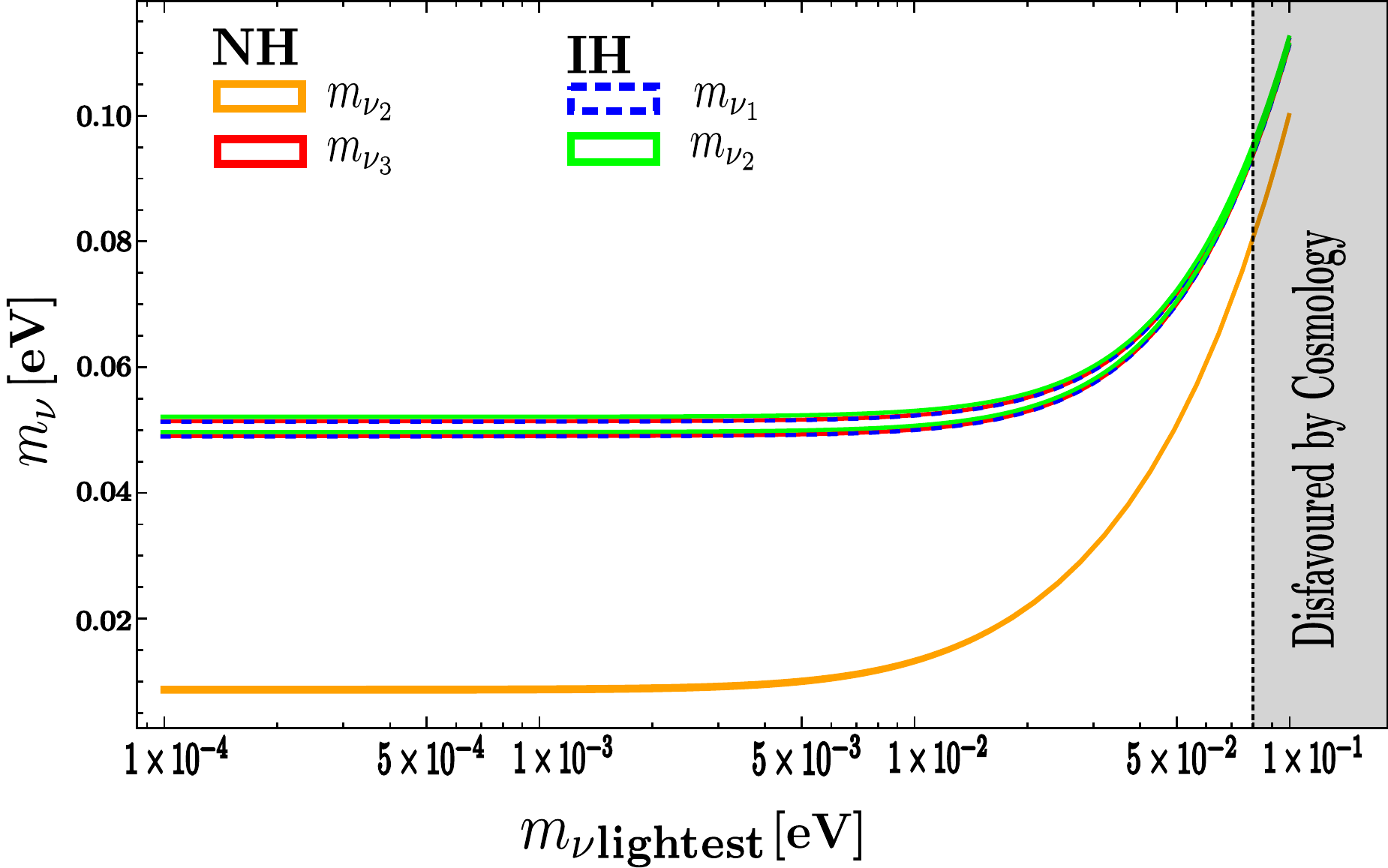} 
  \end{tabular}
  \caption{In the left-panel we show the sum of neutrino masses. 
   The right-panel shows the neutrino masses. The neutrino oscillation parameters 
   $\Delta m_{i j}^{2}$ are taken within the currently allowed $3\sigma$ 
   range~\cite{Esteban:2016qun}. 
   The upper bound on the mass of the lightest neutrino is obtained from the Planck 
   result where $\sum m_{\nu_{i}} < 0.23$~eV at 95\% level~\cite{Ade:2015xua}. }
   \label{Fig:Masa_neutrinos}
 \end{center}
\end{figure}
The equivalence between the PDG and symmetric parameterization may be expressed as 
${\bf U}_{ \mathrm{PDG} } = {\bf K} {\bf U}_{ \mathrm{Sym} }$, 
where 
${\bf K} = \textrm{diag} \left( 1, e^{i \frac{ \alpha_{21} }{2} }, e^{i \frac{ \alpha_{31} }{2} } \right)$
with $\delta_{CP} = \phi_{13} - \phi_{23} - \phi_{12}$, 
$\alpha_{21} = - 2 \phi_{12}$ and $\alpha_{31} = - 2 ( \phi_{12} + \phi_{23} )$.
%
\section{Numerical analysis\label{sec:numericalanal}}
%
%
%
In the three flavor neutrino scheme there are six independent parameters which rule the 
behavior of neutrino oscillation phenomena: flavor mixing angles, the ``Dirac-like'' 
CP-violating phase and squared-mass splitting. 
The latest neutrino oscillation parameter is defined as
$\Delta m_{i j}^{2} \equiv m^{2}_{ \nu_{i} } - m^{2}_{ \nu_{j} }$, 
in agreement with the results obtained in the global fit reported in 
Ref.~\cite{Esteban:2016qun}. These neutrino oscillation parameters have the following 
numerical values (at BFP$\pm 1 \sigma$ and $3\sigma$)\footnote{Here, NH and IH denote the 
normal and inverted hierarchy in neutrino mass spectrum, respectively.}
\begin{equation}\label{Eq:Val_DMij}
 \begin{array}{rl}\vspace{2mm}
  \Delta m_{21}^{2} \; \left( 10^{-5}~\textrm{eV}^{2} \right) = & 
   7.50_{-0.17}^{+0.19}, \; 7.03-8.09, \\ \vspace{2mm}
  \Delta m_{31}^{2} \; \left( 10^{-3} \textrm{eV}^{2} \right) = & 
   2.524_{-0.040}^{+0.039}, \; 2.407-2.643, \; \textrm{for NH},\\
  \Delta m_{23}^{2} \; \left( 10^{-3} \textrm{eV}^{2} \right) = &    
   2.514_{-0.041}^{+0.038}, \; 2.399-2.635, \; \textrm{for IH}.
 \end{array}
\end{equation}

From the definition of squared-mass splitting $\Delta m_{i j}^{2}$ two of neutrino masses 
can be written as: 
\begin{equation}\label{Eq:For_mnu}
 \begin{array}{l}\vspace{2mm}
  m_{ \nu_{3 [2]} } = \sqrt{ m_{\nu_{1 [3]} }^{2} + \Delta m_{31 [23]}^{2} }, 
  \quad \textrm{and} \quad  
  m_{ \nu_{2 [1]} } = \sqrt{ m_{\nu_{1 [3]} }^{2} + \Delta m_{21 [31]}^{2} }.
 \end{array}
\end{equation} 
where~$m_{\nu_{1 [3]} }$ is the lightest neutrino 
mass\footnote{The subscript $i[j]$ denote to normal [inverted] hierarchy in the neutrino 
masses spectrum.}. Also, this neutrino mass is considered as the only one free parameter in 
the above expressions, since the mass-squared differences $\Delta m_{i j}^{2}$ are determined 
by experimental means.

From the results reported by Planck Collaboration for cosmological parameters, the upper 
limit on the active neutrino masses sum is $\sum m_{\nu_{i}} < 0.23$~eV, 
for an active neutrinos number equal to 
$N_{\textrm eff} = 3.15 \pm 0.23$~\cite{Ade:2015xua}. 
This upper bound is independent of  hierachy in the neutrino mass spectrum. 
So, with all the above experimental information and considering the expressions in 
Eq.~(\ref{Eq:For_mnu}), we can obtain the following value range for neutrino masses 
\begin{equation}
 \begin{array}{lll}
  m_{ \nu_{1} } \left( 10^{-2}~\textrm{eV} \right) = 
  \left \{ \begin{array}{l}
   \left[0.00 , 7.10 \right], \\
   \left[4.90 , 8.25 \right],
  \end{array}  \right.  &
  m_{ \nu_{2} } \left( 10^{-2}~\textrm{eV} \right) = 
  \left \{ \begin{array}{l}
   \left[0.84 , 7.13 \right], \\
   \left[4.97 , 8.30 \right], 
  \end{array}  \right.    &
  m_{ \nu_{3} } \left( 10^{-2}~\textrm{eV} \right) = 
  \left \{ \begin{array}{l}
   \left[4.80 , 8.75 \right], \\
   \left[0    , 6.45 \right]. 
  \end{array}  \right.   
 \end{array}
\end{equation}
Here, the oscillation parameters $\Delta m_{ij}^{2}$ are taken within the currently 
allowed $3\sigma$ range~\cite{Esteban:2016qun}. 
The values in the first and second row correspond to a normal and inverted 
hierarchy in the neutrino mass spectrum, respectively.
For both hierarchies there is the possibility that the lightest neutrino mass could be 
zero. Namely in this case, the lightest neutrino is a massless particle. 
In the Fig.~\ref{Fig:Masa_neutrinos}, the behaviour of neutrino masses (right panel), as well 
as of the neutrino masses sum as function of ligntest neutrino mass mnu (left panel), are 
shown.
%
\subsection{The likelihood test $\chi^{2}$}
%
%
{\small
\begin{table}
\begin{center}
 \begin{tabular}{||l||c||c|c|c|c|c||c|c|c|c|c|c||c||}\hline
  & $\Delta m_{ij}^{2}$ at & 
  $\Phi_{\ell 1}~[^{\circ}]$ & 
  $\Phi_{\ell 2}~[^{\circ}]$ & 
  $m_{\nu_{\mathrm{lightest} } }$~[eV] & 
  $\delta_{\ell}$ & 
  $\delta_{\nu}$  & 
  $\theta_{12}^{ ^{\mathrm{th}} }~[^{\circ}]$ &
  $\theta_{23}^{ ^{\mathrm{th}} }~[^{\circ}]$ &
  $\theta_{13}^{ ^{\mathrm{th}} }~[^{\circ}]$ &    
  $\delta_{\mathrm{CP}}~[^{\circ}]$ & 
  $\phi_{12}~[^{\circ}]$ & 
  $\phi_{13}~[^{\circ}]$ &
  $\chi^{2}_{\mathrm{min}}$  \\ \hline \hline
  & 
  3$\sigma$ & 
  $270$ & 
  $195$ &
  $2.57 \times 10^{-3} $ &
  $0.20460$&
  $0.63519$&
  $33.58$&
  $41.60$&
  $8.47$&
  $-68.65$&
  $-5.86$&
  $14.77$& 
  $4.63 \times 10^{-4}$\\
  NH & 
  BFP$\pm 1\sigma$ &
  $270$ & 
  $195$ &
  $2.57 \times 10^{-3} $ &
  $0.22256$&
  $0.64507$&
  $33.59$&
  $41.61$&
  $8.46$&
  $-70.74$&
  $-5.79$&
  $14.67$& 
  $3.13 \times 10^{-4}$\\  
  & 
  BFP & 
  $270$ & 
  $195$ & 
  $2.57 \times 10^{-3} $ &
  $0.21492$ &
  $0.64008$&
  $33.80$ &
  $41.63$ &
  $8.45$ &
  $-69.85$&
  $-5.80$&
  $14.73$& 
  $8.75  \times 10^{-2}$ \\  \hline \hline 
  & 
  3$\sigma$ & 
  $290$&
  $187$&
  $2.49 \times 10^{-2}$& 
  $0.59943$ &
  $0.01999$ &
  $33.67$ &
  $50.08$ &
  $8.48$ &
  $-80.90$ &
  $-5.25$ & 
  $-2.18$ & 
  $2.30 \times 10^{-2}$\\
  IH & 
  BFP$\pm 1\sigma$ & 
  $290$&
  $187$&
  $2.49 \times 10^{-2}$& 
  $0.59888$ &
  $0.01995$ &
  $33.74$ &
  $50.05$ &
  $8.49$ &
  $-80.88$&
  $-5.25$ & 
  $-2.18$ & 
  $4.56 \times 10^{-2}$ \\  
  & 
  BFP & 
  $290$&
  $187$&
  $2.49 \times 10^{-2}$& 
  $0.59798$ &
  $0.01971$ &
  $33.83$ &
  $49.99$ &
  $8.49$ &
  $-80.83$ &
  $-5.25$ &
  $-2.19$ &
  $1.08 \times 10^{-1}$ \\  \hline \hline     
\end{tabular}
\caption{
 Numerical values obtained in the BFP for the five parameters, the lepton mixing angles 
 and the phase factors associated with the CPV. 
 These results were obtained considering to $\Delta m_{ij}^{2}$ at BFP, 
 BFP$\pm \sigma$ and 3$\sigma$ range~\cite{Esteban:2016qun}, 
 and simultaneously  $\Phi_{\ell 1}$,  $\Phi_{\ell 2}$, $\delta_{\ell}$, $\delta_{\nu}$, 
 and $m_{\nu_{1 [3] } }$ are free parameters in the $\chi^{2}$ function. }
 \label{tab:table1}
\end{center}
\end{table}
}
\begin{figure}[!htbp]
 \begin{center}
  \begin{tabular}{cc}
  \subfigure{ \includegraphics[width=8.7cm, height=6.3cm]{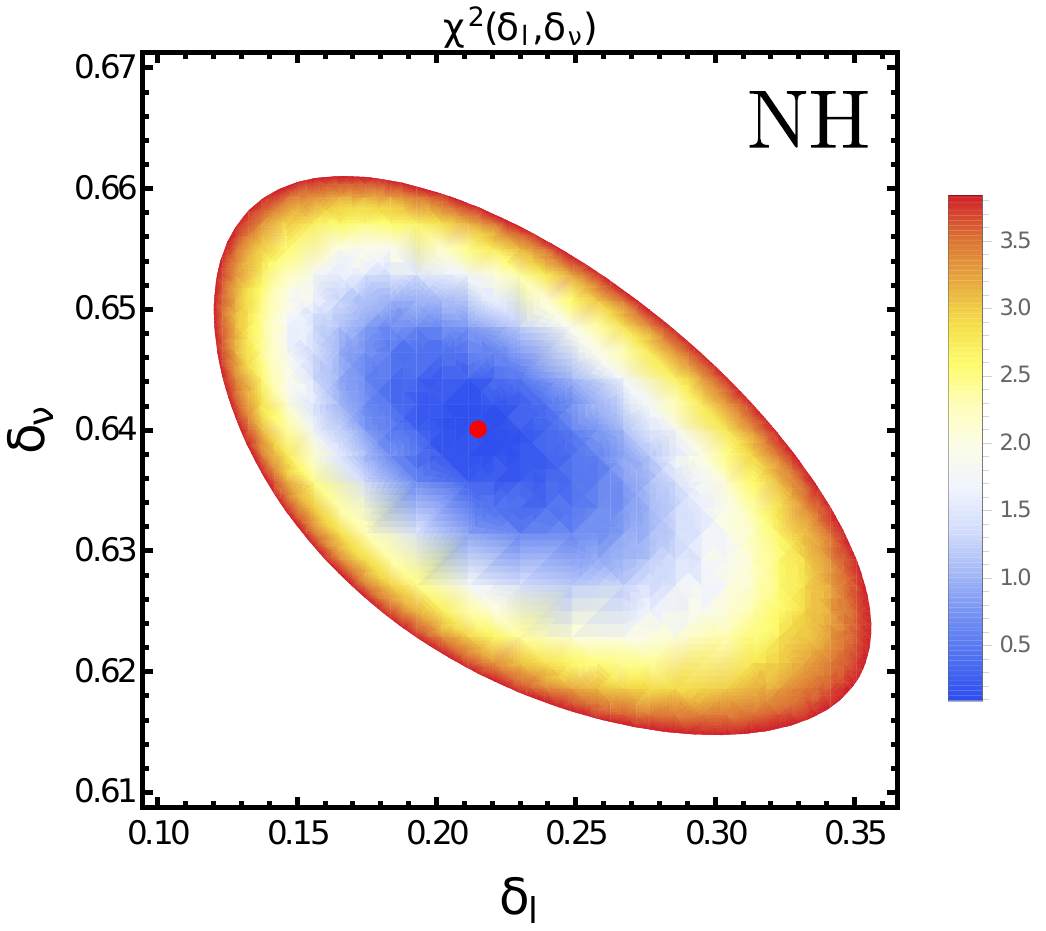}}
  \subfigure{ \includegraphics[width=8.7cm, height=6.3cm]{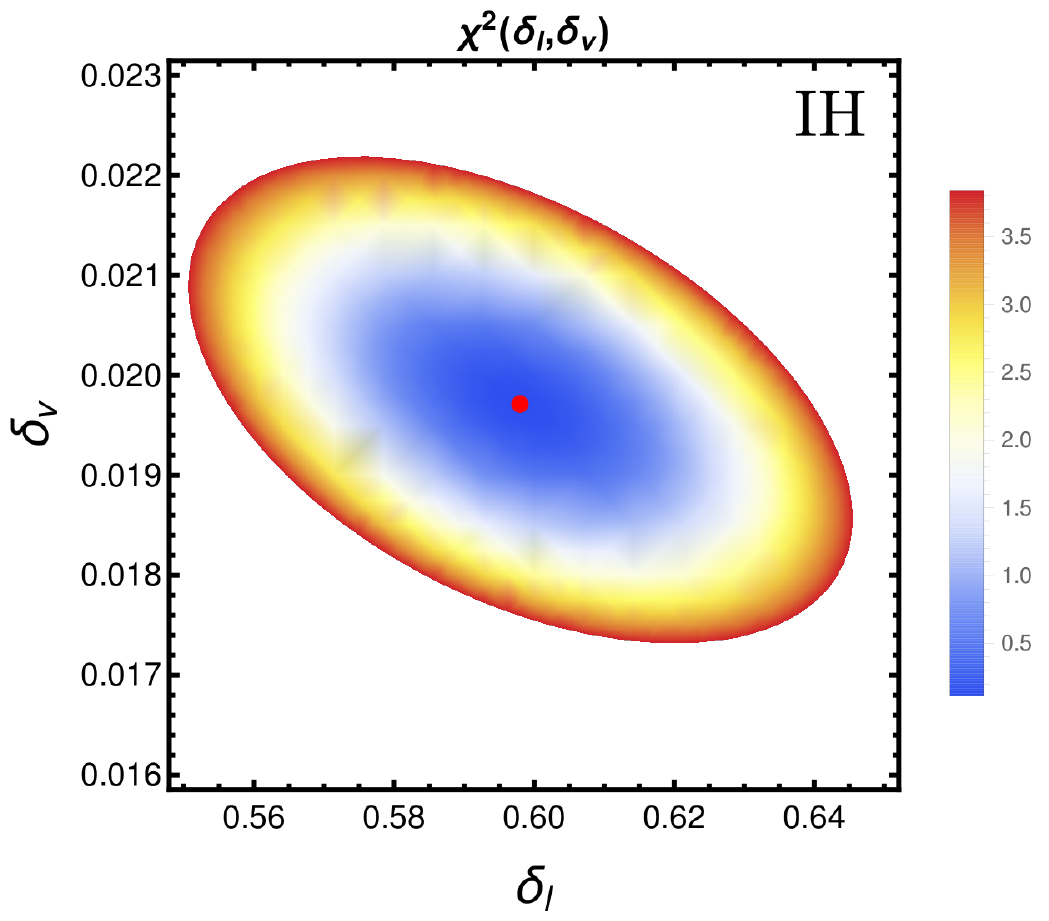}} 
  \end{tabular}
  \caption{The allowed regions in the parameter space for $\delta_l$ and $\delta_\nu$ 
  at 3$\sigma$ C.L., the red point ($\red \bullet$) represents the BFP.
  The left-panel is for normal hierchy, while the right-panel is for the inverted 
  hierarchy. Here, the $m_{\nu_{1[3]}}, \phi_{l1}$, and $\phi_{l2}$ parameters 
  are fixed to the values obtained when the $\Delta m_{ij}^{2}$ neutrino oscillation 
  parameters are given at BFP, see Table~\ref{tab:table1}.}
   \label{Fig:calorchi}
 \end{center}
\end{figure}

To validate our hypothesis where the $S_{3}$ horizontal flavor symmetry is 
explicitly sequential breaking according to the chain 
$S_{ 3L }^{ \texttt{j} } \otimes S_{ 3R }^{ \texttt{j} } 
\supset S_{3}^{\mathrm{diag}} \supset S_{2}^{\mathrm{diag}}$,  
hence all  fermion mass matrices are represented through a matrix with two texture 
zeroes, we make a likelihood test where the $\chi^{2}$ function is defined as: 
\begin{equation}\label{chicuadradadef}
 \chi^{2} = 
  \sum_{i < j}^{3} 
  \frac{ 
   \left( 
    \sin^{2} \theta_{ij}^{ ^{\mathrm{exp} } } 
    - \sin^{2} \theta_{ij}^{ ^{ \mathrm{th} } }  \right)^2 
  }{
   \sigma_{ \theta ij }^{2} 
  }.
\end{equation} 
In this expression, 
the ``$\mathrm{th}$'' superscript is used to denote the  theoretical expressions of 
lepton mixing angles, while the terms with superscript ``$\mathrm{exp}$'' denote to 
the experimental data with uncertainty $\sigma_{\theta_{ij}}$ for lepton mixing angles.
For these latter we consider the following values, at 
BFP$\pm 1 \sigma$~\cite{Esteban:2016qun}:
\begin{equation}\label{Eq:Exp:Thetas}
 \begin{array}{l}
  \sin^{2} \theta_{12}^{ ^{\mathrm{exp}} } (10^{-1}) = 
   3.06 \pm 0.12 , \quad
  \sin^{2} \theta_{23}^{ ^{\mathrm{exp}} } (10^{-1}) = 
   \left \{ \begin{array}{l}
    4.41_{-0.21}^{+0.27} , \\
    5.87_{-0.24}^{+0.20} ,
   \end{array} \right.  \quad
  \sin^{2} \theta_{13}^{ ^{\mathrm{exp}} } (10^{-2}) = 
   \left \{ \begin{array}{l}
    2.166 \pm 0.0075 , \\
    2.179 \pm 0.0076 ,
   \end{array}    \right.  
 \end{array}
\end{equation}
the values in the first and second row correspond to a normal and inverted hierarchy in 
the neutrino mass spectrum, respectively.
From expressions in 
Eqs.~(\ref{eq:Real-O}),~(\ref{Eq:PMNS_th}),~(\ref{Eq:senoscuadrados}), 
and~(\ref{Eq:For_mnu}), it is easy to conclude that the $\chi^{2}$ function depends of five 
free parameters  
$\chi^{2} = 
 \chi^{2} \left( \Phi_{\ell 1}, \Phi_{\ell 2}, \delta_{\ell}, \delta_{\nu}, 
 m_{\nu_{1 [3] } } 
\right)$. 
However, the $\chi^{2}$ function depends only on three experimental data which 
correspond to the leptonic flavor mixing angles. 
Therefore, if simultaneously we consider $\Phi_{\ell 1}$, $\Phi_{\ell 2}$, 
$\delta_{\ell}$, $\delta_{\nu}$, and $m_{\nu_{1 [3] } }$ as free parameters in the 
likelihood test, we can only determine the values of these parameters in the best fit 
point~(BFP).
In accordance with the above, we first seek the BFP by means of a likelihood test where 
the $\chi^{2}$ function have all those the five free parameters 
$\Phi_{\ell 1}$, $\Phi_{\ell 2}$, $\delta_{\ell}$, $\delta_{\nu}$, and 
$m_{\nu_{1 [3] } }$. 
To minimize the $\chi^{2}$ function we have done a scanning of the parameter space where we 
considered the following values for the charged lepton masses~\cite{Olive:2016xmw}
\begin{equation}
 \begin{array}{l}
  m_{e}=0.5109998928 \pm 0.000000011, \quad 
  m_{\mu}=105.6583715\pm 0.0000035, \quad
  m_{\tau}=1776.82\pm 0.16,
 \end{array}
\end{equation}
while in Eq.~(\ref{Eq:Exp:Thetas}) are given the experimental values for leptonic mixing 
angles. 

\begin{figure}[!htbp]
 \begin{center}
  \begin{tabular}{cc}
     \subfigure{ \includegraphics[width=7.7cm, height=5.3cm]{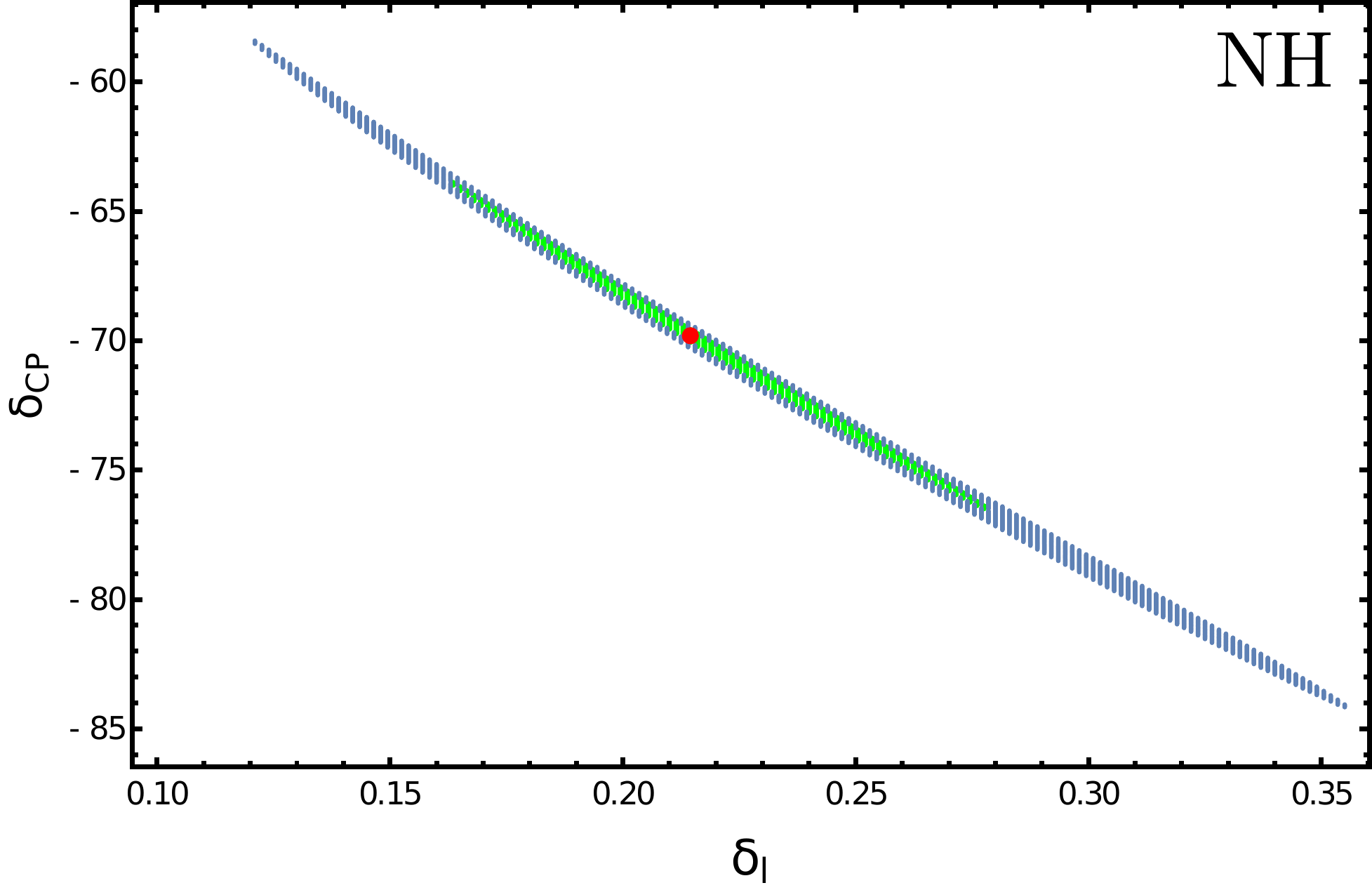} }
   \subfigure{   \includegraphics[width=7.7cm, height=5.3cm]{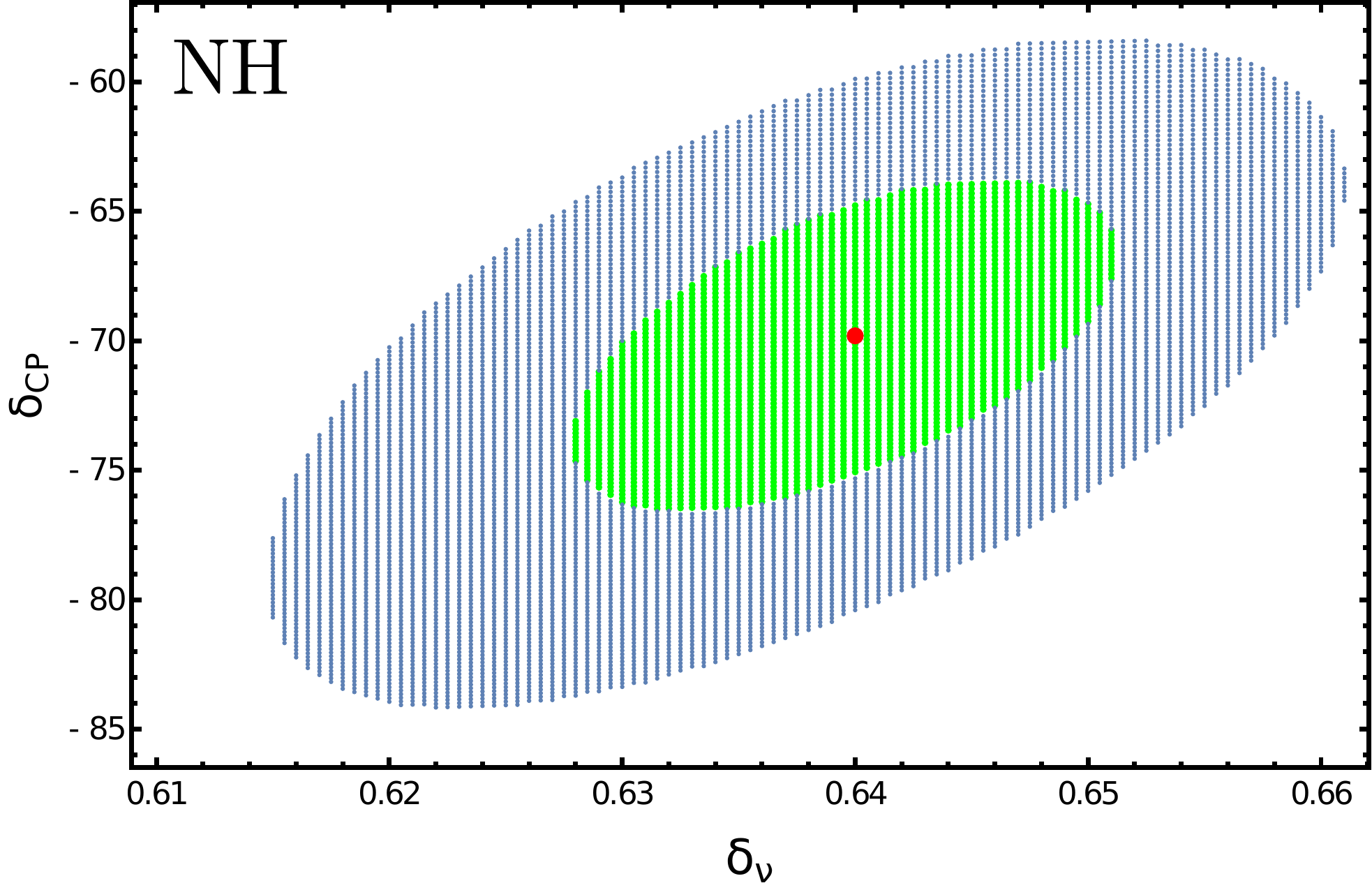}} \\
       \subfigure{ \includegraphics[width=7.7cm, height=5.3cm]{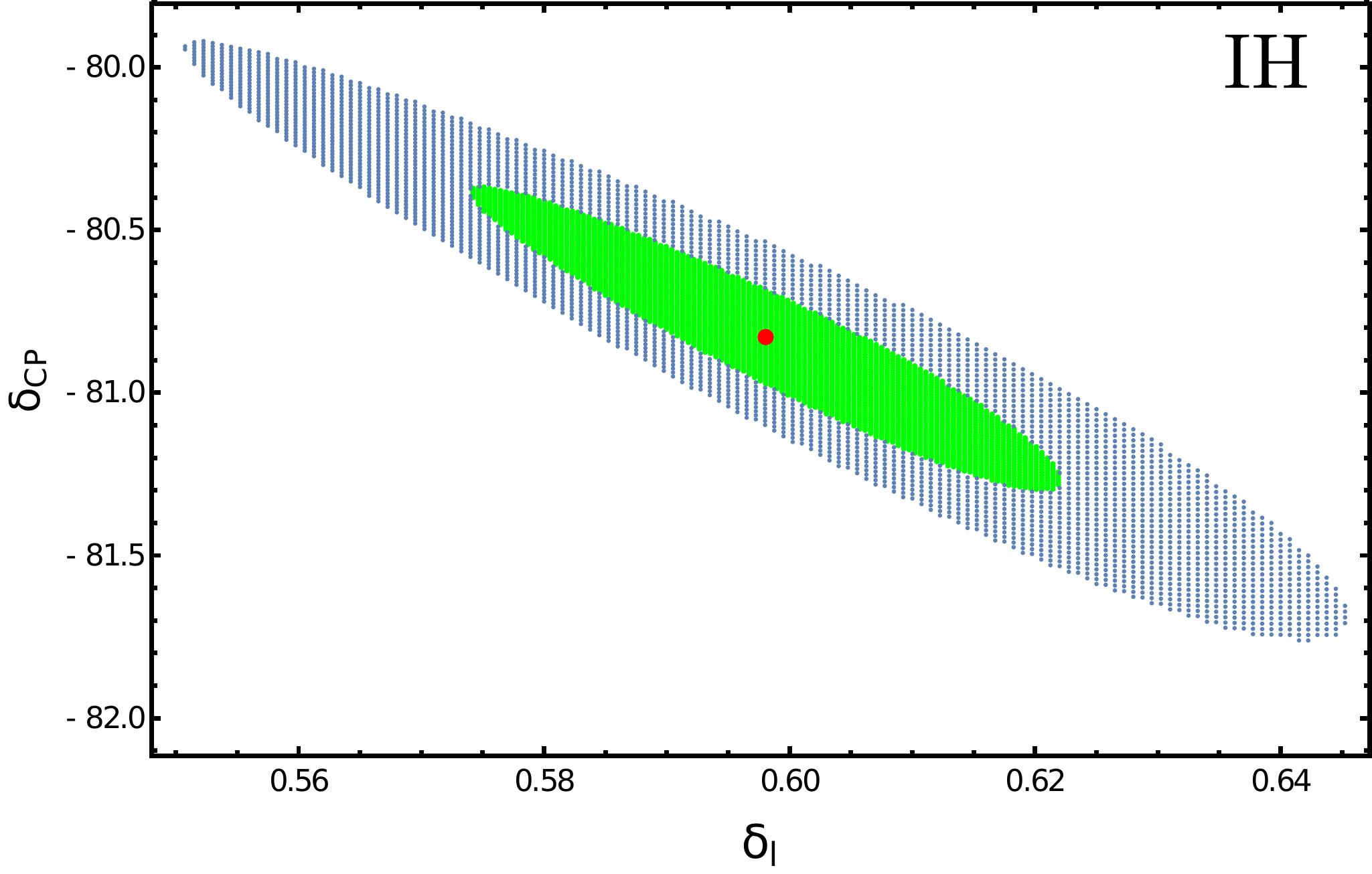} }
   \subfigure{   \includegraphics[width=7.7cm, height=5.3cm]{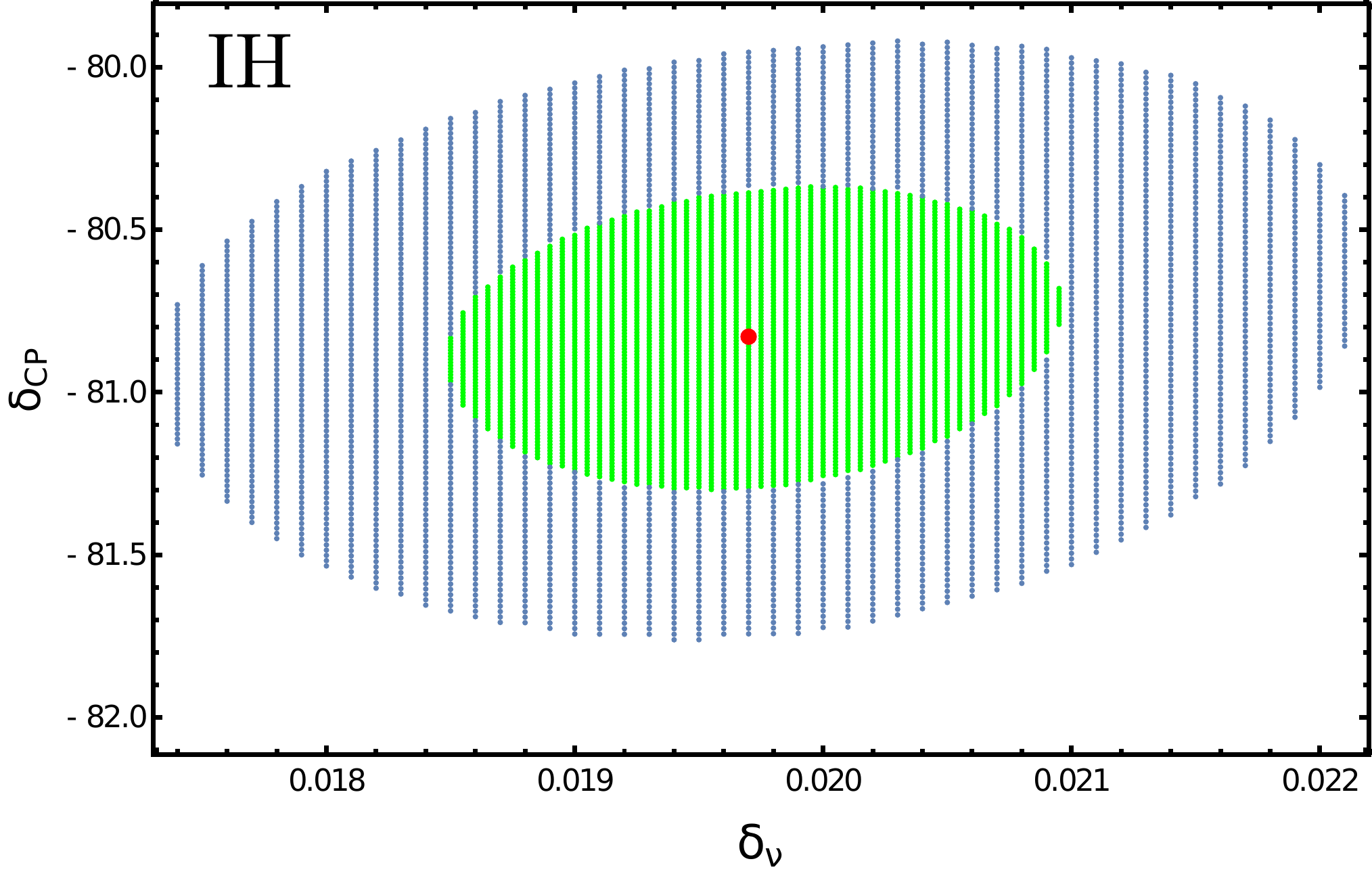}}
  \end{tabular}
  \caption{
   The allowed regions for ``Dirac-like'' CPV phase $\delta_{\mathrm{CP}}$, 
   and the free parameters $\delta_l$ and $\delta_\nu$. 
   The $m_{\nu_{1 [3] } }$,  $\Phi_{\ell 1}$ and $\Phi_{\ell 2}$ parameters are fixed to 
   the values given in Table~\ref{tab:table1} for the BFP. 
   The BFP is denoted by red point ${\color{red}\bullet}$, 
   the green and blue regions are for 1$\sigma$ and 3$\sigma$ C.L., respectively. 
   The upper and lower panels correspond to normal and inverted hierarchy, respectively.}
   \label{Fig:deltacp}
 \end{center}
\end{figure}
In the Table~\ref{tab:table1} we show the numerical values obtained in the BFP for the 
five parameters, the lepton mixing angles and the phase factors associated with the CPV. 
All of these results were obtained considering to $\Delta m_{ij}^{2}$ at BFP, 
BFP$\pm \sigma$ and 3$\sigma$ range, and simultaneously  $\Phi_{\ell 1}$, 
$\Phi_{\ell 2}$, $\delta_{\ell}$, $\delta_{\nu}$, and $m_{\nu_{1 [3] } }$ are free 
parameters in the $\chi^{2}$ function of the likelihood test.

Now, as we know the numerical values of the five free parameters at BFP,  
we perform a new $\chi^{2}$ analysis for the case when the oscillation parameters 
$\Delta m_{ij}^{2}$ take the values at BFP and where we fix $m_{\nu_{1 [3] } }$, 
$\Phi_{\ell 1}$ and $\Phi_{\ell 2}$ parameters to the values given in 
Table~\ref{tab:table1}.

So, $\chi^{2} = \chi^{2}(\delta_l,\delta_\nu)$ function implies one degree of 
freedom.
In the Fig.~\ref{Fig:calorchi}, we show the allowed regions in the parameter space for 
$\delta_l$ and $\delta_\nu$ at 3$\sigma$ C.L., the red point ($\red \bullet$) represent 
the BFP.  The left-panel is for normal hierarchy and we can see that $\delta_\nu$ and 
$\delta_l$ parameters are of the order of $10^{-1}$. The right-panel is for the 
inverted hierarchy and here $\delta_\nu$ parameter is $\sim10^{-1}$, while the $\delta_l$ 
parameter is of the order of $10^{-2}$.

\begin{figure}[!htbp]
 \begin{center}
  \begin{tabular}{cc}
     \subfigure{ \includegraphics[width=7.7cm, height=5.3cm]{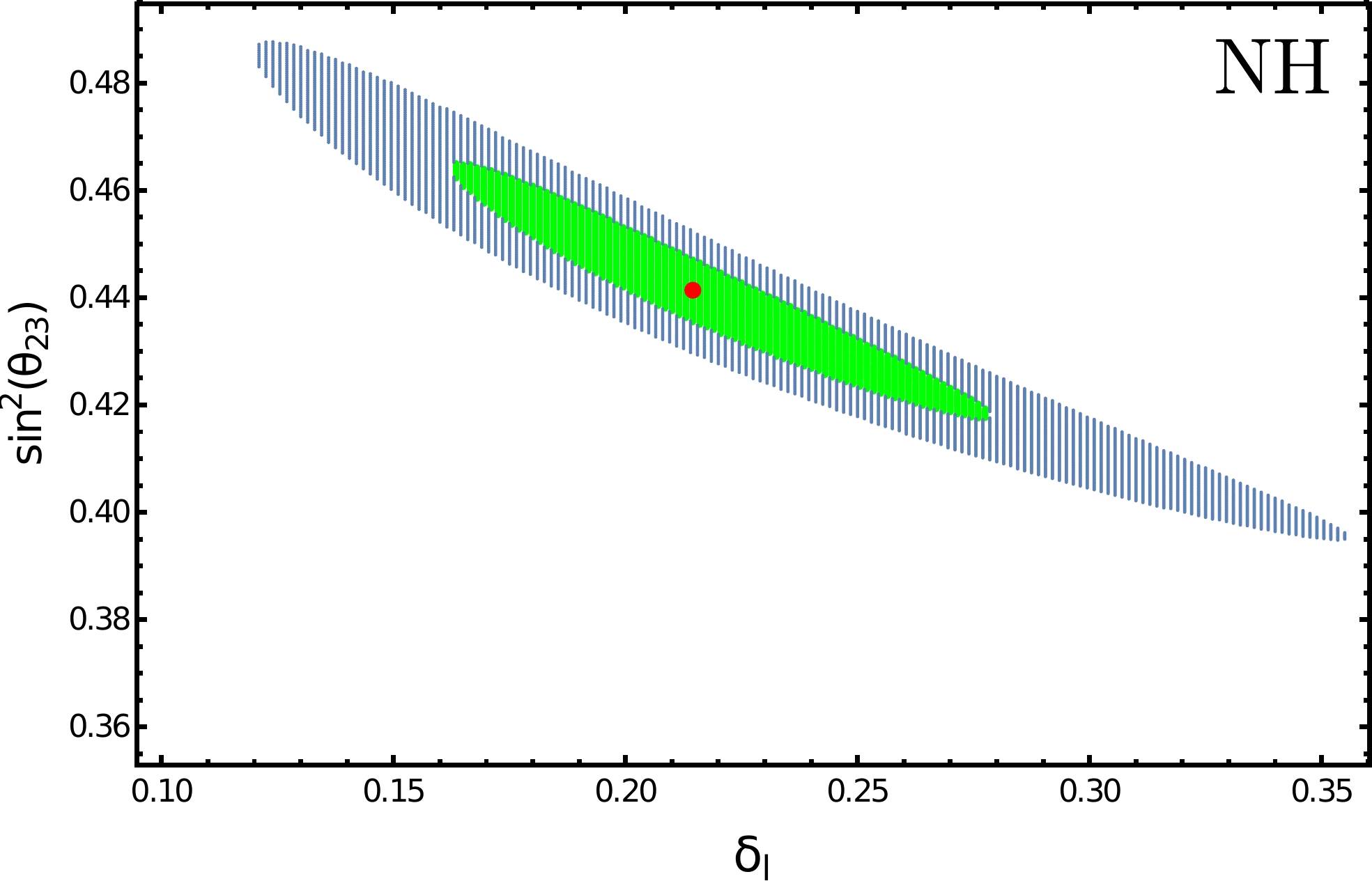} }
    \subfigure{ \includegraphics[width=7.7cm, height=5.3cm]{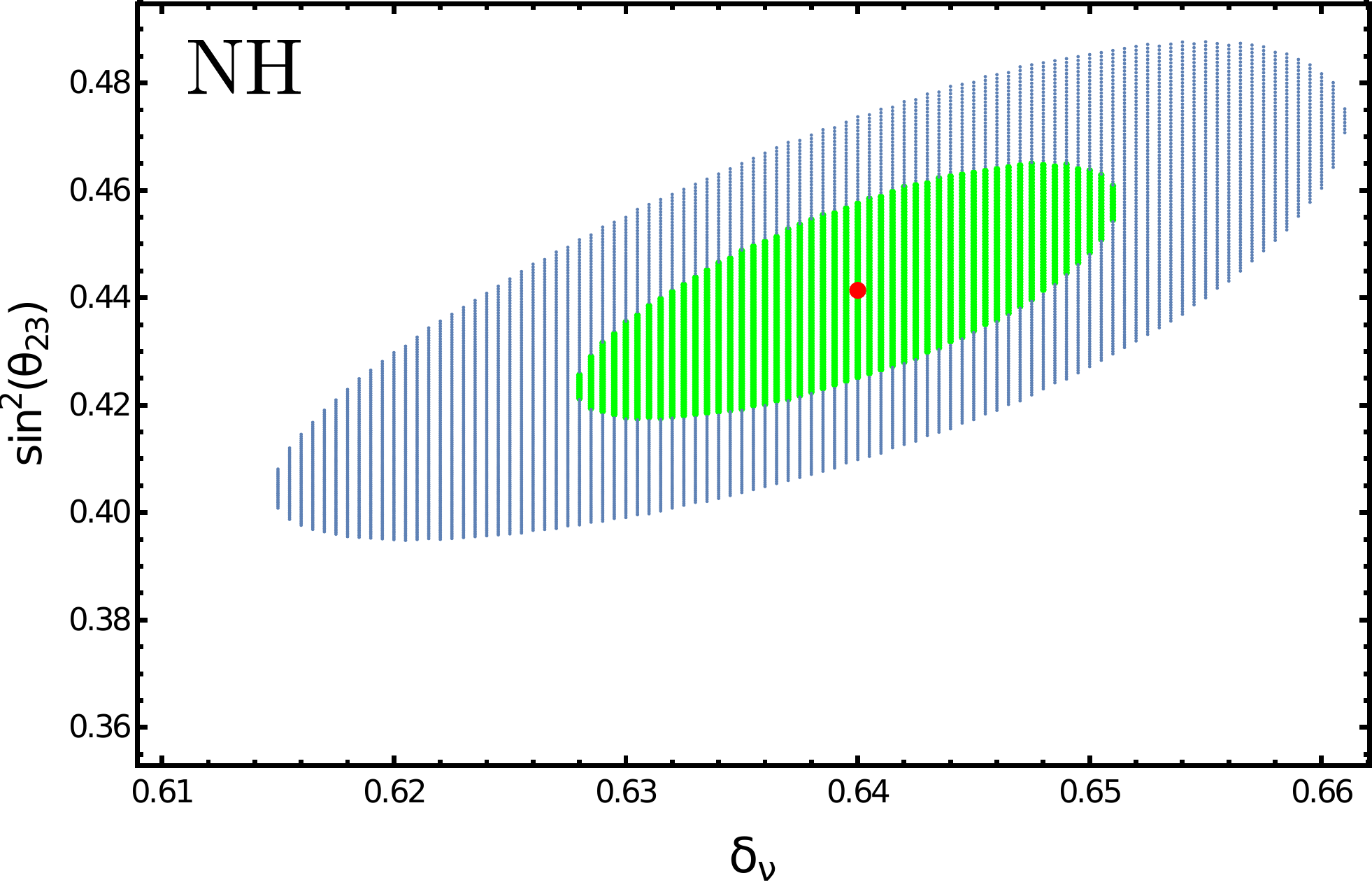}} \\
        \subfigure{ \includegraphics[width=7.7cm, height=5.3cm]{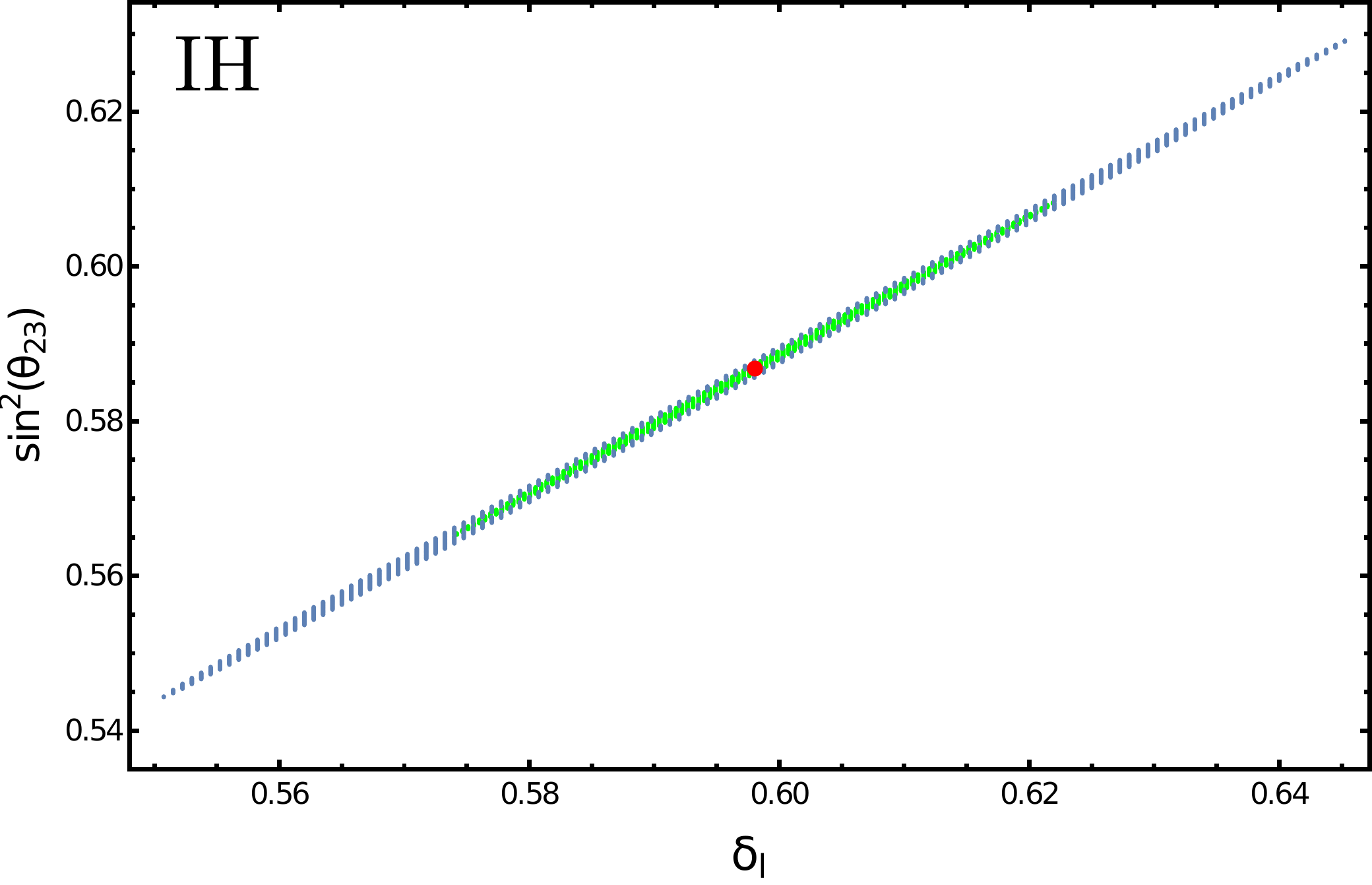} }
    \subfigure{ \includegraphics[width=7.7cm, height=5.3cm]{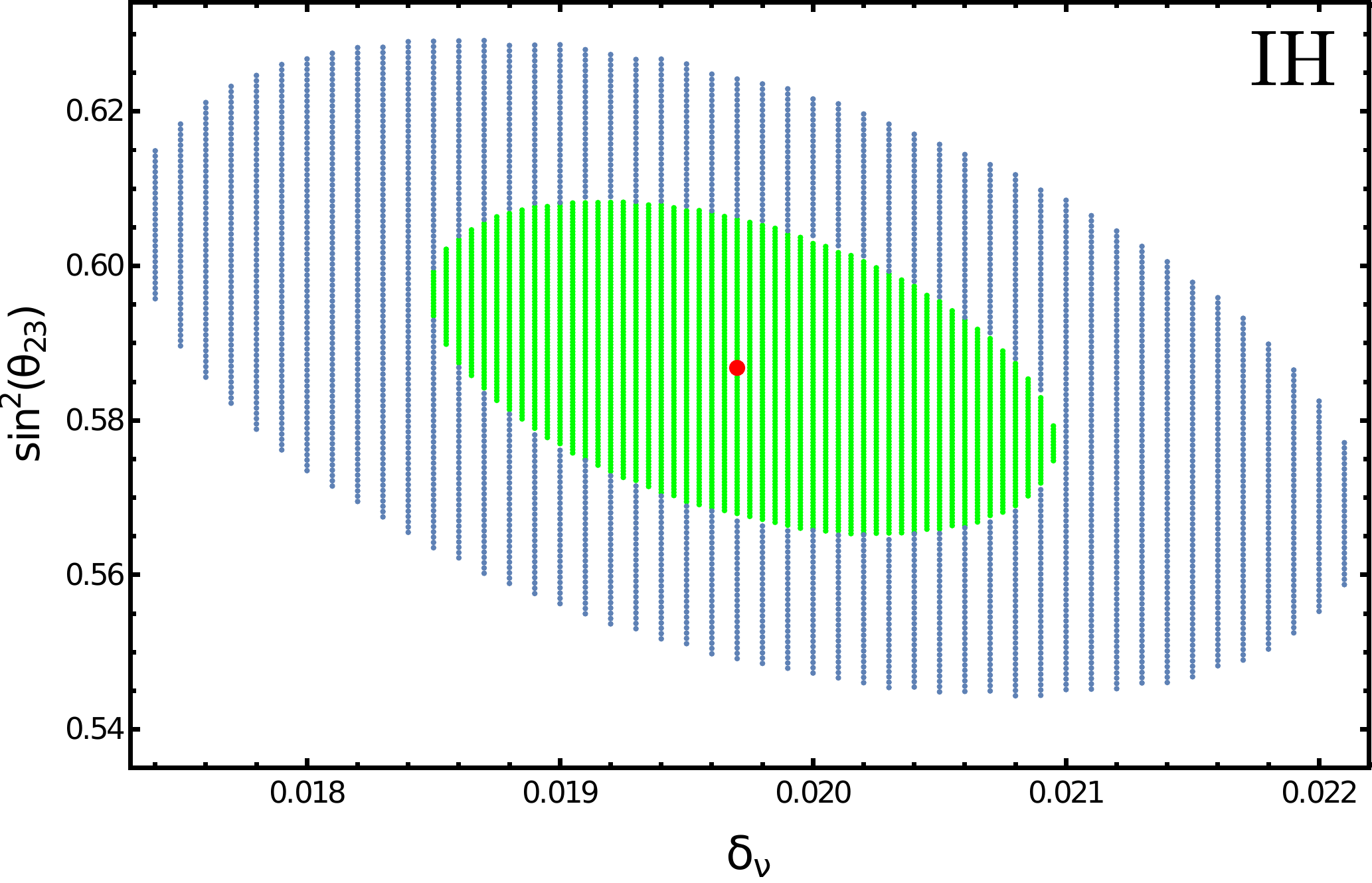}}
  \end{tabular}
  \caption{
   The allowed regions for atmospheric mixing angle $\sin^2 \theta_{23}$
   and the free parameter $\delta_l$ and $\delta_\nu$. 
   The $m_{\nu_{1 [3] } }$,  $\Phi_{\ell 1}$ and $\Phi_{\ell 2}$ parameters are fixed to 
   the values given in Table~\ref{tab:table1}, when the oscillation parameters 
   $\Delta m_{ij}^{2}$ takes the values  at BFP. 
   The BFP is denoted by red point ${\color{red}\bullet}$, 
   the green and blue regions are for 1$\sigma$ and 3$\sigma$ C.L., respectively. 
   The upper and lower panels correspond to normal and inverted hierarchy, respectively correspondingly.
  }  \label{Fig:senos}
 \end{center}
\end{figure}

Associated to the parameter regions of $\delta_{\nu}$ and $\delta_{l}$ given in 
Fig.~\ref{Fig:calorchi} for both hierarchies on neutrino mass spectrum, and based on 
Eq.~(\ref{Eq:CP_Phases}), we found the predicted regions by the 
$\nu\textrm{2HDM}\otimes S_{3}$ for 
``Dirac-like'' phase $\delta_{CP}$. These regions are shown in Fig.~\ref{Fig:deltacp}. 
In concordance with experimental data, plots as function of $\delta_{l}$ are more 
restricted than the other one as function of $\delta_{\nu}$. These  results correspond closely with
allowed regions obtained in the global fit reported in 
Ref.~\cite{Esteban:2016qun}.

In the same way, for both hierarchies, we analyze the three leptonic flavor mixing angles, 
but in Fig.~\ref{Fig:senos} we just show  the allowed regions for the atmospheric 
mixing angle $\theta_{23}$, at BFP, $\pm 1\sigma$ and $3\sigma$ C.L. In order to round 
the above results, from our analysis we obtain the following values for the three mixing 
angles, at BFP$\pm 1\sigma$ C.L.:  
\begin{equation}
\begin{array}{c}
\sin^{2} \theta_{12}^{ ^{\mathrm{th}} } (10^{-1}) = 
\left \{ \begin{array}{l}
3.09_{-0.065}^{+0.066} , \\
3.10_{-0.011}^{+0.011} , 
\end{array} \right.  \quad
\sin^{2} \theta_{23}^{ ^{\mathrm{th}} } (10^{-1}) = 
\left \{ \begin{array}{l}
4.41_{-0.14}^{+0.10}  , \\
5.87_{-0.223}^{+0.224} ,  
\end{array} \right.  \quad
\sin^{2} \theta_{13}^{ ^{\mathrm{th}} } (10^{-2}) = 
\left \{ \begin{array}{l}
2.160 \pm 0.14 ,\\
2.177 \pm 0.12 .
\end{array}    \right. 
\end{array}
\end{equation}
We also obtained the following allowed value ranges at BFP$\pm 1\sigma$ for the 
``Dirac-like'' phase $\delta_{\mathrm{CP}}$, as well as for the two Majorana phase factors 
$\phi_{12}$ and $\phi_{13}$:
\begin{equation}\label{Eq:CP_phases-Fit}
\begin{array}{c}
\delta_{CP} (^{\circ}) = 
\left \{ \begin{array}{l}
-69.8_{-6.110}^{+5.508}  , \\
-80.83_{-0.709}^{+0.652} ,
\end{array} \right.  \quad
\phi_{12} (^{\circ}) = 
\left \{ \begin{array}{l}
-5.800_{-0.150}^{+0.170} , \\
-5.24_{-0.148}^{+0.153}  ,
\end{array} \right.  \quad
\phi_{13} (^{\circ}) = 
\left \{ \begin{array}{l}
14.744_{-1.366}^{+1.266}   ,\\
-2.190_{-0.0005}^{+0.0030} .
\end{array} \right.  
\end{array}
\end{equation}
From Eq.~(\ref{Eq:CP_phases-Fit}) and the Table~\ref{tab:table1} we can conclude 
that values for the $\delta_{\text{CP}}$ phase obtained 
in our scheme are consistent with a maximal CP violation. 

Finally, as a immediate result of the above likelihood analysis, 
the entries magnitude of ${\mathbf U}_{\textrm{PMNS}}$ mixing matrix can numerically 
computed. 
So, at $3\sigma$ C.L., we have that ${\mathbf U}_{\textrm{PMNS}}$ matrix takes the form: 
\begin{equation}
\begin{pmatrix}
0.822^{+0.0044}_{-0.0045} & 0.550^{+0.0055}_{-0.0054} & 0.147^{+0.0047}_{-0.0048} \\
0.395^{+0.0181}_{-0.0154} & 0.642^{+0.0008}_{-0.0001} & 0.657^{+0.0082}_{-0.0111} \\
0.410^{+0.0056}_{-0.0089} & 0.534^{+0.0045}_{-0.0056} & 0.739^{+0.0088}_{-0.0064} \\
\end{pmatrix}, \qquad {\bf {\textrm{Normal Hierarchy}}},
\end{equation}
\begin{equation}
\begin{pmatrix}
0.822^{+0.0012}_{-0.0012} & 0.551^{+0.0007}_{-0.0006} & 0.147^{+0.0041}_{-0.0041} \\
0.355^{+0.0072}_{-0.0071}& 
0.547^{+0.0144}_{-0.0149} & 
0.758^{+0.0138}_{-0.0141} \\
0.446^{+0.0077}_{-0.0080} & 0.630^{+0.0120}_{-0.0122} & 
0.636^{+0.0174}_{-0.0178} \\
\end{pmatrix}, \qquad {\bf {\textrm{Inverted Hierarchy}}}.
\end{equation}
%
%
\section{Phenomenological implications}\label{sec:PenImp}
%
In the above section we have seen that in our theoretical framework, where the $S_{3}$ flavor 
symmetry sets up that the fermion mass matrices should have two texture zeroes, 
we can reproduce the values of oscillation parameters in very good agreement with the last
experimental data. 
In the following, we shall investigate the phenomenological implications of these 
results for the neutrinoless double beta decay ($0\nu \beta \beta $) and the CP violation in 
neutrino oscillations in matter.
%
%
%
\subsection{Neutrinoless double beta decay\label{subsec:neutrinoless}}  
%
%
%
%
\begin{figure}[t]
 \begin{center}
  \begin{tabular}{c}
   \includegraphics[width=0.45\linewidth]{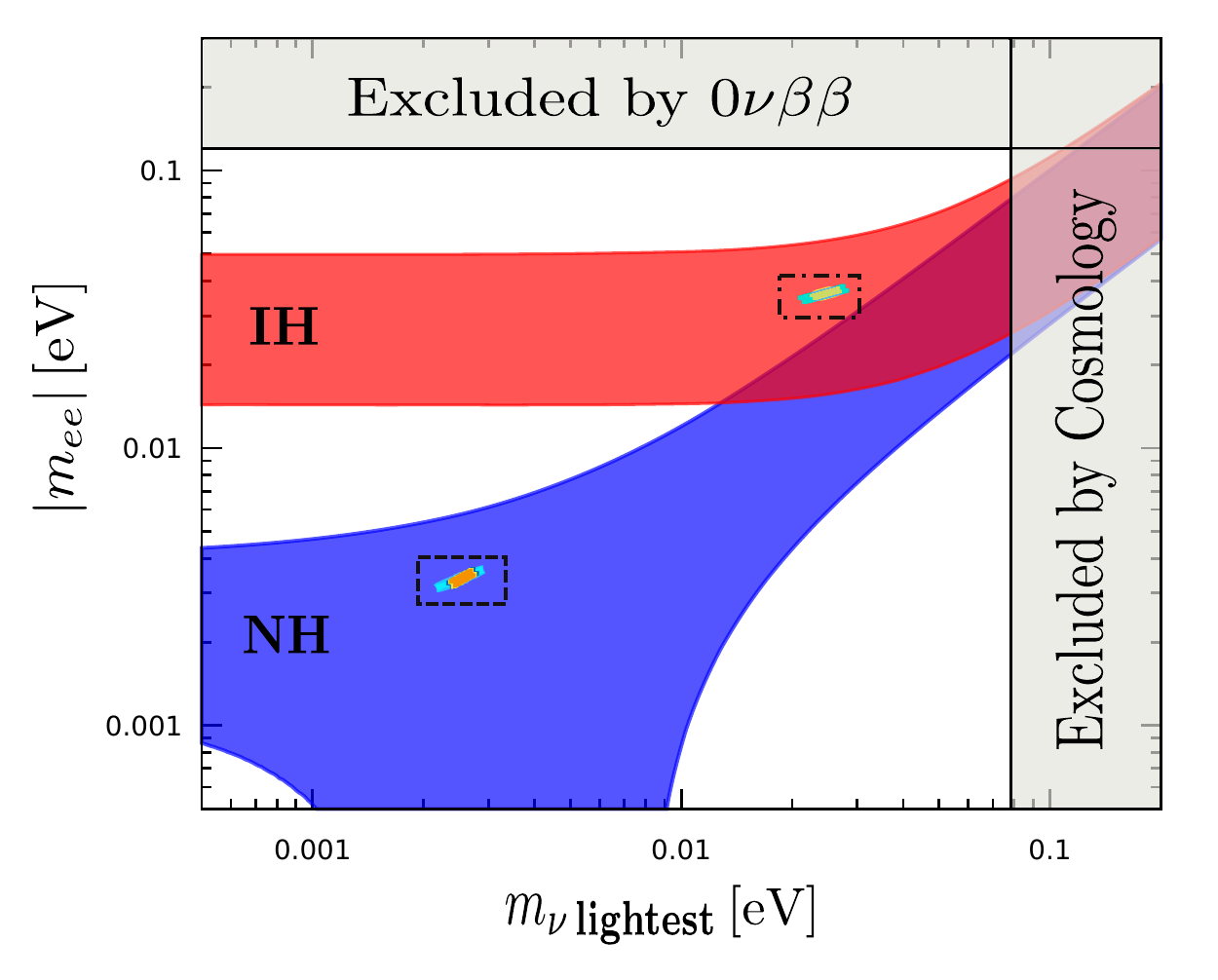} \\
   \includegraphics[width=0.4\linewidth]{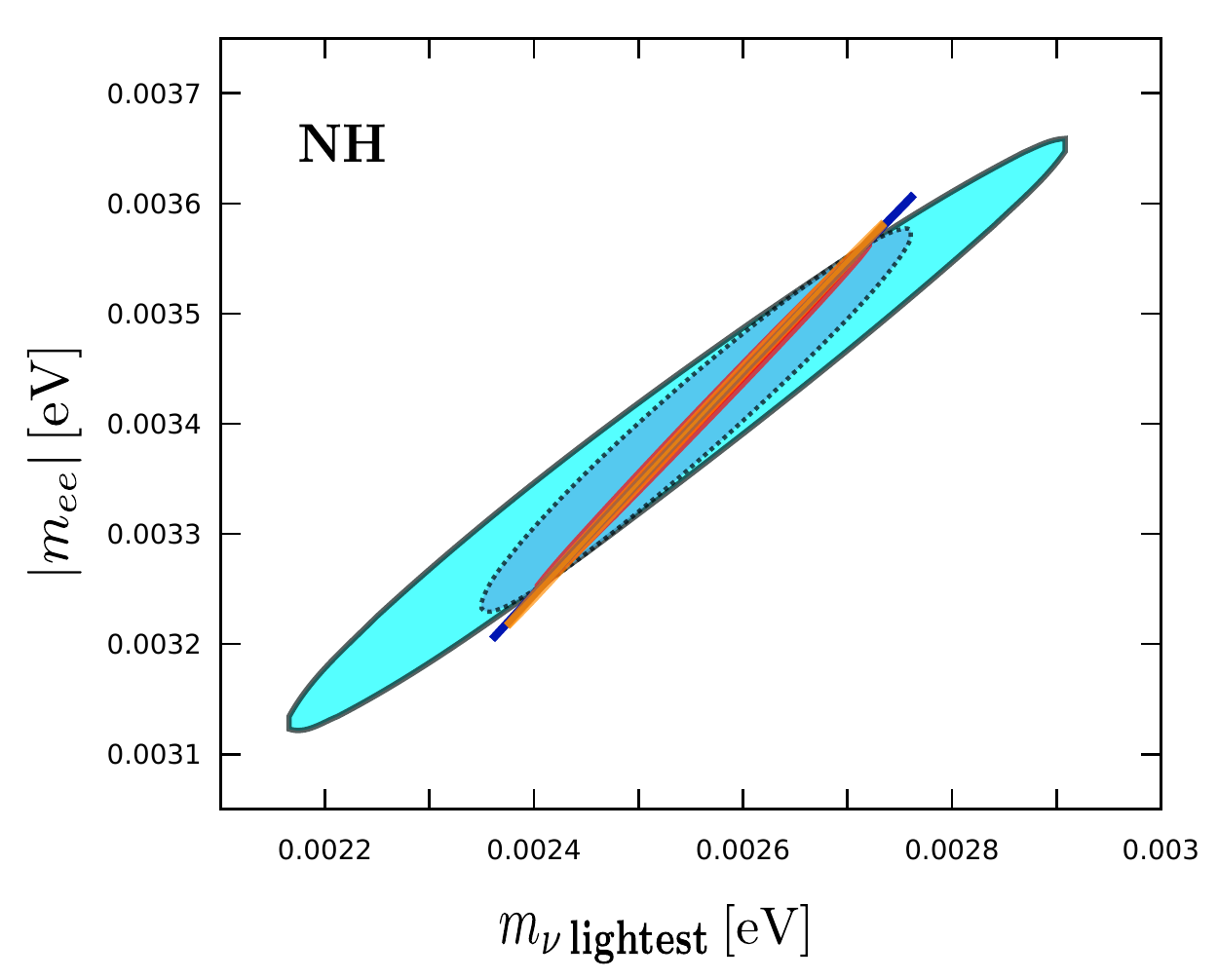} \includegraphics[width=0.4\linewidth]{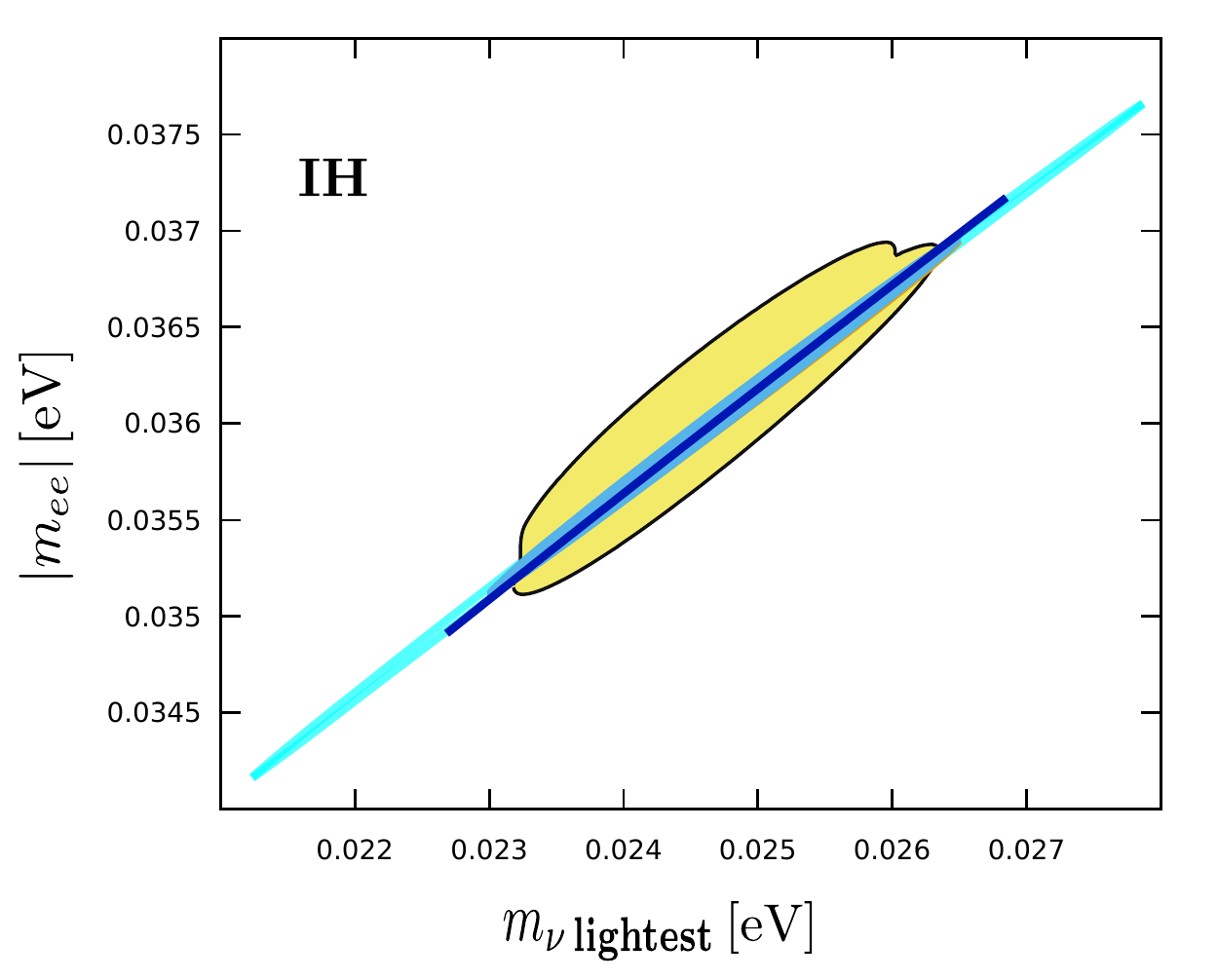} 
  \end{tabular}
  \caption{In the upper panel we show the effective mass $\left| m_{ee} \right|$ which is 
   involved in the $0\nu \beta \beta$ decay.
   The red and blue bands are obtained with the current experimental data on neutrino  
   oscillations, at 3$\sigma$~\cite{Esteban:2016qun}, for an inverted and normal neutrino 
   mass hierarchy,  respectively.
   On the one hand, from the combination of EXO-200~\cite{Auger:2012ar,Albert:2014awa} 
   and KamLAND-ZEN~\cite{Gando:2012zm} results we have the follow upper bound for 
   $\left| m_{ee} \right| < 0.120$eV. 
   On the other hand, from the results reported by Planck Collaboration we have that    
   $\sum_{i} m_{i} < 0.230$eV at $95\%$ level~\cite{Ade:2015xua}, thus an upper bound on 
   the lightest neutrino mass is established.
   In the left- and right-lower panels we show a zoom in of the allowed regions for 
   $\left| m_{ee} \right|$ obtained at 95\% C.L. in the context of 
   $\nu$2HDM$\otimes S_{3}$ for a normal and inverted hierarchy, respectively. 
  }\label{Fig:zbbb}
 \end{center}
\end{figure}

The $0\nu \beta \beta$ is a rare second-order weak process where a nucleus $(A,Z)$ decays 
into another one by the emission of two electrons, whose mode decay is $(A,Z) \rightarrow 
(A, Z + 2) + e^{-} + e^{-}$. 
The observation of this process would establish that neutrino 
are Majorana particles and that total lepton number is not a conserved symmetry in  
nature~\cite{Schechter:1981bd,Duerr:2011zd}.  In the most simple version of the process, 
the amplitude for the decay is proportional to a quantity called the effective mass 
$m_{ee}$~\cite{Beringer:1900zz,Barger:2003vs,King:2013psa}. In the symmetric parametrization of lepton mixing matrix 
the effective mass parameter have the shape~\cite{Schechter:1980gr,Chen:2015siy}
\begin{equation}
 \left| m_{ee} \right| =
  \left| m_{ \nu_{1} } \cos^{2} \theta_{12} \cos^{2} \theta_{13}
   + m_{ \nu_{2} } \sin^{2} \theta_{12} \cos^{2} \theta_{13} e^{ - i 2 \phi_{12} }
   + m_{ \nu_{3} } \sin^{2} \theta_{13} e^{ - i 2 \phi_{13} } \right|\, ,
\end{equation}
where $\phi_{12}$ and $\phi_{13}$ are the Majorana phases given in 
Eq.~(\ref{Eq:CP_Phases}). 
In the Fig.~\ref{Fig:zbbb} we show the allowed regions for the  magnitude of effective 
mass parameter $m_{ee}$, which were obtained in the context of $\nu$2HDM$\otimes S_{3}$.
Each one of these regions was obtained by setting the values of some of the five free 
parameters in the $\chi^{2}$ function, Eq.~(\ref{chicuadradadef}), to the values given in 
the Table~\ref{tab:table1} for $\Delta m_{ij}^{2}$ at BFP. 
Then, for both hierarchies in the lower panels of Fig.~\ref{Fig:zbbb}, the blue lines 
were obtained by means a likelihood test where the values of $\phi_{\ell 1}$, 
$\phi_{\ell 2}$, $\delta_{e}$ and $\delta_{\nu}$ are fixed, while
$m_{\nu_{\mathrm{lightest} } }$ is free parameter. 
The orange bands were obtained by means a likelihood test where the values of 
$\phi_{\ell 2}$, $\delta_{e}$ and $\delta_{\nu}$ are fixed, while
$m_{\nu_{\mathrm{lightest} } }$ and $\phi_{\ell 1}$ are free parameters.  
The yellow bands were obtained by means of a likelihood test where the values of 
$\phi_{\ell 1}$, $\delta_{e}$ and $\delta_{\nu}$ are fixed, while
$m_{\nu_{\mathrm{lightest} } }$ and $\phi_{\ell 2}$ are free parameters.  
The sky blue bands were obtained through a likelihood test where the values of 
$\phi_{\ell 1}$, $\phi_{\ell 2}$  and $\delta_{\nu}$ are fixed, while
$m_{\nu_{\mathrm{lightest} } }$ and  $\delta_{e}$ are free parameters. 
Finally, the turquoise bands were obtained via a likelihood test where the values of 
$\phi_{\ell 1}$, $\phi_{\ell 2}$  and $\delta_{e}$ are fixed, while
$m_{\nu_{\mathrm{lightest} } }$ and  $\delta_{\nu}$ are free parameters. 

To round the previous results, in the Table~\ref{tab:table2} we show the 
allowed numerical ranges, at 95\% C.L., for the magnitude of effective mass parameter 
$m_{ee}$ and the lightest neutrino mass $m_{\nu_{\mathrm{lightest} } }$. 
From the results in the Table~\ref{tab:table2} it is easy conclude that for the normal 
hierarchy $m_{ \nu_{1} } \sim 2 \times 10^{-3}$~eV and 
$|m_{ee}| \sim 3 \times 10^{-3}$~eV, while for the inverted hierarchy  
$m_{ \nu_{3} } \sim 2 \times 10^{-2}$~eV and 
$|m_{ee}| \sim 3 \times 10^{-2}$~eV. 
%
\begin{table}
\begin{center}
 \begin{tabular}{||l||c||c|c||}\hline
   & 
  Fixed parameters & 
  $m_{\nu_{\mathrm{lightest} } }~[ 10^{-2}$~eV] & 
  $\left| m_{ee} \right|~[ 10^{-2}$~eV]  \\ \hline \hline
   & 
   $\phi_{\ell 1}$, $\phi_{\ell 2}$, $\delta_{e}$, $\delta_{\nu}$  
  & [ 0.2360 , 0.2768 ] &  [ 0.3204 , 0.3608 ] \\ 
  & $\phi_{\ell 2}$, $\delta_{e}$, $\delta_{\nu}$  
  & [ 0.2374 , 0.2735 ] &  [ 0.3215 , 0.3583 ]  \\ 
 NH  & $\phi_{\ell 1}$, $\delta_{e}$, $\delta_{\nu}$  
  & [ 0.2404 , 0.2711 ] & [ 0.3251 , 0.3563 ] \\  
  & $\phi_{\ell 1}$, $\phi_{\ell 2}$, $\delta_{\nu}$  
  & [ 0.2349 , 0.2761 ] & [ 0.3229 , 0.3577 ]  \\ 
  & $\phi_{\ell 1}$, $\phi_{\ell 2}$, $\delta_{e}$  
  & [ 0.2164 , 0.2908 ] & [ 0.3121 , 0.3659 ] \\ \hline \hline 
   & 
   $\phi_{\ell 1}$, $\phi_{\ell 2}$, $\delta_{e}$, $\delta_{\nu}$  
  & [ 2.268 , 2.685 ]  &  [ 3.491 , 3.717 ] \\ 
  & $\phi_{\ell 2}$, $\delta_{e}$, $\delta_{\nu}$  
  & [ 2.317 , 2.635 ] &  [ 3.511 , 3.694 ]  \\ 
 IH  & $\phi_{\ell 1}$, $\delta_{e}$, $\delta_{\nu}$  
  & [ 2.311 , 2.650 ] & [ 3.515 , 3.696 ] \\  
  & $\phi_{\ell 1}$, $\phi_{\ell 2}$, $\delta_{\nu}$  
  & [ 2.301 , 2.648 ] & [ 3.512 , 3.695 ] \\ 
  & $\phi_{\ell 1}$, $\phi_{\ell 2}$, $\delta_{e}$  
  & [ 2.123 , 2.787 ] & [ 3.416 , 3.767 ] \\ \hline \hline    
 \end{tabular}
 \caption{The allowed numerical ranges, at 95\% C.L., for the effective mass parameter 
  magnitude $m_{ee}$ and the lightest neutrino mass $m_{\nu_{\mathrm{lightest} } }$.}\label{tab:table2}
\end{center}
\end{table}
%
%
%
\subsection{CP violation in neutrino oscillations in matter\label{subsec:CPV}}  
%
%
%
\begin{figure}[!htbp]
 \begin{center}
  \begin{tabular}{cc}
 \subfigure{ \includegraphics[width=7.7cm, height=5.3cm]{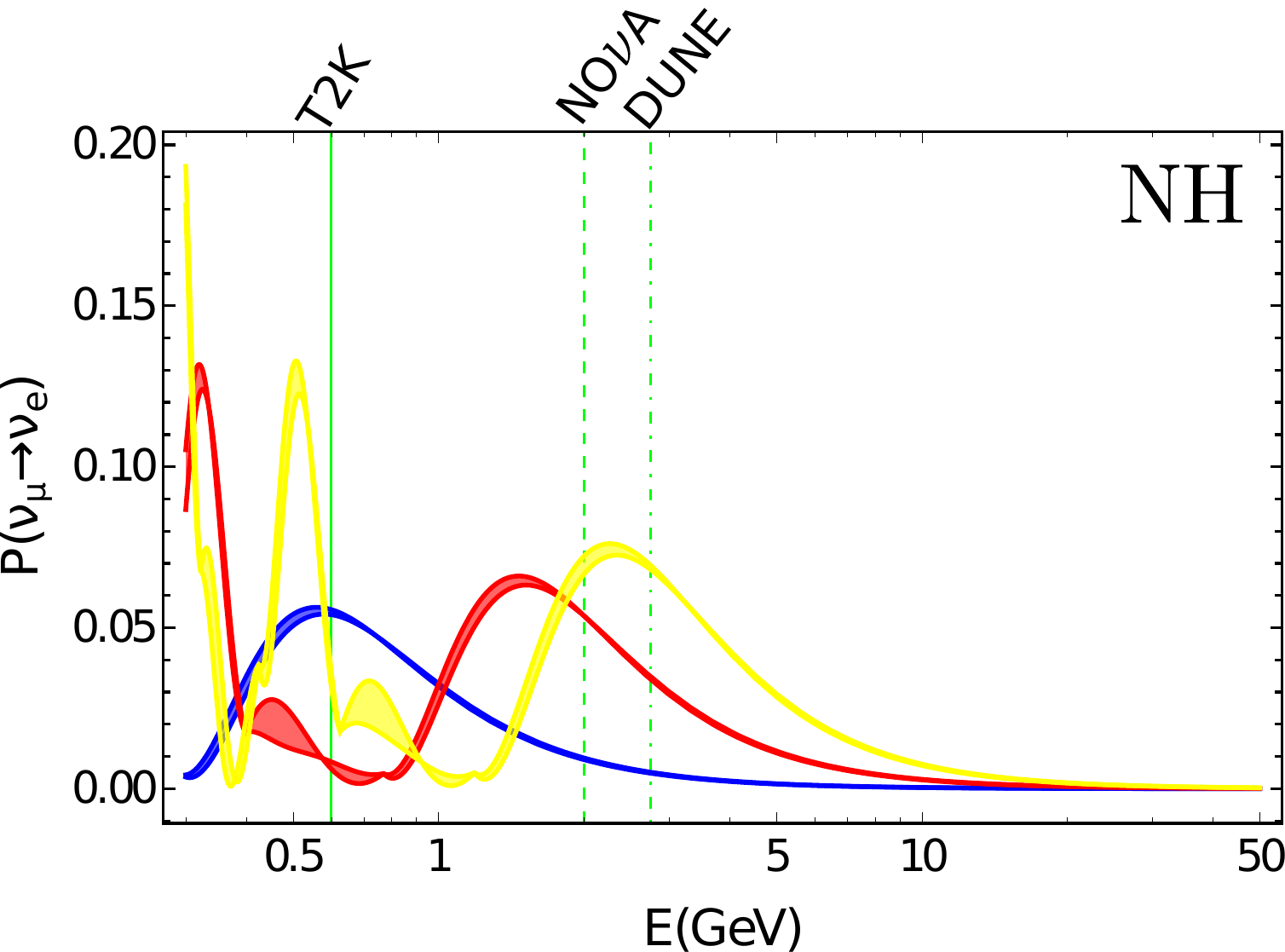} } 
   \subfigure{   \includegraphics[width=7.7cm, height=5.3cm]{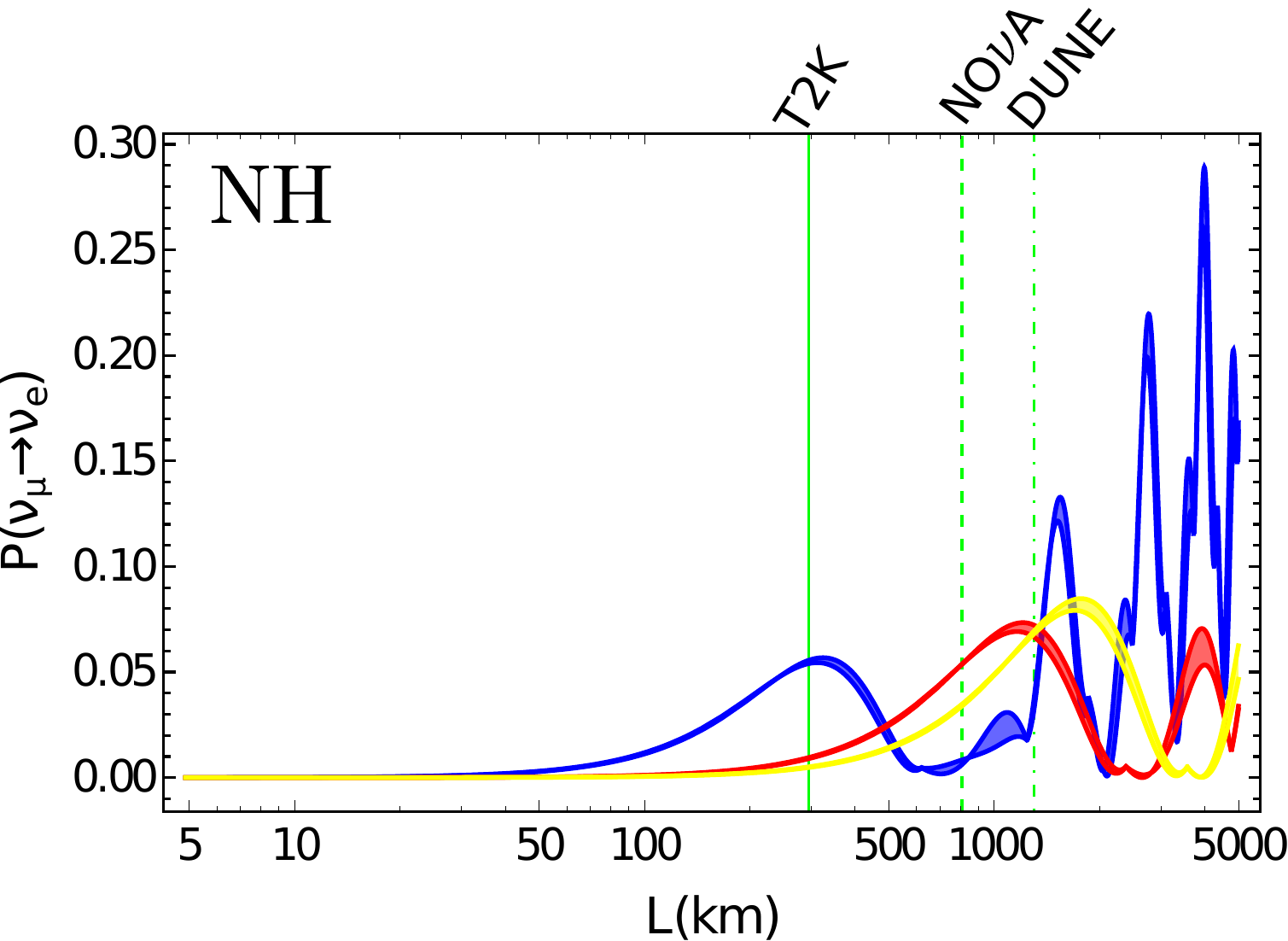}} \\
       \subfigure{ \includegraphics[width=7.7cm, height=5.3cm]{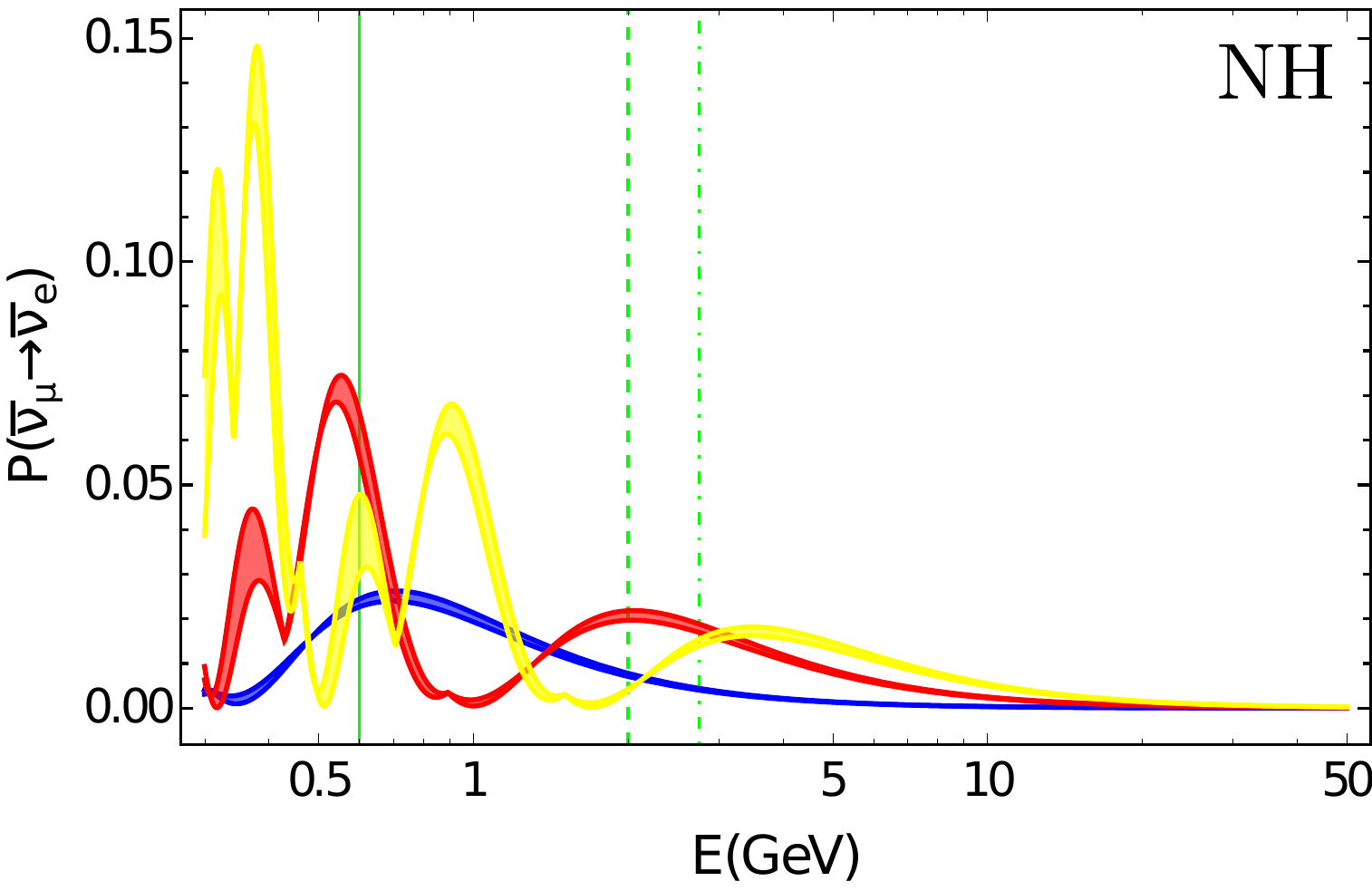} }
   \subfigure{   \includegraphics[width=7.7cm, height=5.3cm]{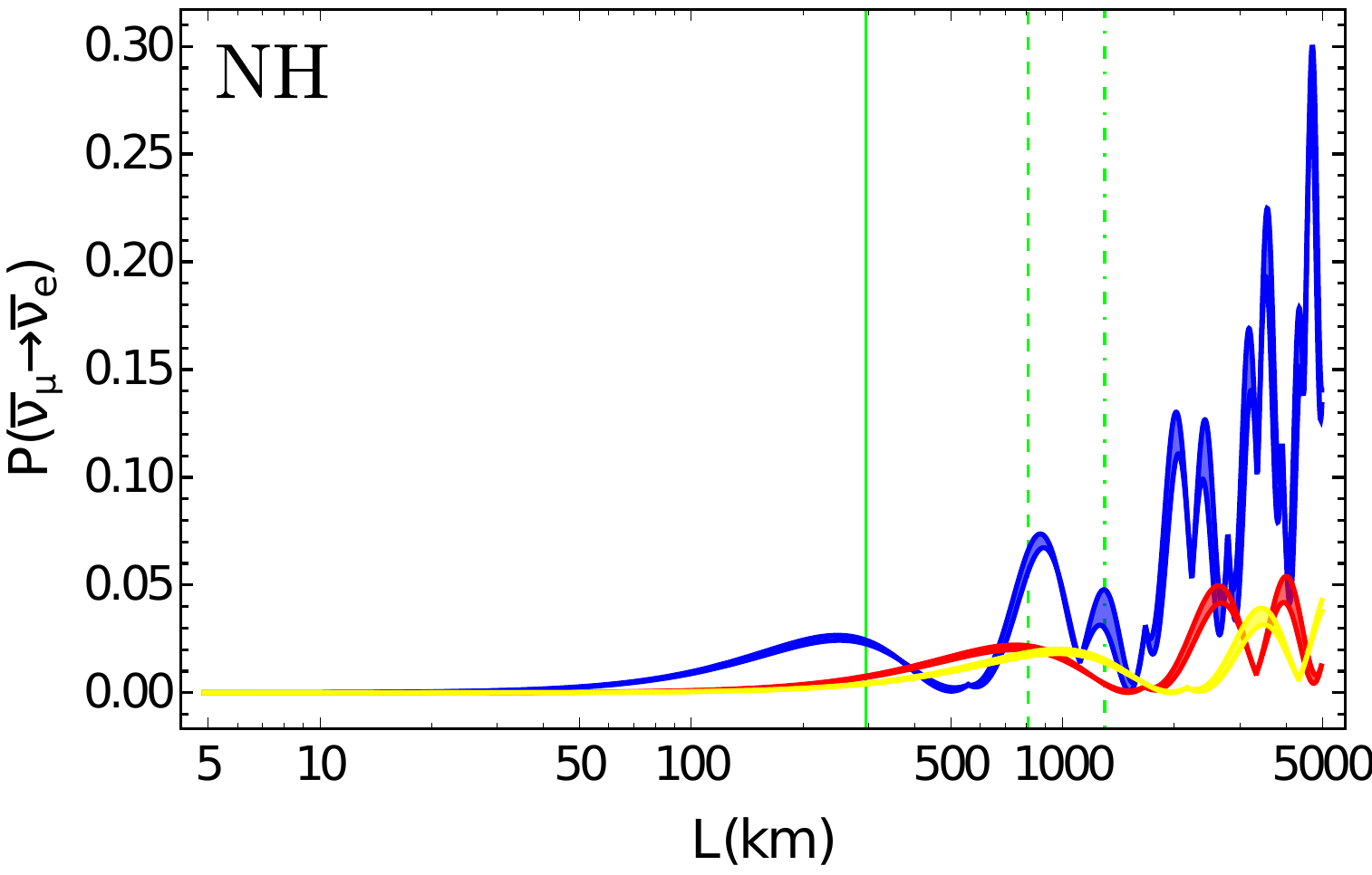}}\\
         \subfigure{ \includegraphics[width=7.7cm, height=5.3cm]{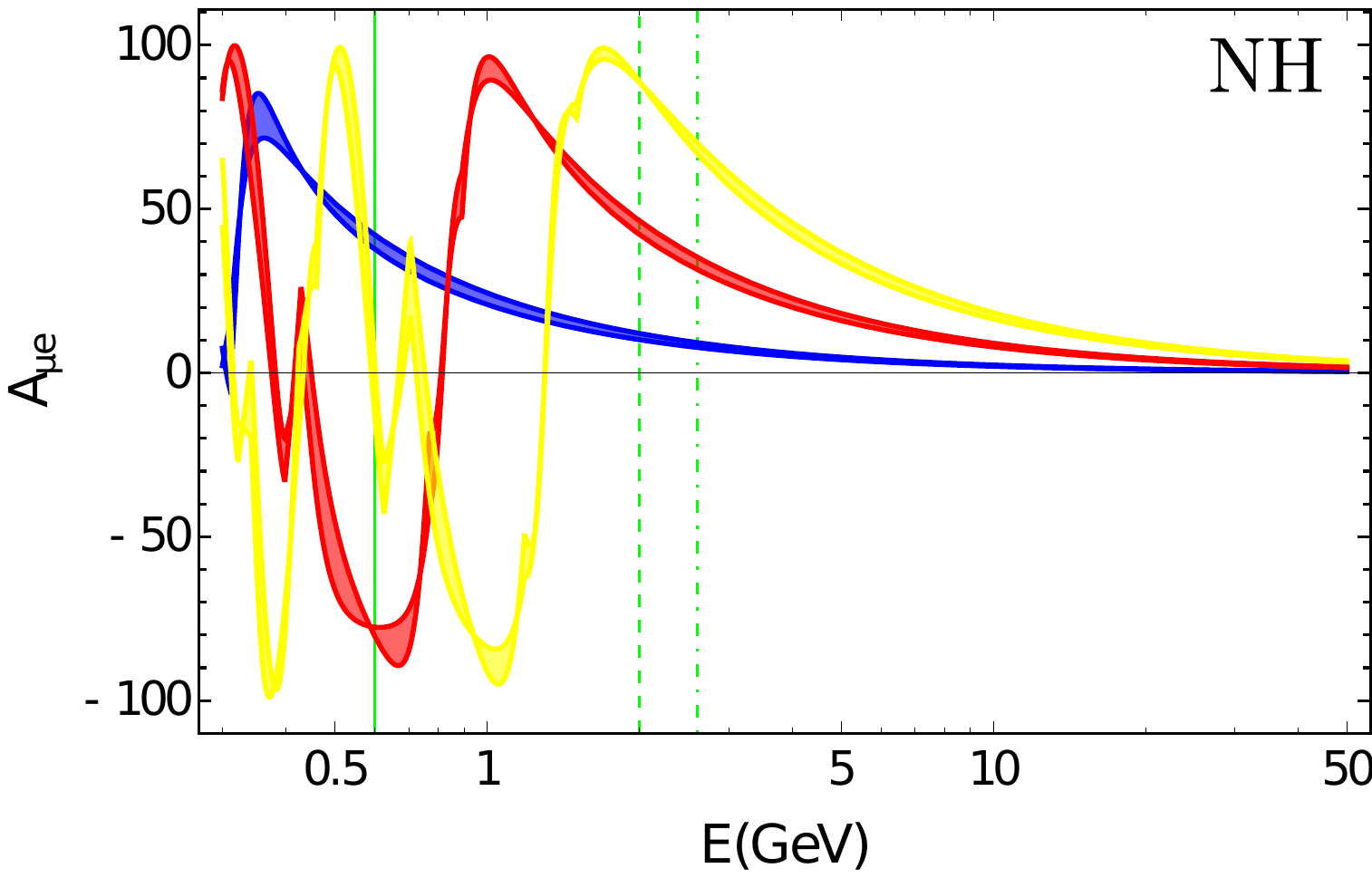} }
   \subfigure{   \includegraphics[width=7.7cm, height=5.3cm]{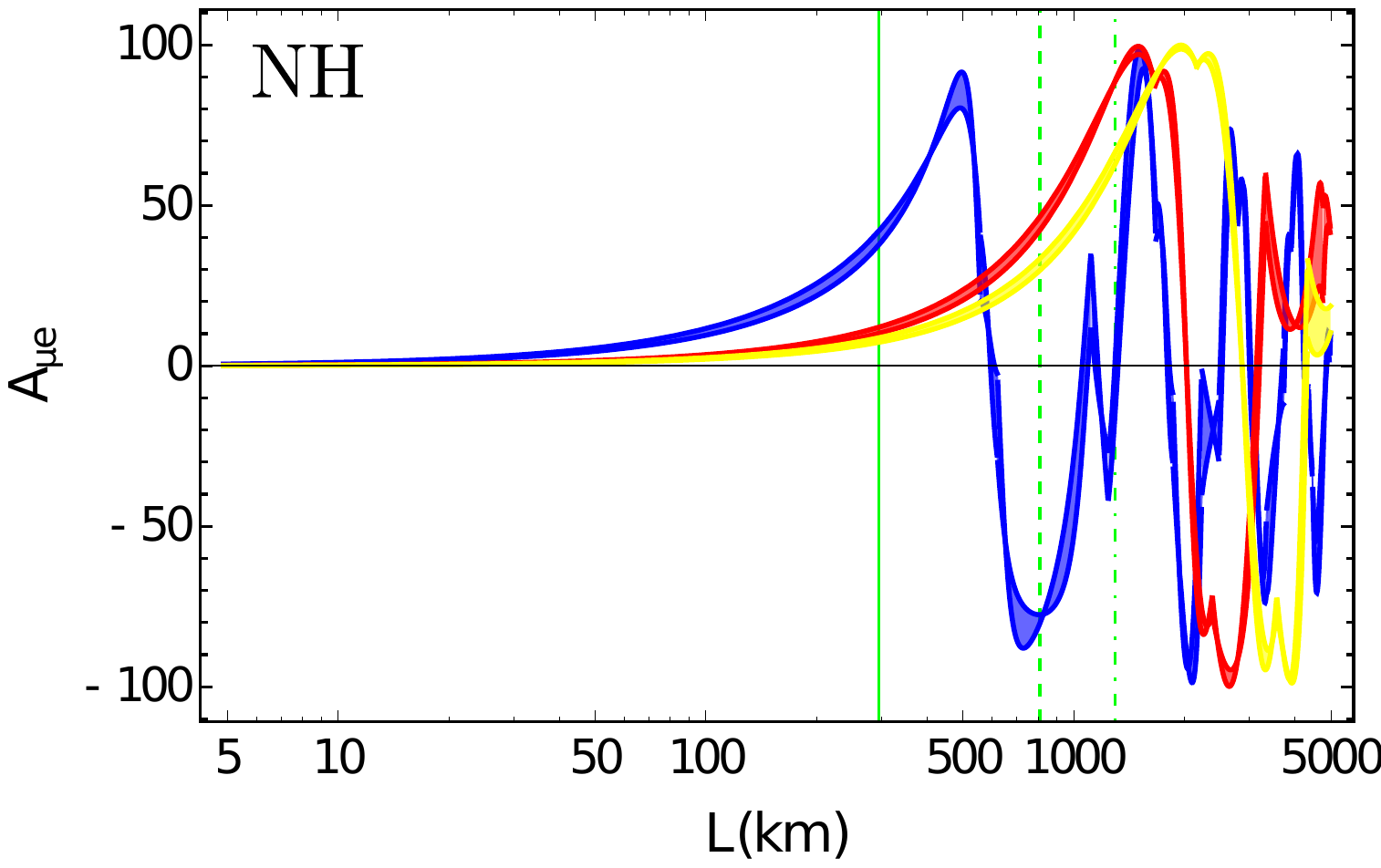}}
  \end{tabular}
  \caption{
   The $P(\nu_\mu \rightarrow \nu_e)$ and $P({\bar \nu}_\mu \rightarrow {\bar \nu}_e)$ 
   transition probabilities, and the asymmetry $\textrm{A}_{\mu e}$ between them, for a 
   normal hierarchy in the neutrino mass spectrum. 
   The blue, red and yellow bands are obtained for a Base-Line energy $L$ of 295, 810 and 
   1300 km for left panels. For the right panels these bands belong to a 
   neutrino energy $E$ of  0.3, 2 and 2.8 GeV, 
   which correspond to the T2K, NO$\nu$A and DUNE experiment, respectively.
   Here, the $\delta_{\mathrm{CP}}$ phase takes values within 1$\sigma$ C.L. range given 
   in Eq.~(\ref{Eq:CP_phases-Fit}) . The remaining parameters are fixed to the values 
   obtained at BFP, which are given in Eq.~(\ref{Eq:Val_DMij}) for $\Delta m_{ij}^{2}$ 
   and Table~\ref{tab:table1} for flavor mixing angles.
   }
   \label{Fig:probabilitiesNH}
 \end{center}
\end{figure}
\begin{figure}[!htbp]
 \begin{center}
  \begin{tabular}{cc}
 \subfigure{ \includegraphics[width=7.7cm, height=5.3cm]{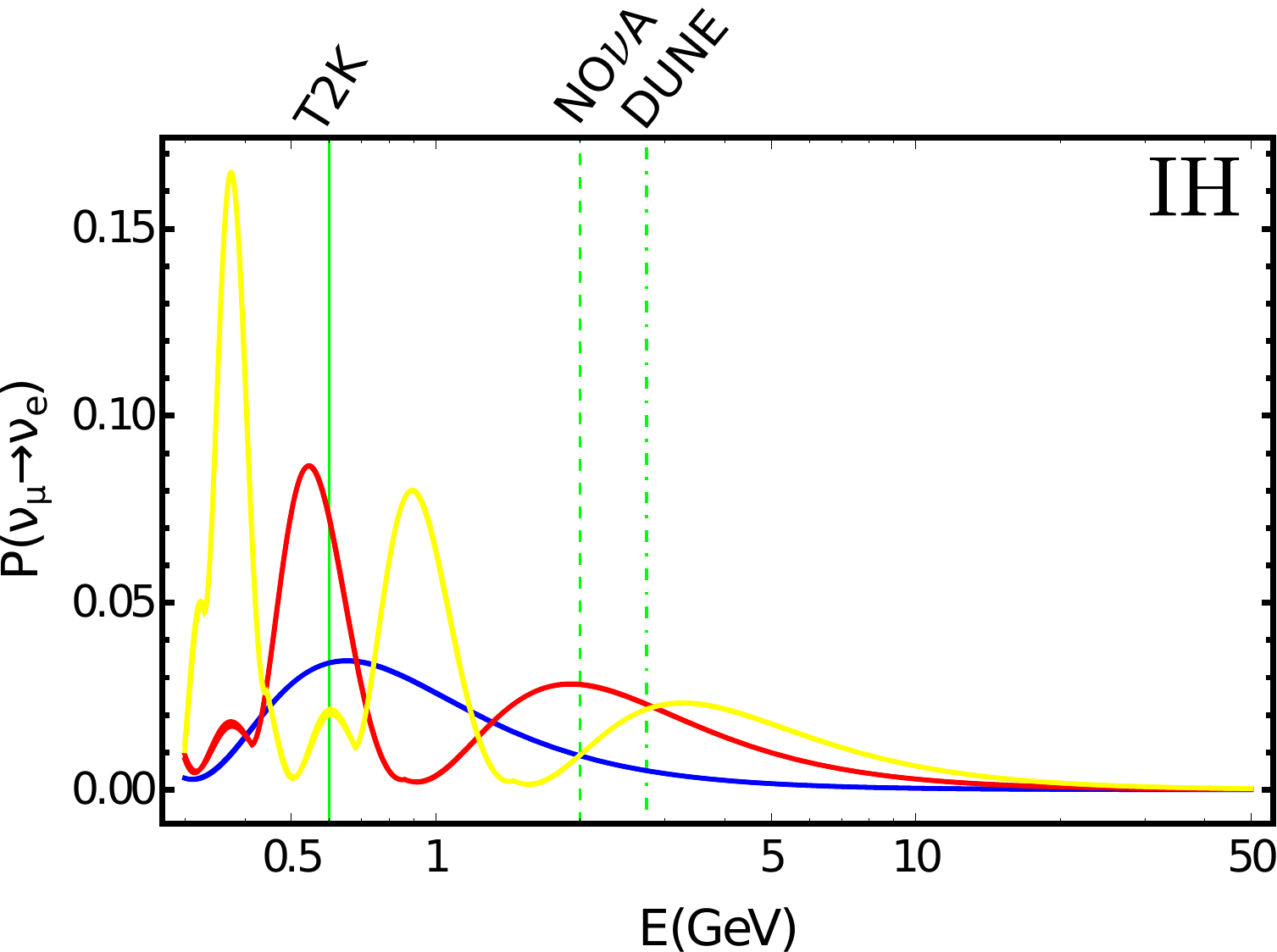} } 
   \subfigure{   \includegraphics[width=7.7cm, height=5.3cm]{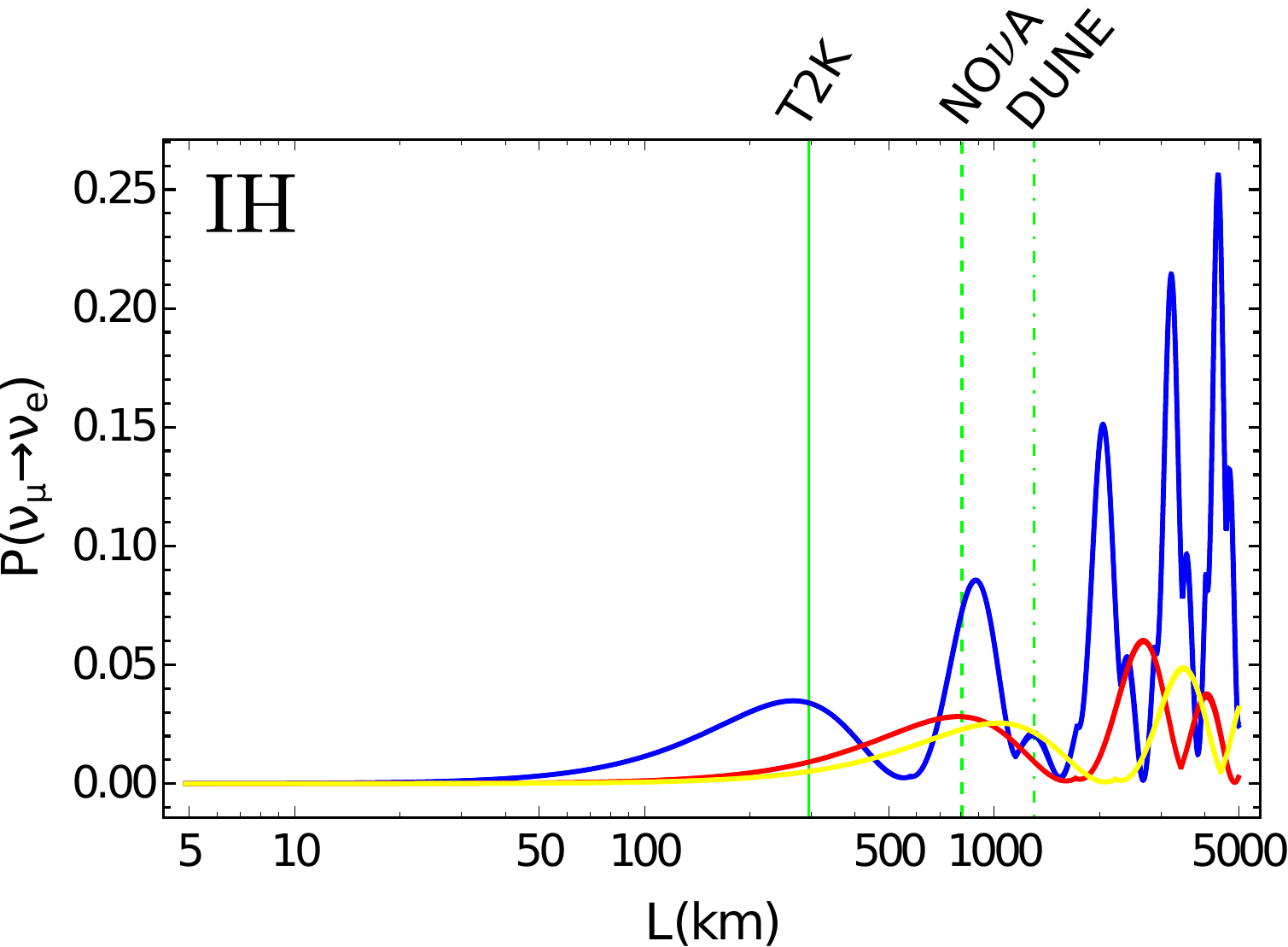}} \\
       \subfigure{ \includegraphics[width=7.7cm, height=5.3cm]{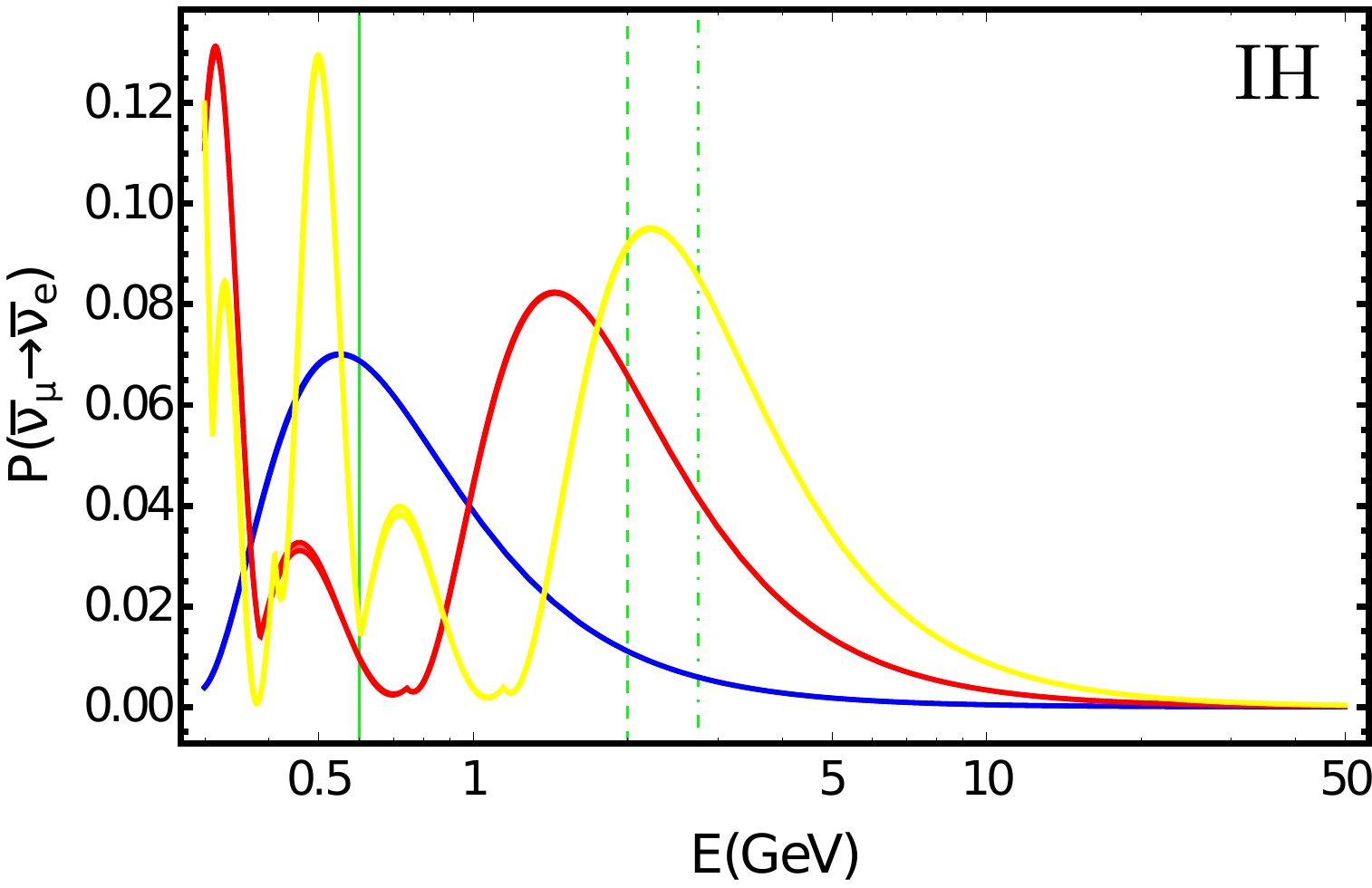} }
   \subfigure{   \includegraphics[width=7.7cm, height=5.3cm]{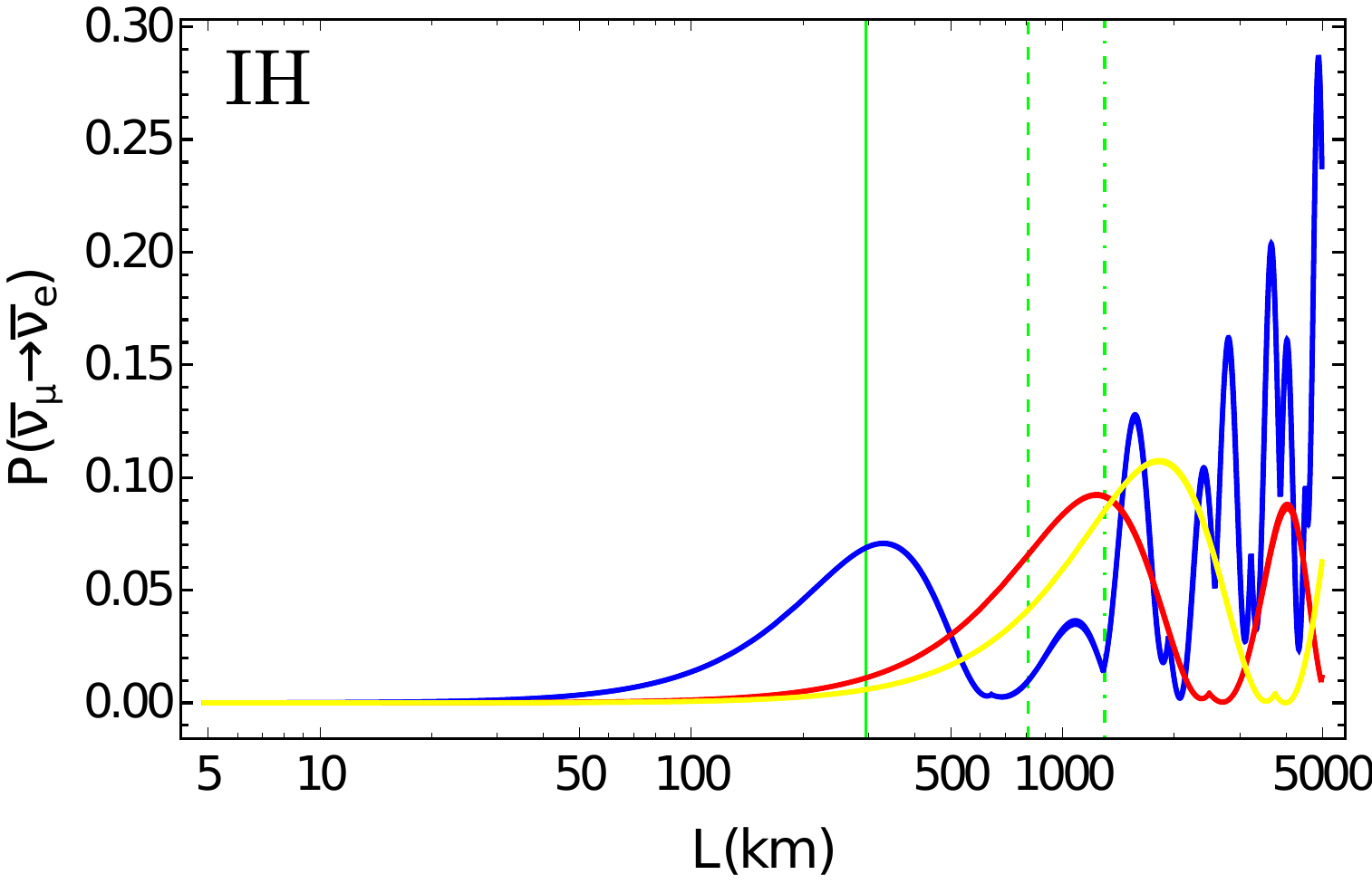}}\\
         \subfigure{ \includegraphics[width=7.7cm, height=5.3cm]{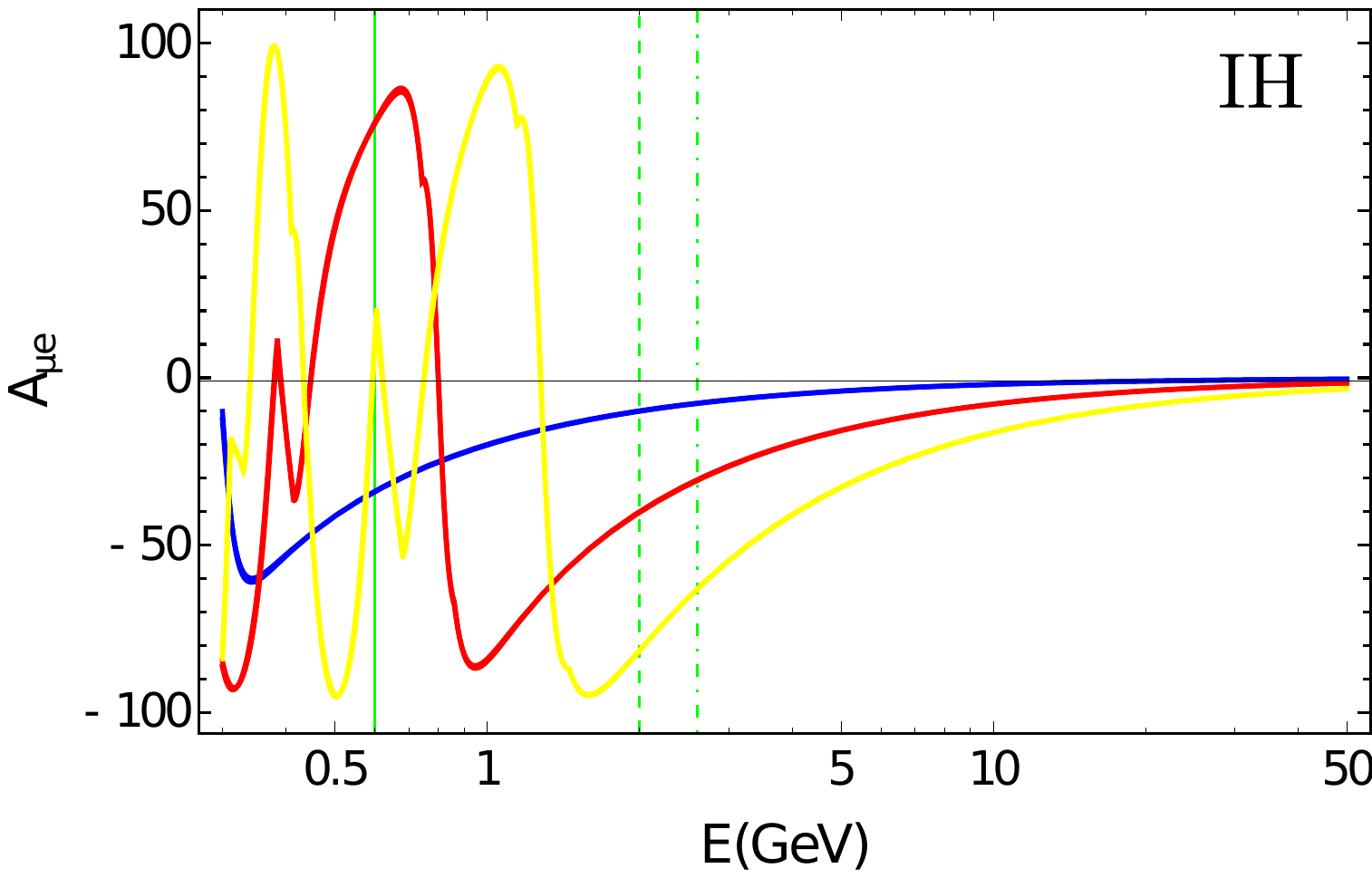} }
   \subfigure{   \includegraphics[width=7.7cm, height=5.3cm]{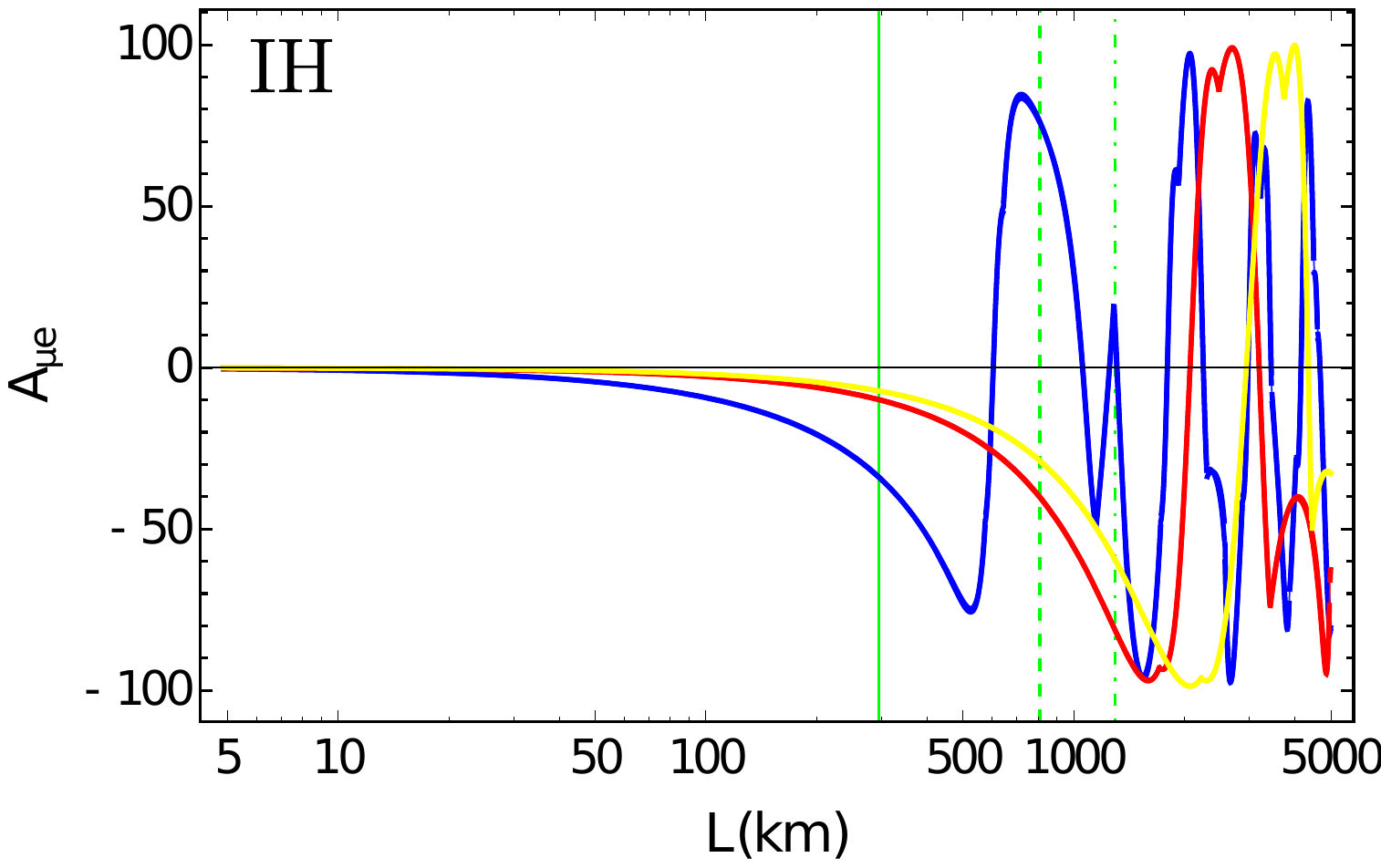}}
  \end{tabular}
  \caption{
   The $P(\nu_\mu \rightarrow \nu_e)$ and $P({\bar \nu}_\mu \rightarrow {\bar \nu}_e)$ 
   transition probabilities, and the asymmetry $\textrm{A}_{\mu e}$ between them, for an 
   inverted hierarchy in the neutrino mass spectrum. 
   The blue, red and yellow bands are obtained for a Base-Line energy $L$ of 295, 810 and 
   1300 km for left-panels. For the right-panels these bands  belong to a 
   neutrino energy $E$ of  0.3, 2 and 2.8 GeV, 
   which correspond to the T2K, NO$\nu$A and DUNE experiment, respectively.
   Here, the $\delta_{\mathrm{CP}}$ phase takes values within 1$\sigma$ C.L. range given 
   in Eq.~(\ref{Eq:CP_phases-Fit}). The remaining parameters are fixed to the values 
   obtained at BFP, which are given in Eq.~(\ref{Eq:Val_DMij}) for $\Delta m_{ij}^{2}$ 
   and Table~\ref{tab:table1} for flavor mixing angles.}
   \label{Fig:probabilitiesIH}
 \end{center}
\end{figure}
In the recent years, we have entered into a precision era in the determination of flavor 
leptonic mixing angles.
However, it is not the same situation for the CP violation in this sector, since has yet 
to be determined experimentally the numerical value of CP violation phase. But we 
have a hunch of where to look: 
the neutrino oscillations with matter effects~\cite{Diwan:2016gmz}. 
One of the aims of the LBL neutrino experiments such as 
T2K~\cite{Abe:2011ks} and NO$\nu$A~\cite{Adamson:2016tbq}, as well as the proposed experiment 
DUNE~\cite{Acciarri:2015uup}, it is determination of the ``Dirac-like'' CP violation 
phase and other parameters that rule the neutrino oscillations $\nu_{\mu} \to \nu_{e}$ and 
$\bar{\nu}_{\mu} \to \bar{\nu}_{e}$.
The transition probability in matter for the oscillation between electron and muon 
neutrinos, as well as for the oscillation between electron and muon antineutrinos, have the 
form~\cite{Nunokawa2008338,Chen:2015siy,Gonzalez-Canales:2016xia}:
\begin{equation}\label{Eq:P_mu-e}
 \begin{array}{l} \vspace{2mm}
  P \left( \nu_{\mu} \rightarrow \nu_{e} \right) \simeq 
   P_{ \textrm{atm} } + P_{ \textrm{sol} } 
   + 2 \sqrt{ P_{ \textrm{atm} } } \sqrt{ P_{ \textrm{sol} } } 
   \cos \left( \Delta_{32}  + \delta_{\textrm{CP}} \right)  , \\
  P \left( \bar{\nu}_{\mu} \rightarrow \bar{\nu}_{e} \right) \simeq 
   {\cal P}_{ \textrm{atm} } + P_{ \textrm{sol} } 
   + 2 \sqrt{ {\cal P}_{ \textrm{atm} } } \sqrt{ P_{ \textrm{sol} } } 
   \cos \left( \Delta_{32}  - \delta_{\textrm{CP}} \right) ,
 \end{array} 
\end{equation} 
where 
\begin{equation}
 \begin{array}{l} \vspace{2mm}
  \sqrt{ P_{ \textrm{sol} } } = 
    \cos \theta_{23} \sin 2 \theta_{12} \frac{ \sin a L }{ a L } \Delta_{21}, 
    \\ \vspace{2mm}   
  \sqrt{ P_{ \textrm{atm} } } = 
   \sin \theta_{23} \sin 2 \theta_{13} \frac{ \sin \left( \Delta_{31} - a L \right) }{ 
   \left( \Delta_{31} - a L \right) } \Delta_{31} , \\ \vspace{2mm}
  \sqrt{ {\cal P}_{ \textrm{atm} } } = 
   \sin \theta_{23} \sin 2 \theta_{13} \frac{ \sin \left( \Delta_{31} + a L \right) }{ 
   \left( \Delta_{31} + a L \right) } \Delta_{31} .  
 \end{array}
\end{equation}
In the above expressions, $L$ is the Base-Line,
\begin{equation}
 \begin{array}{l}
  \Delta_{ij} = \frac{ \Delta m_{ij}^{2} L }{ 4E }, \quad 
  \Delta m_{ij}^{2} = m_{i}^{2} - m_{j}^{2} 
  \quad \textrm{and} \quad
  a = \frac{ G_{F} N_{e} }{ \sqrt{2}}.
 \end{array}
\end{equation} 
Here, $E$ is the energy of neutrino beam, $G_{F}$ is the Fermi constant and 
$N_{e}$ is the density of electrons.
The $a$ parameter is $a \approx ( 3500~\textrm{km})^{-1}$ for the Earth 
crust~\cite{Nunokawa2008338} .
The asymmetry between $P \left( \nu_{\mu} \rightarrow \nu_{e} \right)$ and 
$P \left( \bar{\nu}_{\mu} \rightarrow \bar{\nu}_{e} \right)$ in matter 
is~\cite{Gonzalez-Canales:2016xia}
\begin{equation}\label{Eq:A_mu-e}
 \begin{array}{rl}\vspace{2mm}
  {\cal A}_{\mu e} & 
  = 
  \frac{ 
   P \left( \nu_{\mu} \rightarrow \nu_{e} \right) 
   -
   P \left( \bar{\nu}_{\mu} \rightarrow \bar{\nu}_{e} \right) 
  }{
   P \left( \nu_{\mu} \rightarrow \nu_{e} \right) 
   + 
   P \left( \bar{\mu}_{\nu} \rightarrow \bar{\nu}_{e} \right)
  } \\ 
  & 
  =
  \frac{ 
   \left( P_{ \textrm{atm} } -{\cal P}_{ \textrm{atm} } \right) 
   + 2 \sqrt{ P_{ \textrm{sol} } } 
    \left( 
     \sqrt{ P_{ \textrm{atm} } } 
     \cos \left( \Delta_{32}  + \delta_{\textrm{CP}} \right)   
     - 
     \sqrt{ {\cal P}_{ \textrm{atm} } } 
     \cos \left( \Delta_{32}  - \delta_{\textrm{CP}} 
    \right) 
   \right) 
  }{
   \left( P_{ \textrm{atm} }  + {\cal P}_{ \textrm{atm} } \right) 
   + 2 \sqrt{ P_{ \textrm{sol} } } 
    \left( 
     \sqrt{ P_{ \textrm{atm} } } 
     \cos \left( \Delta_{32}  + \delta_{\textrm{CP}} \right)   
     + 
     \sqrt{ {\cal P}_{ \textrm{atm} } } 
     \cos \left( \Delta_{32}  - \delta_{\textrm{CP}} 
    \right) 
   \right) 
   + 2 P_{ \textrm{sol} } 
  }.
 \end{array}
\end{equation}
The above asymmetry ${\cal A}_{\mu e}$ is basically due to the absence of positrons in 
the journey of neutrino (anti-neutrino) through the earth. 
Hence, a neutrino experiment with a LBL would be more sensitive to measure this asymmetry. 

The T2K neutrino oscillation experiment has a LBL of 295~km, while the energy 
of its neutrino beam has a peak around to 0.6~GeV and width of 
$\sim$0.3~GeV~\cite{Abe:2011ks}. 
In Fig.~\ref{Fig:probabilitiesNH} we show the transition probability 
$\nu_{\mu} ( \bar{\nu}_{\mu} ) \rightarrow \nu_{e} (\bar{\nu}_{e})$, as well 
as the asymmetry ${\cal A}_{\mu e}$ for T2K experiment.

The NO$\nu$A neutrino oscillation experiment has a LBL of 810~km, while the 
energy~ of its neutrino beam has a peak around to 2~GeV~\cite{Adamson:2016tbq}. 
In Fig.~\ref{Fig:probabilitiesNH} we show the transition probability 
$\nu_{\mu} ( \bar{\nu}_{\mu} ) \rightarrow \nu_{e} (\bar{\nu}_{e})$, as well 
as the asymmetry ${\cal A}_{\mu e}$ for NO$\nu$A experiment.

Finally, the future neutrino oscillation experiment DUNE will have a LBL of 
$\sim1300$~km, while the energy of its neutrino beam will have a peak around to 
$2.5-3.0$~GeV~\cite{Acciarri:2015uup}. 
In Fig.~\ref{Fig:probabilitiesNH} the transition probability 
$\nu_{\mu} ( \bar{\nu}_{\mu} ) \rightarrow \nu_{e} (\bar{\nu}_{e})$, and 
the asymmetry ${\cal A}_{\mu e}$ for DUNE experiment are shown. 
%
%
\section{Conclusions\label{sec:conclusions}}
%
%
%
We have studied the theory of neutrino masses, mixings and CPV as the 
realization of an $S_{3}$ flavor symmetry in the framework of the Two Higgs Doublet Model 
type-III.
In this $\nu$2HDM$\otimes S_{3}$ extension of Standard Model, on the one hand, 
the active neutrinos acquire their little masses via the type-I seesaw mechanism. 
On the other hand, the explicit sequential breaking of flavor symmetry according the chain 
$S_{ 3L }^{ \texttt{j} } \otimes S_{ 3R }^{ \texttt{j} } \supset 
S_{3}^{\mathrm{diag}} \supset S_{2}^{\mathrm{diag}}$, allow us to represent in the flavor 
basis the Yukawa matrices with an Hermitian matrix with two texture zeroes. 
Consequently, we obtained an unified treatment for all fermion mass matrices in the 
model, which are represented through of a matrix with two texture zeroes.

The unitary matrices that diagonalize the mass matrices are expressed in terms of 
fermion mass ratios. 
Then, the entries of the Yukawa matrices in the mass basis naturally acquire the form of 
the so-called {\it Cheng-Sher ansatz}. 
Also, the lepton flavor mixing matrix PMNS is expressed as function of the masses of 
charged leptons and neutrinos, two phases associated with the CP-violation, and two 
parameters associated with the flavor symmetry breaking.  
The unitary  matrix that allows us to pass from the weak basis to the flavor symmetry 
adapted basis, is unobservable in the Higgs-fermions couplings and lepton flavor 
mixing matrix.

To validate our hypothesis where the $S_{3}$ horizontal flavor symmetry is 
explicitly breaking according to the chain 
$S_{ 3L }^{ \texttt{j} } \otimes S_{ 3R }^{ \texttt{j} } 
\supset S_{3}^{\mathrm{diag}} \supset S_{2}^{\mathrm{diag}}$,  
all  fermion mass matrices are represented through a matrix with two texture 
zeroes. Furthermore, we make a likelihood test where we compare the theoretical 
expressions of the flavor mixing angles with the current experimental data on 
masses and flavor mixing of leptons. The results obtained in this $\chi^{2}$ 
analysis are in very good agreement with the current experimental data. 

We also obtained the following allowed value ranges, at BFP$\pm 1\sigma$ for the 
``Dirac-like'' phase factor, as well as for the two Majorana phase factors:
\begin{equation}
\begin{array}{c}
\delta_{CP} (^{\circ}) = 
\left \{ \begin{array}{l}
-69.8_{-6.110}^{+5.508}  ,\\
-80.83_{-0.709}^{+0.652} ,
\end{array} \right.  \quad
\phi_{12} (^{\circ}) = 
\left \{ \begin{array}{l}
-5.800_{-0.150}^{+0.170}  ,\\
-5.24_{-0.148}^{+0.153}   ,
\end{array} \right.  \quad
\phi_{13} (^{\circ}) = 
\left \{ \begin{array}{l}
14.744_{-1.366}^{+1.266}   ,\\
-2.190_{-0.0005}^{+0.0030} .
\end{array} \right. 
\end{array}
\end{equation}
The upper (lower) row corresponds to the normal (inverted) hierarchy in the neutrino mass 
spectrum. These values of the phase factors are in agreement with a maximum CPV 
in the neutrino oscillation in matter. Finally, we also analyzed the phenomenological 
implications of the above numerical values of the CP-violation phases on the neutrinoless 
double beta decay, as well as for LBL neutrino oscillation experiments such as 
T2K, NO$\nu$A, and DUNE.

%
\appendix
%
%
\section{Three dimensional representation of $S_{3}$}
%
The permutations of symmetry group $S_{3}$ can be represented on the reducible 
triplet as~\cite{Georgi:1999wka,Ishimori:2012zz}: 
\begin{equation}\label{eq:A-1}
 \begin{array}{ccc} \vspace{2mm}
  {\bf D}^{(3)} \left( E \right) = 
   \left( \begin{array}{ccc}
    1 & 0 & 0 \\
    0 & 1 & 0 \\
    0 & 0 & 1
   \end{array} \right), &
  {\bf D}^{(3)} \left( A_{1} \right) =
   \left( \begin{array}{ccc}
    0 & 1 & 0 \\
    1 & 0 & 0 \\
    0 & 0 & 1
   \end{array}  \right), &
  {\bf D}^{(3)} \left( A_{2} \right) =
   \left( \begin{array}{ccc}
    0 & 0 & 1 \\
    1 & 0 & 0 \\
    0 & 1 & 0
   \end{array}  \right) , \\
  {\bf D}^{(3)} \left( A_{3} \right) =
   \left( \begin{array}{ccc}
    1 & 0 & 0 \\
    0 & 0 & 1 \\
    0 & 1 & 0
   \end{array}  \right) , &
  {\bf D}^{(3)} \left( A_{4} \right) =
   \left( \begin{array}{ccc} 
    0 & 0 & 1 \\
    1 & 0 & 0 \\
    0 & 1 & 0
   \end{array}  \right) , &    
  {\bf D}^{(3)} \left( A_{5} \right) =
   \left( \begin{array}{ccc}
    0 & 1 & 0 \\
    0 & 0 & 1 \\
    1 & 0 & 0
   \end{array}  \right) .   
 \end{array}
\end{equation}
In this representation the projection operators take the form:
\begin{equation}\label{eq:A-P1}
 \begin{array}{ll}
  \textrm{Symmetric singlet}, & 
  {\bf P}_{\bf 1 } =
  \frac{1}{3} 
  \left( \begin{array}{ccc}
   1 & 1 & 1 \\
   1 & 1 & 1 \\
   1 & 1 & 1 
  \end{array} \right) =
  | v_{\bf 1} \rangle \langle v_{\bf 1} | . \\
  \textrm{Anti-symmetric singlet}, & 
  {\bf P}_{\bf 1' } = 0. \\
  \textrm{Doublet}, & 
  {\bf P}_{\bf 2 } =
  \frac{1}{3} 
  \left( \begin{array}{ccc}
   2  & -1 & -1 \\
   -1 &  2 & -1 \\
   -1 & -1 &  2 
  \end{array} \right) =
  | v_{\bf 2A} \rangle \langle v_{\bf 2A} | + 
  | v_{\bf 2S} \rangle \langle v_{\bf 2S} |. \\  
 \end{array}
\end{equation}
Here, the vector 
$| v_{\bf 1} \rangle = \frac{1}{\sqrt{3}} \left( 1, 1, 1 \right)^{\top}$ is 
associated with the symmetric singlet. In the projection operator ${\bf P}_{\bf 2 }$, 
we have the vectors 
$| v_{\bf 2A} \rangle = \frac{1}{ \sqrt{2} } \left( -1, 1, 0 \right)^{\top}$ and  
$| v_{\bf 2S} \rangle = \frac{1}{ \sqrt{2} } \left(  1, 1, -2 \right)^{\top}$, which 
are associated with the doublet.

Correspondingly, the vectors $| v_{\bf 2A} \rangle$ and 
$| v_{\bf 2S} \rangle$ are antisymmetric and symmetric, under the permutation of 
first two elements. With the previous three vectors we can construct some tensors 
that can be helpful. 
Then,
\begin{equation}
 \begin{array}{ccc}
  | v_{\bf 2S} \rangle \langle v_{\bf 2A} | = 
  \frac{ 1 }{ \sqrt{12} }
  \left( \begin{array}{ccc}
   -1 &  1 & 0 \\
   -1 &  1 & 0 \\
    2 & -2 & 0 
  \end{array}  \right)  & 
  \textrm{and} &
  | v_{\bf 2A} \rangle \langle v_{\bf 2S} | = 
  \frac{ 1 }{ \sqrt{12} }
  \left( \begin{array}{ccc}
   -1 & -1 &  2 \\
    1 &  1 & -2 \\
    0 &  0 & 0 
  \end{array}  \right).
 \end{array}
\end{equation}
If we define the tensors 
${\bf T}_{x}^{+} = | v_{\bf 2S} \rangle \langle v_{\bf 2A} | + | v_{\bf 2A} \rangle 
\langle v_{\bf 2S} |$ and
${\bf T}_{x}^{-} = i \left( | v_{\bf 2A} \rangle \langle v_{\bf 2S} | - 
| v_{\bf 2S} \rangle \langle v_{\bf 2A} | \right) $, we obtain
\begin{equation}\label{eq:Ty+-}
 \begin{array}{ccc}
  {\bf T}_{x}^{+} = \frac{1}{\sqrt{3}}
  \left( \begin{array}{ccc}
   -1 &  0 &  1 \\
    0 &  1 & -1 \\
    1 & -1 &  0 
  \end{array}  \right) & 
  \textrm{and} &
 {\bf T}_{x}^{-} = \frac{1}{\sqrt{3}}
  \left( \begin{array}{ccc}
    0 &   i & -i \\
   -i &   0 &  i \\
    i &  -i &  0 
  \end{array}  \right). 
 \end{array} 
\end{equation}
The terms proportional to tensors ${\bf T}_{x}^{+}$ and ${\bf T}_{x}^{-}$ mix the 
components of the doublet representation each other. Now,
\begin{equation}
 \begin{array}{ccc}
  | v_{\bf 1} \rangle \langle v_{\bf 2A} | = 
  \frac{ 1 }{ \sqrt{6} }
  \left( \begin{array}{ccc}
   -1 &  1 & 0 \\
   -1 &  1 & 0 \\
   -1 &  1 & 0 
  \end{array}  \right)  & 
  \textrm{and} &
  | v_{\bf 2A} \rangle \langle v_{\bf 1} | = 
  \frac{ 1 }{ \sqrt{6} }
  \left( \begin{array}{ccc}
   -1 & -1 & -1 \\
    1 &  1 &  1 \\
    0 &  0 &  0 
  \end{array}  \right).
 \end{array}
\end{equation}
If we define the tensors 
${\bf T}_{y}^{+} = | v_{\bf 1} \rangle \langle v_{\bf 2A} | + | v_{\bf 2A} \rangle 
\langle v_{\bf 1} |$ and
${\bf T}_{y}^{-} = i \left( | v_{\bf 2A} \rangle \langle v_{\bf 2A} | - 
| v_{\bf 2A} \rangle \langle v_{\bf 2S} | \right) $, we obtain
\begin{equation}
 \begin{array}{ccc}
  {\bf T}_{y}^{+} = \frac{1}{\sqrt{6}}
  \left( \begin{array}{ccc}
   -2 &  0 & -1 \\
    0 &  2 &  1 \\
   -1 &  1 &  0 
  \end{array}  \right) & 
  \textrm{and} &
 {\bf T}_{y}^{-} = \frac{1}{\sqrt{6}}
  \left( \begin{array}{ccc}
    0 & 2i &  i \\
   -2i &  0 & -i \\
   -i &  i &  0 
  \end{array}  \right). 
 \end{array} 
\end{equation}
The terms proportional to tensors ${\bf T}_{y}^{+}$ and ${\bf T}_{y}^{-}$ mix the antisymmetric component of doublet with the singlet. Finally, 
\begin{equation}
 \begin{array}{ccc}
  | v_{\bf 1} \rangle \langle v_{\bf 2S} | = 
  \frac{ 1 }{ 3\sqrt{2} }
  \left( \begin{array}{ccc}
    1 &  1 & -2 \\
    1 &  1 & -2 \\
    1 &  1 & -2 
  \end{array}  \right)  & 
  \textrm{and} &
  | v_{\bf 2S} \rangle \langle v_{\bf 1} | = 
  \frac{ 1 }{ 3\sqrt{2} }
  \left( \begin{array}{ccc}
    1 &  1 &  1 \\
    1 &  1 &  1 \\
   -2 & -2 & -2 
  \end{array}  \right) .
 \end{array}
\end{equation}
If we define the tensors 
${\bf T}_{z}^{+} = | v_{\bf 1} \rangle \langle v_{\bf 2S} | + | v_{\bf 2S} \rangle 
\langle v_{\bf 1} |$ and
${\bf T}_{y}^{-} = i \left( | v_{\bf 2A} \rangle \langle v_{\bf 2A} | - 
| v_{\bf 2A} \rangle \langle v_{\bf 2S} | \right) $, then
\begin{equation}
 \begin{array}{ccc}
  {\bf T}_{z}^{+} = \frac{1}{3 \sqrt{2}}
  \left( \begin{array}{ccc}
    2 &   2 & -1 \\
    2 &   2 & -1 \\
   -1 &  -1 & -4 
  \end{array}  \right) & 
  \textrm{and} &
 {\bf T}_{z}^{-} = \frac{1}{\sqrt{2}}
  \left( \begin{array}{ccc}
    0 & 0 & -i \\
    0 & 0 & -i \\
    i & i &  0 
  \end{array}  \right). 
 \end{array} 
\end{equation}
The terms proportional to tensors ${\bf T}_{z}^{+}$ and ${\bf T}_{z}^{-}$ mix the 
symmetric component of doublet with the singlet. The tensor ${\bf T}_{z}^{+}$ 
can be written as a linear combination of two independent matrices, 
\begin{equation}
 {\bf T}_{z}^{+} = 
  \sqrt{ \frac{2}{3} } {\bf T}_{z1}^{+} - \frac{1}{3 \sqrt{2}} {\bf T}_{z2}^{+},
\end{equation}
where
\begin{equation}\label{eq:Tz12+}
 \begin{array}{ccc}
  {\bf T}_{z1}^{+} = 
  \left( \begin{array}{ccc}
    1 &  1 & 0 \\
    1 &  1 & 0 \\
    0 &  0 & -2 
  \end{array}  \right) & 
  \textrm{and} &
 {\bf T}_{z2}^{+} =
  \left( \begin{array}{ccc}
    0 & 0 & 1 \\
    0 & 0 & 1 \\
    1 & 1 & 0 
  \end{array}  \right). 
 \end{array}
\end{equation}
%
%
\section{Cheng-Sher Parameters}\label{App:B}
%
%
\begin{equation}
 \begin{array}{l}\vspace{2mm}
  \widetilde{\bf Y}_{k}^{ \texttt{j} } = 
  {\bf U}_{ \texttt{j} }^{\dagger} 
  {\bf Y}_{k}^{ \mathrm{w}, \texttt{j} } {\bf U}_{ \texttt{j} } = 
  {\bf O}_{ \texttt{j} }^{ \top } {\bf P}_{ \texttt{j} }^{ \dagger } 
   {\bf U}_{ \mathrm{s} }^{\dagger} 
  {\bf Y}_{k}^{ \mathrm{w}, \texttt{j} } 
  {\bf U}_{ \mathrm{s} } {\bf P}_{ \texttt{j} } {\bf O}_{ \texttt{j} } , \\
    \left( \widetilde{\bf Y}_{k}^{ \texttt{j} } \right)_{ \texttt{rt} } =  
   \dfrac{ \sqrt{ m_{ \texttt{jr}  } m_{ \texttt{js}  } } }{ v }
   \left( \widetilde{\bf \chi}_{k}^{ \texttt{j} } \right)_{ \texttt{rt} } ,
   \quad \texttt{r,t} = 1,2,3
 \end{array} 
\end{equation}
where 
\begin{equation}
 \begin{array}{ll}\vspace{2mm}
  \left( \widetilde{\bf \chi}_{k}^{ \texttt{j} } \right)_{ 11 } = & 
   2 \frac{ 
    \xi_{ \texttt{J} 1} 
   }{ 
    D_{\texttt{j}1 } 
   }  
   \sqrt{ \frac{
    \widehat{m}_{ \texttt{j} 2 } 
   }{
    \widehat{m}_{ \texttt{j} 1 }
   }
   \left( 1 - \delta_{ \texttt{j} } \right)  
   } 
   \cos \left( \phi_{k}^{ \texttt{j} } - \phi_{ \texttt{j} } \right)
   \widetilde{a}_{k}^{ \texttt{j} } 
   + 
   \frac{ 
    \left( 1 - \delta_{\texttt{j}} \right) \xi_{ \texttt{j} 1} 
   }{
    D_{ \texttt{j} 1} 
   }
   \widetilde{b}_{k}^{ \texttt{j} }
   -
   2 \frac{ 
    \sqrt{ \delta_{ \texttt{j} } \left( 1 - \delta_{ \texttt{j} } \right)
    \xi_{ \texttt{j} 1 } \xi_{ \texttt{j} 2 } } 
   }{
    D_{ \texttt{j} 1 } 
   }
   \widetilde{c}_{k}^{ \texttt{j} }   
   +
   \frac{   
    \delta_{ \texttt{j} } \xi_{ \texttt{j} 2 } 
   }{
    D_{ \texttt{j} 1 } 
   } 
   \widetilde{d}_{k}^{ \texttt{j} } , \\ \vspace{2mm}
  \left( \widetilde{\bf \chi}_{k}^{ \texttt{j} } \right)_{ 12 } = &
   \sqrt{
    \frac{
     \left( 1 - \delta_{ \texttt{j} } \right) \xi_{ \texttt{j}1 } 
     \xi_{ \texttt{j} 2 } 
    }{
     D_{ \texttt{j} 1 } D_{ \texttt{j} 2 } 
    } 
   }
   \left(
    \sqrt{
     \frac{
      \widehat{m}_{ \texttt{j} 2 } 
     }{
      \widehat{m}_{ \texttt{j} 1 }
     }
    } 
    e^{ i \left( \phi_{k}^{ \texttt{j} } - \phi_{ \texttt{j} } \right) } 
    -
    \sqrt{
     \frac{ 
      \widehat{m}_{ \texttt{j} 1 } 
     }{
      \widehat{m}_{ \texttt{j} 2 }
     } 
    } 
    e^{-i  \left( \phi_{k}^{ \texttt{j} } - \phi_{ \texttt{j} } \right) } 
    \right) \widetilde{a}_{k}^{ \texttt{j} }
   +
   \left( 1 - \delta{ \texttt{j} } \right) 
   \sqrt{ 
    \frac{ 
     \xi_{ \texttt{j} 1 } \xi_{ \texttt{j} 2 }  
    }{
     D_{ \texttt{j} 1 } D_{ \texttt{j} 2 } 
    } 
   }
   \widetilde{b}_{k}^{ \texttt{j} } \\ \vspace{2mm}
   &   
   - 
   \left( \xi_{ \texttt{j}1 } + \xi_{ \texttt{j} 2 } \right) 
   \sqrt{
    \frac{
     \delta_{ \texttt{j} } \left( 1 - \delta_{ \texttt{j} } \right) 
    }{
     D_{ \texttt{j} 1 } D_{ \texttt{j} 2 } 
    }
   } \widetilde{c}_{k}^{ \texttt{j} }
   + 
   \delta_{ \texttt{j} } 
   \sqrt{ 
    \frac{
     \xi_{ \texttt{j} 1 } \xi_{ \texttt{j} 2 } 
    }{
     D_{ \texttt{j} 1 } D_{ \texttt{j} 2} 
    } }
  \widetilde{d}_{k}^{ \texttt{j} }, \\ \vspace{2mm} 
  \left( \widetilde{\bf \chi}_{k}^{ \texttt{j} } \right)_{ 13 } = &
   \sqrt{ 
    \frac{ \widehat{m}_{ \texttt{j} 2 } }{ \widehat{m}_{ \texttt{j} 1 } }
    \frac{
     \left( 1- \delta_{ \texttt{j} } \right) \delta_{ \texttt{j} } \xi _{ \texttt{j} 1 }
    }{
     D_{ \text{j} 1 } D_{ \texttt{j} 3 }  
    }
   } 
   \left( 
    \widehat{m}_{ \texttt{j} 1} 
    e^{ - i \left( \phi_{k}^{ \texttt{j} } - \phi_{ \texttt{j} } \right) } 
    + e^{ i \left( \phi_{k}^{ \texttt{j} } - \phi_{ \texttt{j} } \right) } 
   \right)
  \widetilde{a}_{k}^{ \texttt{j} }  
  +
  \left( 1 - \delta_{ \texttt{j} } \right)
   \sqrt{
    \frac{ 
     \delta_{ \texttt{j} } \xi_{ \texttt{j} 1 } 
    }{
     D_{ \texttt{j} 1 } D_{ \texttt{j} 3 } 
    }
   } \widetilde{b}_{k}^{ \texttt{j} }  \\ \vspace{2mm}
    & +  
   \left( \xi _{ \texttt{j} 1 } - \delta_{ \texttt{j} } \right)
   \sqrt{
    \frac{
     \left( 1 - \delta_{ \texttt{j} } \right) \xi_{ \texttt{j} 2 } 
    }{
     D_{ \texttt{j} 1} D_{ \texttt{j} 3 } 
    } 
   }  
   \widetilde{c}_{k}^{ \texttt{j} }  
   -
   \xi _{ \texttt{j} 2 } 
   \sqrt{
    \frac{
     \delta_{\texttt{j}} \xi_{ \texttt{j} 1} 
    }{
     D_{ \texttt{j} 1 } D_{ \texttt{j} 3 } 
    }
   }   
   \widetilde{d}_{k}^{ \texttt{j} }  , \\ \vspace{2mm}
  \left( \widetilde{\bf \chi}_{k}^{ \texttt{j} } \right)_{ 22 } = &
   - 2
   \sqrt{ 
    \frac{ 
     \widehat{m}_{ \texttt{j} 1 } 
    }{ 
     \hat{m}_{ \texttt{j} 2 } 
    }
    \left( 1 - \delta_{ \texttt{j} } \right)
   } 
   \frac{ 
    \xi_{ \texttt{j} 2 } 
   }{
    D_{\texttt{j}2  } 
   } 
   \cos \left( \phi_{k}^{ \texttt{j} } - \phi_{ \texttt{j} } \right)  
   \widetilde{a}_{k}^{ \texttt{j} }  
   +
   \frac{  
    \left( 1 - \delta_{ \texttt{j} } \right) \xi_{ \texttt{j} 2 } 
   }{
    D_{ \texttt{j} 2} 
   }
   \widetilde{b}_{k}^{ \texttt{j} }  
   - 2 
   \frac{ 
    \sqrt{ 
     \delta_{ \texttt{j} } \left( 1 - \delta_{ \texttt{j} } \right)
     \xi_{ \texttt{j} 1 } \xi_{ \texttt{j} 2 } 
    }
   }{
    D_{ \texttt{j} 2 }
   }
   \widetilde{c}_{k}^{ \texttt{j} }  
   +
   \frac{ 
    \delta _j \xi_{ \texttt{j} } 
   }{
    D_{ \texttt{j} 2 }
   } 
   \widetilde{d}_{k}^{ \texttt{j} } , \\ \vspace{2mm}
  \left( \widetilde{\bf \chi}_{k}^{ \texttt{j} } \right)_{ 23 } = &
   - \sqrt{ 
    \frac{ \widehat{m}_{ \texttt{j} 1 } }{ \widehat{m}_{ \texttt{j} 2} }
    \frac{ \delta_{ \texttt{j} } \left( 1 - \delta_{ \texttt{j} } \right) 
     \xi_{ \texttt{j} 2 }
    }{
     D_{ \texttt{j} 2 } D_{ \texttt{j} 3 }   
    }
   }
   \left( 
    e^{ i \left( \phi_{k}^{ \texttt{j} } - \phi_{ \texttt{j} } \right) } 
    - \widehat{m}_{ \texttt{j} 2 } 
    e^{ - i \left( \phi_{k}^{ \texttt{j} } - \phi_{ \texttt{j} } \right) } 
   \right)
   \widetilde{a}_{k}^{ \texttt{j} }  
   +
   \left( 1 - \delta_{ \texttt{j} } \right) 
   \sqrt{ 
    \frac{ 
     \delta_{ \texttt{j} } \xi_{ \texttt{j} 2 } 
    }{ 
     D_{ \texttt{j} 2 } D_{ \texttt{j} 3 }
    }
   }
   \widetilde{b}_{k}^{ \texttt{j} }  \\ \vspace{2mm}
   & 
   +
   \left( \xi_{ \texttt{j} 2 } - \delta_{ \texttt{j} } \right) 
   \sqrt{
    \frac{
     \left( 1 - \delta_{ \texttt{j} } \right) 
     \xi_{ \texttt{j} 1 } 
    }{ 
     D_{ \texttt{j} 2 } D_{ \texttt{j} 3 } 
    }
   }
   \widetilde{c}_{k}^{ \texttt{j} } 
   -
   \xi_{ \texttt{j} 1 } 
   \sqrt{
    \frac{
     \delta_{ \texttt{j} } \xi_{ \texttt{j} 2 }
    }{
     D_{ \texttt{j} 2 } D_{ \texttt{j} 3 } 
    }
   } 
   \widetilde{d}_{k}^{ \texttt{j} } , \\ \vspace{2mm}
  \left( \widetilde{\bf \chi}_{k}^{ \texttt{j} } \right)_{ 33 } = &
   2 
   \sqrt{
    \left( 1 - \delta_{ \texttt{j} } \right) 
    \widehat{m}_{ \texttt{j} 1 } \widehat{m}_{ \texttt{j} 2 } 
   } 
   \frac{ \delta_{ \texttt{j} } }{ D_{ \texttt{j} 3 } } 
   \cos \left( \phi_{k}^{ \texttt{j} } - \phi_{ \texttt{j} } \right)
   \widetilde{a}_{k}^{ \texttt{j} }  
   +
   \frac{ 
    \delta_{ \texttt{j} } \left( 1 - \delta_{ \texttt{j} } \right)
   }{
    D_{ \texttt{j} 3 }
   }
   \widetilde{b}_{k}^{ \texttt{j} }  
   +
   2
   \frac{  
    \sqrt{
     \delta_{ \texttt{j} } \left( 1 - \delta_{ \texttt{j} } \right)
     \xi_{ \texttt{j} 1 } \xi_{ \texttt{j} 2 }
    }
   }{
    D_{ \texttt{j} 3 }
   }
   \widetilde{c}_{k}^{ \texttt{j} }  
   +
   \frac{d_{\text{jk}}
   \xi _{\text{j1}} \xi _{\text{j2}}}{D_{\text{j3}}}
   \widetilde{d}_{k}^{ \texttt{j} } ,
   \\ \vspace{2mm}            
 \end{array}
\end{equation}
with
\begin{equation}
 \begin{array}{l}
  \widetilde{a}_{k}^{ \texttt{j} } = \frac{ v }{ m_{ \texttt{j} 3 } }
   \left | A_{k}^{ \texttt{j} } \right|, \quad
  \widetilde{b}_{k}^{ \texttt{j} } = \frac{ v }{ m_{ \texttt{j} 3 } }
   B_{k}^{ \texttt{j} }, \quad 
  \widetilde{c}_{k}^{ \texttt{j} } = \frac{ v }{ m_{ \texttt{j} 3 } }
   C_{k}^{ \texttt{j} }, 
  \quad \textrm{and} \quad   
  \widetilde{d}_{k}^{ \texttt{j} } = \frac{ v }{ m_{ \texttt{j} 3 } }
   D_{k}^{ \texttt{j} }.     
 \end{array}
\end{equation}
%
%
\section{Mixing Matrix}\label{App:C}
%
%
The lepton flavor mixing matrix is 
\begin{equation}
 \begin{array}{l}
  {\bf U}_{\mathrm{PMNS}} = {\bf U}_{l}^{\dagger} {\bf U}_{\nu} =
   {\bf O}_{l}^{\top} 
   {\bf P}_{l}^{ \dagger }
   {\bf U}_{\mathrm{s}}^{ \dagger }
  {\bf U}_{\mathrm{s}} {\bf P}_{\nu} 
  {\bf O}_{\nu}^{\mathrm{n[i]}}
  = {\bf O}_{l}^{\top} {\bf P}^{ ( \nu - l ) } {\bf O}_{\nu}^{\mathrm{n[i]}}.
 \end{array}
\end{equation}
The explicit form of entries of previous matrix are:
\begin{equation}
 \begin{array}{l} \vspace{2mm}
  \left( {\bf U}_{\mathrm{PMNS}} \right)_{11} = 
   \sqrt{ 
    \frac{ 
     \hat{m}_{\mu} \hat{m}_{\nu 2[1]} \xi_{l1} \xi _{\nu 1[3]} 
    }{ 
     D_{l1} D_{ \nu 1[3] } 
    } 
   }
   +
   \sqrt{ 
    \frac{ 
     \hat{m}_{e} \hat{m}_{\nu 1[3]} 
    }{ 
     D_{l1} D_{ \nu 1[3] } 
    } 
   } 
   \left( 
    \sqrt{ 
     \left( 1 - \delta_{\nu} \right) \left( 1 - \delta_{l} \right) 
     \xi_{l1} \xi _{\nu 1[3]}  } e^{ i \phi_{ l1 } }
    +  
    \sqrt{ 
     \delta_{\nu} \delta_{l} \xi_{l2} \xi_{\nu 2[1] } } 
     e^{ i \phi_{ l2 } } 
  \right) , \\ \vspace{2mm}
  \left( {\bf U}_{\mathrm{PMNS}} \right)_{12} = 
   - \sqrt{ 
    \frac{ 
     \hat{m}_{\mu} \hat{m}_{\nu 1[3]} \xi_{l1} \xi _{\nu 2[1] } 
    }{ 
     D_{l1} D_{\nu 2[1]} 
    } 
   }
   + 
   \sqrt{ 
    \frac{ 
     \hat{m}_{e} \hat{m}_{\nu 2[1]} 
    }{ 
     D_{l1} D_{\nu 2[1]} 
    } 
   } 
   \left( 
    \sqrt{ 
     \left( 1 - \delta_{\nu} \right) \left( 1 - \delta_{l} \right) \xi_{l1}  
     \xi _{\nu 2[1] } } e^{ i \phi_{ l1 } } 
    + 
    \sqrt{ 
     \delta_{\nu} \delta_{l} \xi_{l2} \xi _{\nu 1[3]} } e^{ i \phi_{ l2 } } 
   \right) , \\ \vspace{2mm}
  \left( {\bf U}_{\mathrm{PMNS}} \right)_{13} = 
   \sqrt{ 
    \frac{ 
     \widehat{m}_{\mu} \widehat{m}_{\nu 1[3]} \widehat{m}_{\nu 2[1]} 
     \delta_{\nu} \xi_{l1} 
    }{ 
     D_{l1} D_{\nu 3[2]} 
    }
   }
   +
   \sqrt{ 
    \frac{ 
     \widehat{m}_{e} 
    }{ 
     D_{l1} D_{\nu 3[2]} 
    } 
   } 
   \left( 
    \sqrt{ 
     \left( 1 -\delta_{\nu} \right) \delta_{\nu} \left( 1 - \delta_{l} \right) 
      \xi_{l1} } e^{ i \phi_{ l1 } } 
      - 
      \sqrt{ \delta_{l} \xi_{l2} \xi_{\nu 1[3]} \xi _{\nu 2[1] } } 
      e^{ i \phi_{ l2 } }  
     \right) , \\ \vspace{2mm} 
  \left( {\bf U}_{\mathrm{PMNS}} \right)_{21} = 
   - \sqrt{ 
    \frac{ 
     \widehat{m}_{e} \widehat{m}_{\nu 2[1]} \xi_{l2} \xi _{\nu 1[3]} 
    }{ 
     D_{l2} D_{ \nu 1[3] } 
    } 
   }
   +
   \sqrt{ 
    \frac{ 
     \widehat{m}_{\mu} \widehat{m}_{\nu 1[3]} 
    }{ 
     D_{l2} D_{ \nu 1[3] } 
    } 
   } 
   \left( 
    \sqrt{ \left( 1 - \delta_{\nu} \right) \left( 1 - \delta_{l} \right) \xi_{l2} 
    \xi _{\nu 1[3]} } e^{ i \phi_{l1} } 
    +
    \sqrt{ \delta_{\nu} \delta_{l} \xi_{l1} \xi _{\nu 2[1] } } 
    e^{ i \phi_{ l2 } }
   \right), \\ \vspace{2mm}
  \left( {\bf U}_{\mathrm{PMNS}} \right)_{22} = 
   \sqrt{ 
    \frac{ 
     \widetilde{m}_{e} \widetilde{m}_{\nu 1 [3]} 
     \xi_{l2} \xi_{\nu 2 [1]}
    }{ 
     D_{l2} D_{ \nu 2[1] } 
    } 
   }  
   + 
   \sqrt{ 
    \frac{ 
     \widetilde{m}_{\mu} \widetilde{m}_{\nu 2 [1]} 
    }{ 
     D_{l2} \, D_{\nu 2 [1]} 
    } 
   } 
   \left( 
    \sqrt{ \left( 1 - \delta_{\nu} \right) \left( 1 - \delta_{l} \right) 
    \xi_{l2} \xi_{\nu 2 [1]} }  e^{i \phi_{l1} } 
    + 
    \sqrt{ \delta_{\nu} \delta_{l} \xi_{l1} \xi_{\nu 1 [3]} } 
    e^{i \phi_{ l2 } } 
   \right), \\ \vspace{2mm}
  \left( {\bf U}_{\mathrm{PMNS}} \right)_{23} = 
   - \sqrt{ 
    \frac{ 
     \widetilde{m}_{e} \widetilde{m}_{\nu 1[3]} \widetilde{m}_{\nu 2[1]} \delta_{\nu} 
     \xi_{l2} 
    }{ 
     D_{l2} D_{\nu 3[2]} 
    } 
   } 
   + 
   \sqrt{ 
    \frac{ 
     \widetilde{m}_{\mu} 
    }{ 
     D_{l2} D_{\nu 3[2]} 
    } 
   } 
   \left( 
    \sqrt{ \delta_{\nu} \left( 1 - \delta_{\nu } \right)  
    \left( 1 - \delta_{l} \right) \xi_{l2} } e^{i \phi_{l1} }  
    - 
    \sqrt{ \delta_{l} \xi_{l1} \xi_{\nu 1[3]} \xi_{\nu 2[1]} } 
    e^{i \phi_{ l2 } } 
   \right) , \\ \vspace{2mm}
  \left( {\bf U}_{\mathrm{PMNS}} \right)_{31} = 
   \sqrt{ 
    \frac{ 
     \widetilde{m}_{e} \widetilde{m}_{\mu} \widetilde{m}_{\nu 2[1]} \delta_{l} 
     \xi_{\nu 1[3]} 
    }{ 
     D_{l3} D_{\nu 1[3]}
    } 
   } 
   + 
   \sqrt{ 
    \frac{ 
     \widetilde{m}_{\nu 1[3]} 
    }{ 
     D_{l3} D_{\nu 1[3]} 
    }
   } 
   \left( 
    \sqrt{ \delta_{l} \left( 1 - \delta_{\nu } \right)  
    \left( 1 - \delta_{l} \right) \xi_{\nu 1[3]} } e^{i \phi_{l1} }
    - 
    \sqrt{ \delta_{\nu} \, \xi_{l1} \, \xi_{l2} \, \xi_{\nu 2[1]} } 
    e^{i \phi_{ l2 } } 
   \right) ,\\ \vspace{2mm}
  \left( {\bf U}_{\mathrm{PMNS}} \right)_{32} = 
   - 
   \sqrt{ 
    \frac{ 
     \widetilde{m}_{e} \widetilde{m}_{\mu} \widetilde{m}_{\nu 1[3]} \delta_{l} 
     \xi_{\nu 2[1]} 
    }{ 
     D_{l3} D_{\nu 2[1]} 
    } 
   } 
   + 
   \sqrt{ 
    \frac{ 
     \widetilde{m}_{\nu 2[1]} 
    }{ 
     D_{l3} D_{\nu 2[1]} 
    } 
   } 
   \left( 
    \sqrt{ \delta_{l} \left( 1 - \delta_{\nu } \right)  
    \left( 1 - \delta_{l} \right) \xi_{\nu 2[1]} } e^{i \phi_{l1} } 
    - 
    \sqrt{ \delta_{\nu} \xi_{l1} \xi_{l2} \xi_{\nu 1[3]} } 
    e^{i \phi_{ l2 } } 
   \right) , \\ \vspace{2mm}
  \left( {\bf U}_{\mathrm{PMNS}} \right)_{33} =  
   \sqrt{ 
    \frac{ 
     \widetilde{m}_{e} \widetilde{m}_{\mu} \widetilde{m}_{\nu 1[3]} 
     \widetilde{m}_{\nu 2[1]} \delta_{l} \delta_{\nu} 
    }{ 
     D_{l3} \, D_{\nu 3[2]} 
    } 
   } 
   + 
   \sqrt{ \frac{ 1 }{ D_{l3} \, D_{\nu 3[2]} } } 
   \left( 
    \sqrt{ \delta_{l} \delta_{\nu } \left( 1 - \delta_{\nu } \right) 
    \left( 1 - \delta_{l} \right) } e^{i \phi_{l1} } 
    - \sqrt{ \xi_{l1} \xi_{l2} \xi_{\nu 1[3]} \xi_{\nu 2[1]} } 
    e^{i \phi_{ l2 } }
   \right) .    
 \end{array}
\end{equation}

\begin{acknowledgments}
This work has been partially supported by \textit{CONACYT-SNI (M\'exico)}. 
The authors thankfully acknowledge the computer resources, technical expertise and support 
provided by the Laboratorio Nacional de Superc\'omputo del Sureste de M\'exico 
CONACYT network of national laboratories."
FGC acknowledges the financial support from {\it CONACYT} and {\it PRODEP} under Grant 
No.~511-6/17-8017.
\end{acknowledgments}


\end{document}